\documentclass[prd,showpacs,aps,floatfix] {revtex4}
\input epsf

\makeatletter
\fussy
\def \with respect to { with respect to }

\def\L {\Lambda }
\def\l {\lambda } 
 
\def \t {\theta }
\def \vt {\vartheta }
\def \T {\Theta }

\def\a {\alpha }
\def\dh {\partial }
\def \d {\delta }
\def \D {\Delta }

\def \g {\gamma }
\def \G {\Gamma }
\def \O {\Omega }
\def \o {\omega }
\def \b {\beta }

\def \s {\sigma }
\def \e {\epsilon }

\def \ud { {1 \over 2} }

\def \cddd { {\cal D } }
\def \cala { {\cal A } }
\def \calc { {\cal C } }
\def \calp { {\cal P } }
\def \caly { {\cal Y } }
\def \calf { {\cal F } }

\def \caln { {\cal N } }
\def \call { {\cal L } }

\def \calT { {\cal T}  }

\def \cali { {\cal I } }
\def \calj { {\cal J } }

\def \calr { {\cal R } }
\def \cals { {\cal S } }

\def \cals { {\cal S } }

\def \Eslash {E \kern-.5em\slash}
\def \pslash {p \kern-.5em\slash}
\def \kslash {k \kern-.5em\slash}
\def \Dslash {D \kern-.5em\slash}
\def \hslash {h \kern-.5em\slash}
\def \dslash {\partial \kern-.5em\slash}
\def \vslash {v \kern-.5em\slash}

\def \bfA { {\bf A} }
\def \bfB { {\bf B} }
\def \bfS { {\bf S} }

\def\NPB#1#2#3 {{\rm Nucl.~Phys.}  {\bf{B#1}}, {#3} (#2)}
\def\NPA#1#2#3 {{\rm Nucl.~Phys.}  {\bf{A#1}}, {#3} (#2)}
\def\PLB#1#2#3 {{\rm Phys.~Lett.}  {\bf{B#1}}, {#3} (#2)}
\def\PR#1#2#3 {{\rm Phys.~Rep.}  {\bf#1}, {#3} (#2)} 
\def\PRD#1#2#3 {{\rm Phys.~Rev.}  {\bf{D#1}}, {#3} (#2)} 
\def\PRL#1#2#3 {{\rm Phys.~Rev.~Lett.}  {\bf{#1}}, {#3} (#2)} 
\def\ZPC#1#2#3 {{\rm Z.~Phys.}  {\bf C#1}, {#3} (#2)} 
\def\JHEP#1#2#3 {{\rm JHEP} {\bf C#1}, {#3}  (#2)}     
\def\IJMP#1#2#3 {{\rm Int. J. Mod. Phys.}  {\bf A#1}, {#3} (#2)} 
\def\JMP#1#2#3 {{\rm J. Math. Phys.}  {\bf #1}, {#3} (#2)}
\def\PTP#1#2#3 {{\rm Prog. Theor. Phys.}  {\bf{#1}}, {#3} (#2)}

\@addtoreset{equation}{section}

\newcommand{\be}{\begin{equation}} 
\newcommand{\ee}{\end{equation}} 
\newcommand{\ba}{\begin{array}}
\newcommand{\ea}{\end{array}}
\newcommand{\bea}{\begin{eqnarray}} 
\newcommand{\eea}{\end{eqnarray}} 
\newcommand{\bsea}{\begin{subeqnarray}} 
\newcommand{\esea}{\end{subeqnarray}}

\begin{document} 
\title{Nucleon decay in gauge unified models with intersecting
$D6$-branes} 

\author{M. Chemtob} \email{marc.chemtob@cea.fr}
\affiliation{ Service de Physique Th\'eorique, CEA-Saclay F-91191
Gif-sur-Yvette Cedex FRANCE } \thanks {\it Laboratoire de la Direction
des Sciences de la Mati\`ere du Commissariat \`a l'Energie Atomique }
\date{\today}



\begin{abstract}
Baryon number violation is discussed in gauge unified orbifold models
of type II string theory with intersecting Dirichlet branes.  We
consider setups of $D6$-branes which extend along the flat Minkowski
space-time directions and wrap around 3-cycles of the internal 6-d
manifold. Our study is motivated by the enhancement effect of
low energy amplitudes anticipated for M-theory and type $II$ string
theory models with matter modes localized at points of the internal
manifold.  The conformal field theory formalism is used to evaluate
the open string amplitudes at tree level.  We study the single baryon
number violating processes of dimension $6 $ and $ 5$, involving four
quarks and leptons and in supersymmetry models, two pairs of matter
fermions and superpartner sfermions.
The higher order processes associated with the baryon number violating
operators of dimension $7$ and $9$ are also examined, but in a
qualitative way. We discuss the low energy representation of string theory amplitudes
in terms of infinite series of poles associated to exchange of string
Regge resonance and compactification modes.  The comparison of string
amplitudes with the equivalent field theory amplitudes is first
studied in the large compactification radius limit.  Proceeding next
to the finite compactification radius case, we present a numerical
study of the ratio of string to field theory amplitudes based on
semi-realistic gauge unified non-supersymmetric and supersymmetric
models employing the $Z_3 $ and $Z_2\times Z_2$ orbifolds.  We find a
moderate enhancement of string amplitudes which becomes manifest in
the regime where the gauge symmetry breaking mass parameter exceeds the
compactification mass parameter, corresponding to a gauge unification
in a seven dimensional space-time.
\end{abstract}

\maketitle

\section{INTRODUCTION}
\label{sectintro}

By suggesting the possibility that the Standard Model (SM) gauge and
gravitational interactions unify in a higher dimensional space-time,
string theory~\cite{witten85,hetunif} has opened up novel perspectives
for the gauge unification proposal~\cite{gsw}.  Based on the approach
of Calabi-Yau (CY) manifold compactification, the string theory
applications focused initially on the heterotic string in the regimes
of weak coupling~\cite{arno8594} and strong
coupling~\cite{Mtheory96,witten96,kapcac97,ovrut99}.  These studies
were soon followed by explicit constructions using orbifold
compactification~\cite{aldaz95,kakush97} and free
fermion~\cite{ellisfar97} models of the heterotic string, which were
later extended to orbifold compactification models of the type $II$
string theories with single $Dp$-branes~\cite{lykken98} or the $
Dp/Dp' , \ [p'\ne p ] $ branes-inside-branes type
backgrounds~\cite{tyegut98}.

 The chief distinctive features of string gauge unification reside in
the discrete gauge symmetry breaking scheme by Wilson flux
lines~\cite{ovrut85} and the heavy string threshold
effects~\cite{kap88}.  Regarding, however, the issue of matter
instability caused by baryon number violation, it is fair to say that
no specific stringy effect has emerged from the earlier studies using
the weakly coupled heterotic and type $II$ string theories.  A
different situation seems to take place in the 11-d M-theory
supersymmetric compactification on 7-d internal manifolds $X_7$ of
$G_2$ holonomy~\cite{achar04}, as discussed by Friedman and
Witten~\cite{friedman02}.  The non-Abelian gauge symmetries in these
models are supported on 7-d sub-manifolds, $ B= M_4 \times Q $, loci
for $ R^4/\G $ orbifold type singularities in the directions
transverse to $B$, while the chiral massless fermions are supported at
isolated singularities of the 3-d sub-manifolds, $Q$.  This causes the
existence of a natural regime where the grand unification occurs in a
7-d dimensional space-time with localized matter fermions.  In
comparison to the nucleon decay amplitudes of the equivalent unified
gauge field theories, the string amplitudes are enhanced by a power $
\a _X ^{- {1\over 3} } $ of the unified gauge coupling constant, which
reflects on the short distance singularities from summing over the
momentum modes propagating in the sub-manifold $Q$.  As to the size of
the enhancement effect, however, no definite conclusions could be
drawn because of the poor understanding of M-theory perturbation
theory, not mentioning the greater complexity of $G_2$ holonomy
manifolds~\cite{revachar04}.

Fortunately, it is possible to examine the enhancement effect in a
controlled way by considering the weak coupling dual models based on
type $II a $ string theory orientifold compactification on $ M_4
\times X _6 $ with $D6$-branes wrapped around intersecting
three-cycles of the internal Calabi-Yau complex threefold, $X_6$.
Using a toy model realizing $SU(5)$ gauge unification, Klebanov and
Witten~\cite{KW03} showed that the four fermion string amplitude for
localized matter modes in the gauge group representations, $10 \cdot
{10} ^\dagger \cdot 10 \cdot {10} ^\dagger $, featured a power
dependence on the unified gauge coupling constant of same form as in
the M-theory model, namely, $ \cala \propto \a _X ^{2/3}$.  In the
most favorable case where the fermion modes sit at coincident
intersection points of the $D$-branes, the enhancement due to the
gauge coupling constant dependence was found to be offset by certain
constant factors which resulted in a rather modest net effect.

Our goal in the present work is to document the enhancement of string
amplitudes anticipated in M-theory~\cite{friedman02} by developing
further semiquantitative calculations in models with intersecting
$D$-branes.  The initiating discussion in Ref.~\cite{KW03} made use of
the large compactification radius limit in which predictions are
largely insensitive to the Wilson line mechanism responsible for the
unified gauge symmetry breaking.  We extend this study in three main
directions.  First, we consider semi-realistic orbifold-orientifold
models realizing the $SU(5)$ gauge unification with a chiral spectrum
of massless matter modes that closely reproduces the Standard Model
spectrum.  Second, we evaluate the nucleon decay four point amplitudes
in the two independent configurations of gauge group matter
representations, $10 \cdot {10} ^\dagger \cdot 10 \cdot {10} ^\dagger
$ and $ 10 \cdot \bar 5 \cdot 10 ^\dagger \cdot \bar 5 ^\dagger $.
which contribute to the proton two-body decay channels with emission
of left and right helicity positrons, $p\to \pi ^0+e^+_ {L,R} $.  This
allows us to quantitatively assess
the M-theory prediction for the ordering of partial decay rates, $ \G
(p\to \pi ^0+e^+_L ) >> \G (p\to \pi ^0+e^+_R) $.  Third, we discuss
the string amplitudes at finite compactification radii, in order to
weigh the importance of the string momentum and winding excitations
relative to the string Regge resonances.
Studying the interplay between the compactification and unified
symmetry breaking mass scale parameters, $ M_c$ and $ M_X$, proves
crucial in assessing the size of the enhancement effect.

The contributions to nucleon decay processes from physics at high
energy scales are conveniently represented by non-renormalizable local
operators in the quark and lepton fields~\cite{weinbergs,wilczee}
which violate the baryon and lepton numbers, $ B,\ L$.  In unified
gauge theories, the exchange of massive gauge bosons with leptoquark
quantum numbers induces in the effective action, $(B+L)$ violating,
$(B-L)$ conserving operators of dimension $\cddd =6 $.  In
supersymmetry models, the exchange of massive color triplet matter
higgsino like modes also induces dangerous operators in the quark and
lepton superfields of dimension $\cddd =5 $.  Higher dimension
operators initiating the exotic nucleon decay processes can possibly
arise from mass scales significantly lower than the gauge interactions
unification scale, $ M_X \simeq 3.\times 10^{16} $ GeV.  Of special
interest are the $B-L$ violating operators of dimension $\cddd \geq 7
$, and the double baryon number violating operators of dimension
$\cddd \geq 9 $, responsible for nucleon-antinucleon oscillation.

The 2-d conformal quantum field theory~\cite{polchb} provides a
powerful approach for calculating the on-shell string theory
amplitudes.  The construction of string amplitudes~\cite{cftopen} from
vacuum correlators of open string vertex operators inserted on the
world sheet disk boundary is readily applied to the 4-d amplitudes
describing the nucleon decay processes.  The tree level contributions
to the dimension $\cddd =6 , \ 5 $ operators involve four fermions, $
\psi ^4 $, or two pairs of fermions and bosons, $ \psi ^2 \phi ^2$.
For the intersecting brane models with matter modes localized in the
internal directions, the calculations are made non-trivial by the need
to account for the twisted like boundary conditions of the world sheet
fields.  Fortunately, the energy source approach, which was initially
invented~\cite{bersh87,dixon87,dixontasi,cohn86,zam87} and
subsequently developed~\cite{burwick,orbifhet,bailin93} in the context
of closed string twisted sectors of orbifolds, can be readily extended
to the localized open string sectors of intersecting brane models.
Following previous studies devoted to the discussion of open string
modes with mixed ND boundary conditions in $Dp/Dp'$-brane
models~\cite{hashi96,gava97,fro099,antbengier01}, the implementation
of this approach for intersecting brane models has been clarified in
discussions by Cvetic and Papadimitriou~\cite{Cvetic:2003ch}, Abel and
Owen~\cite{owen03}, Klebanov and Witten~\cite{KW03}, Jones and
Tye~\cite{jones03}, L\"ust et al.,~\cite{luststie04} and Antoniadis
and Tuckmantel~\cite{antontel04}.  A comprehensive review of the first
quantization and conformal field theory formalisms for the open string
string sectors of $D$-brane models is currently under
preparation~\cite{chemnd05}.  We should also mention here the studies by Billo et
al.,~\cite{billo02}, Bertolini et al.,~\cite{bertol05}, and Russo and
Sciuto,~\cite{russo07} which develop the alternative approach based on
the operator formalism.


An intense activity has been devoted in recent years to the discussion
type $ II $ string theory compactification using intersecting
$D$-brane backgrounds~\cite{berk96,balasub97,arfaei97}.  To be
complete, we should mention the parallel development on T-dual
$D$-brane models using magnetised
backgrounds~\cite{bachas95,antonti00} and on the magnetic field
deformations of the heterotic string~\cite{witten84}.  Important
efforts towards building satisfactory models have been spent in works
by Aldazabal et
al.,~\cite{aldaisb01,iban0,iban1,ibanyuk,iban2,cimmag04},
Kokorelis~\cite{kokosusy02,kokogut02,kokor02}, Blumenhagen et
al.,~\cite{Blumenhagen:1999md,Blumenhagen:1999db,Blumenhagen:1999ev,Blumenhagen:2000ea,Blumenhagen:2000fp,Blumenhagen:2000wh,Blumenhagen:2001te,Blumenhagen:2001mb},
and Cvetic et
al.,~\cite{Cvetic:2001nr,Cvetic:2001tj,Cvetic:2002pj,Cvetic:2003xs,Cvetic:2004ui,Cvetic:2005lll,cvet06}.
A useful review is presented in Ref.~\cite{blumshiu05}.  To develop
our applications in the present work, we consider semi-realistic
models already discussed in the literature.  First, we should note
that the minimal toroidal orientifold models developed by Cremades et
al.,~\cite{iban1,ibanyuk,iban2} are of little use to us in the present
work, because these realize direct compactifications to the Standard
Model (SM), the minimal supersymmetric Standard Model
(MSSM)~\cite{iban1,ibanyuk,iban2}, or the Pati-Salam type unified
model~\cite{kokogut02}, which all accommodate a built-in $ U(1)_{B +L}
$ global symmetry which guarantees $B$ and $ L $ number conservation.
A better answer to our needs is provided by the orbifold
models~\cite{Blumenhagen:1999md,Blumenhagen:1999db,Blumenhagen:2001te,Blumenhagen:2001mb,Cvetic:2001nr,Cvetic:2001tj,Cvetic:2002pj,Cvetic:2003xs,Cvetic:2004ui,honeck01,honeck02,ellis03,blumz402,blumlust02,blumlonsu04,honec03,honec04,axeni03,kokor04}
owing to their richer structure and more constrained character.  We
have selected the two classes of non-supersymmetric and supersymmetric
gauge unified models realizing a minimal type $SU(5)$ gauge
unification.  The first class relates to the works of Blumenhagen et
al.,~\cite{Blumenhagen:2001te} and Ellis et al.,~\cite{ellis03} using
the $Z_3$ orbifold, and the second class to the work of Cvetic et
al.,~\cite{Cvetic:2001nr} using the $Z_2\times Z_2 $ orbifold.  It is
important to emphasize at this point that the calculation of string
amplitudes relies on features of the non-chiral mass spectrum and the
wave functions of low-lying modes which are not directly addressed in
these works.  For instance, the application to $Z_3 $ orbifold models
rely on data involving the real wrapping numbers rather than the
effective ones which are defined by summing over the orbifold group
orbits.

The discussion in the present paper is organized into five sections.
In Section~\ref{sect1}, we review the first quantization and conformal
field theory formalisms for type $II$ orientifold models with
intersecting $D6$-branes. The open string sectors are discussed first
for tori and next for orbifolds.  A review of the energy source
approach for calculating the correlators of coordinate twist field
operators is provided in Appendix~\ref{apptwist}. In
Section~\ref{sect2}, we discuss the calculation of string amplitudes
for nucleon decay processes in the gauge unified models.  The world
sheet disk amplitudes are considered in succession for the processes
involving four massless fermions, two pairs of massless fermions and
bosons, four massless fermions and a scalar, and six massless
fermions.  These are described by operators of dimension, $ \cddd = 6
$ and $ 5$, $ \cddd = 7, $ and $ \cddd = 9$, obeying the selection
rules, $\D B= \D L =-1$, $\D B=- \D L =-1 $ and $ \D B=-2 ,\ \D L =0
$.
In Section~\ref{sect3}, we discuss the relations linking the low
energy gauge and gravitational interaction coupling constants to the
fundamental string theory parameters $g_s $ and $ m_s$, and to the
mass parameters representing the average compactification radius of
wrapped three-cycles, $ r = 1/M_c$, and the unified gauge symmetry
breaking, $M_X$, arising as an infrared mass cutoff.  We consider in
turn two distinct regularization procedures of the string
amplitudes. The first uses the subtraction prescription replacing the
massless pole terms by massive ones, and the second uses the
displacement prescription splitting the unified $D$-brane into
separated stacks.  To illustrate the dependence of four fermion string
amplitudes on the $D$-branes intersection angles, an initial numerical
application is considered within the large compactification radius
limit.  However, the main thrust of the present work bears on the
study of baryon number violating string amplitudes at finite
compactification radius.  The results illustrating the enhancement of
nucleon decay string amplitudes are presented in Section~\ref{sect4}.
We consider first the non-supersymmetric $SU(5)$ unified models of the
$ Z_3 $ orbifold due to Blumenhagen et al.,~\cite{Blumenhagen:2001te}
and Ellis et al.,~\cite{ellis03} and next the supersymmetric $SU(5)$
unified model of the $Z_2\times Z_2$ orbifold due to Cvetic et
al.,~\cite{Cvetic:2001nr}.  In Section~\ref{secconcl}, we summarize
our main conclusions.  For completeness, we provide a brief review of
the $ Z_3 $
orbifold models in Appendix~\ref{appz3} and a brief review of baryon
number violating processes for gauge unified theories in
Appendix~\ref{bnvnd}.

\section{Review   of type $II$  string  orientifolds in intersecting
$D6$-brane backgrounds}
\label{sect1}

We present in this section a brief review of the open string sector of
type $II$ string theory orientifolds with intersecting $D6$-branes.
The first-quantized and vertex operators formalisms are discussed for
toroidal compactification in Subsections~\ref{subsect11} and
\ref{sub12} and for orbifold compactification in
Subsection~\ref{sub13}.


\subsection{Toroidal compactification} 
\label{subsect11}

\subsubsection{World sheet  field theory}

The Neveu-Schwarz-Ramond type $ II a $ string theory deals with 2-d
world sheet Riemann surfaces on which live the coordinate and
Majorana-Weyl 2-d spinor fields parameterizing the 10-d target
space-time with signature $(-+ \cdots +) $.  For a flat space-time,
one uses the diagonal representation of the metric tensor, $ - \eta
_{00} = \eta _{\mu \mu } = \eta _{ mm } =1$, associated to the
orthogonal field basis, $ X ^M ,\ \psi ^M ,\ [M=(\mu , m) =0, 1 \cdots
, 9, \ \mu =0, 1,2, 3 ; \ m=4, 5,6,7,8, 9 ] $ with the complexified
basis of coordinate fields, $ X ^A ,\ \bar X ^A, \ [A = 0, 1,\cdots ,
4]$ defined by \bea && X ^M = [X ^\mu ; \ X ^m ] = [X ^\mu ; \ X ^A,
\bar X^{A} ], \ [A= (a, 0)= (I, 4, 0) ,\
I =1,2,3 ] \cr && \bigg [ X ^ \pm \equiv (X ^ {0} , X ^{\bar 0} ) = {
X ^{0} \pm X ^1 \over \sqrt 2 } ,\ (X ^ {4}, X ^ {\bar 4}) = {X ^2 \pm
i X ^3 \over \sqrt 2 } ;\cr && (X ^{I } ,\ \bar X ^{I } ) \equiv { X
^I_1\pm i X ^I_2 \over \sqrt 2 } = { X ^{2I +2} \pm i X ^{2I +3} \over
\sqrt 2 } \bigg ] , \eea with similar linear combinations for the
complexified basis of spinor fields, $\psi ^{A} ,\ \bar \psi ^{A } $.
We consider the orientifold toroidal compactification on $ M_4\times
T^6/ (\O \calr (-1)^ {F_L} ) $, with factorisable 6-d tori, $ T^6 =
\prod _{I=1} ^3 T^2_I $, symmetric under the product of parities
associated with the world sheet orientation twist, $\O $, the
antiholomorphic space reflection, $\calr ,$ and the left sector
space-time fermion number, $F_L$.  For definiteness, we choose the
orientifold reflection about the real axes of the three complex
planes, $ \calr = \calr _y : \ X^I \to \bar X ^I $.  The $ T^2_I$ tori
may also be parameterized by the pairs of hatted real lattice
coordinates, $ \hat X _a ^I =(\hat X ^I_1, \hat X ^I_2) , \ [a =1,2] $
of periodicities $2\pi $ each, $ \hat X _a ^I = \hat X _a ^I + 2\pi $.
Except in special instances where the dependence on $\a ' $ is made
explicit, we generally use units for the string length scale such
that, $\a ' =1$.

We focus our discussion on the open string perturbation theory tree
level with the world sheet surface given by the disk.  The complex
plane and strip parameterizations of the disk are described by the
variables, $z \in C_+ , \ \bar z \in C_- ,$ and $\s \in [0, \pi] ,\ t
\in [-\infty , + \infty ] $, related as, $z \equiv e ^{iw}= e ^{i\s
+t} , \ \bar z \equiv e ^{-i \bar w}= e ^{-i\s +t} $, with the
derivatives replaced as, $\dh _\s \to i (\dh -\bar \dh ) ,\ \dh _t \to
- (\dh +\bar \dh ) $.  The string equations of motion are solved by
writing the coordinate and spinor fields in terms of independent
holomorphic and antiholomorphic (left and right moving) functions \bea
X^M(z,\bar z) = X^M(z) + \tilde X^M(\bar z)
, \ \psi ^M (z,\bar z) = (\psi ^M (z),\ \tilde \psi ^M (\bar z) )^T
, \eea living in the upper and lower halves of the complex plane.  The
complexified spinor fields are also equivalently represented by the
complex boson fields, $ H^{A} (z), \ \tilde H ^{A} (\bar z), [A= (a,
0) = (1,2,3,4,0) ] $ through the exponential map, $ [\psi ^ A (z) ,
\bar \psi ^ A (z) ] \sim e ^{ \pm i H^ A (z) } $ and $ [\tilde \psi ^
{A } (\bar z) , \bar {\tilde \psi }^{A} (\bar z) ] \sim e ^{ \pm i
\tilde H ^{A} (\bar z) } $.

The coordinate fields obey one of the two possible boundary conditions
at the open string end points, $ \s = (0, \pi ) $: Neumann or free
type, $\dh _\s X ^ {M} = 0$, and Dirichlet or fixed type, $\dh _{ t }
X ^ {M} =0$.  The Neumann and Dirichlet boundary conditions
(abbreviated as N and D) for the coordinate and the spinor fields read
in the complex plane formalism as, $ (\dh - \bar \dh ) X ^{M} =0,\
(\psi ^{M} \mp \tilde \psi ^{M} ) = 0 ,$ and $ (\dh + \bar \dh ) X
^{M} =0,\ (\psi ^{M} \pm \tilde \psi ^{M } ) = 0 $, with the upper and
lower signs corresponding to the R and NS sectors.  In the doubling
trick representation of the open string sector, one joins together the
holomorphic and antiholomorphic coordinate and spinor fields into
single holomorphic fields, $ X ^M (z) ,\ \psi ^M (z) $, living in the
full complex plane $C$, by means of boundary conditions along the real
axis, $ x = \Re (z) = \Re (\bar z) \in R $.  These are implemented by
having the holomorphic fields coincide with the left moving fields $
X^M(z) , \ \psi ^M(z) $ on $C_+$ and identifying them with the right
moving fields, up to a sign change for D directions, on $C _-$.
Specifically, $ X^M (z) = \pm \tilde X ^M (z) , \ \psi ^M(z) = \pm
\tilde \psi ^M(z) , \ [z \in C_- ]$ where the upper and lower signs
are in correspondence with N and D boundary conditions.  The
holomorphic superconformal generators are defined in the lower half
complex plane as, $ T(\bar z) = \tilde T(\bar z) ,\ T _F(\bar z) =
\tilde T _F(\bar z) .$ The boundary conditions on the coordinate and
spinor fields exactly combine to leave invariant the linear
combination of supersymmetry generators, $ T_F (z) + \tilde T_F(\bar
z) $, corresponding to the 2-d supersymmetry transformation, $\d X =i
(\e \tilde \psi - \tilde \e \psi ),\ \d \psi = \e \dh X ,\ \d \tilde
\psi = - \tilde \e \bar \dh X , $ with $ \e = -\tilde \e $.  For
completeness, we note that in the presence of constant fluxes of NSNS
and magnetic two-form fields, $ B _{MN} $ and $ F _{MN } $, the
doubling prescription for N
directions~\cite{callan87,bachpor92,seib99} with the constant metric
tensor, $ G_{MN}$, takes the matrix notation form, $ X^M (\bar z) =
(D^{-1} ) _{MN} \tilde X^N (\bar z) ,\ [D = - G^{-1} + 2 (G + \calf )
^{-1} ,\ \calf _{MN} = B _{MN} + 2\pi \a ' F _{MN } ,\ \bar z \in C
_-] .  $

\subsubsection{Geometric data} 

We consider general non-orthogonal (oblique) 2-d tori symmetric under
the reflection $\calr _y $. For definiteness, we choose the upwards
tilted $ T^2 _I$ tori with lattice generated by the pairs of complex
cycles or vielbein vectors, $ e_1^I = r_1^I e ^{i ( {\pi \over 2} - \a
^I ) } / \sin \a ^I , \ e_2 ^I= i r_2^I$, where $\a ^I$ denote the
opening angles and $ r_1 ^{I} ,\ r_2 ^{I}$ the projections of the
cycles radii along the real and imaginary axes of the complex planes.
The diagonal complex structure and K\"ahler moduli parameters of the
$T^2_I$ tori are then expressed as \bea && U^I\equiv U_1 ^I + i U_2 ^I
= -{ e_1 ^I \over e_2^I} = \hat b ^I + i { r_1 ^I \over r_2 ^I },\cr
&& T^I \equiv T_1 ^I + i T_2 ^I = B ^I _{12} + i \sqrt { Det G ^I } =
b^I + i r_1^I r_2 ^I , \ [ {r_1 ^{I} \over r_2 ^I } \cot \a ^I \equiv
- \hat b ^I ] \eea where the $\O \calr $ symmetry restricts the 2-d
tori tilt parameters to the two discrete values, $ \hat b ^I =0, \ \ud
$.  The hat symbol on $ \hat b ^I$ is meant to remind us that this
parameter identifies with the NSNS field VEV or flux in the factorized
T-dual picture~\cite{giveon94}. The parameters $ \hat b ^I$ should not
hopefully be confused with the continuous angle parameter, $ b ^I \sim
b^I +1$, in the K\"ahler moduli $T^I$.
The complexified orthogonal and lattice bases for the $T^2_I$
coordinates, $ X^I _{1,2} $ and $\hat X ^I_{1,2}$, are related by the
formulas \bea && X ^I \equiv {X ^I_1 + i X ^I_2 \over \sqrt 2} = i
\sqrt {T ^I_2\over 2 U ^I_2} ( - U ^I \hat X ^I _1 + \hat X ^I_2),\cr
&& \bar X^{ I} \equiv {X ^I_1 - i X ^I_2 \over \sqrt 2} = - i \sqrt {T
^I_2\over 2 U ^I_2} ( - {\bar U } ^I \hat X ^I _1 + \hat X ^I_2), \eea
with similar formulas holding for spinor fields.

The fixed sub-manifolds under $\calr $ define the orientifold
$O6$-hyperplanes which are sources for the dual pair of closed string
sector RR form fields, $C_7,\ C_1$. With the  choice   $\calr =  \calr
_y$, the $O6$-planes extend along the three flat space directions of $M_4$ and wrap around
the three one-cycles $\hat a ^I $ along the real axes of $T^2_I$.  
The need to neutralize the net RR charges present inside the internal
compact manifold $X_6$ is what motivates introducing $D6$-branes
similarly extended along $M_4$ and wrapped around three-cycles of $
T^6 $.  Recall that the $Dp$-branes arise as soliton solutions of the
type $II$ string equations of motion for the closed string massless
NSNS (supergravity) modes associated to the metric tensor, $G_{MN} $,
and dilaton, $\Phi $, and for the massless RR dual antisymmetric form
fields, $C_{p+1} \sim C_{7-p}$.  The linkage to open strings is
realized by the characteristic property of the $Dp$-branes to serve as
boundaries or topological defect sub-manifolds, immersed in the 10-d
space-time, which support the open string end points.  Since the RR
charges enter as central terms in the supersymmetry algebra, the
supersymmetric $Dp$-branes ($p=0,2,4,6 $ for $ IIa$ and $p=1,3,5 $ for
$ IIb$) preserving a fraction $1/2 $ of the $32$ supersymmetry
charges in the bulk, satisfy a Bogomolnyi
type bound on their mass which guarantees them stability against
decay.

The $(p-3)$-cycles $\Pi _\mu $ of $ T^6$ wrapped by $Op$-planes and
$Dp$-branes solve the string equations of motion as equivalence
classes for closed sub-manifolds modulo boundaries, hence as elements
of the homology vector space, $[\Pi _\mu ] \in H_ {p-3} (X, R) $.  The
dual relationship with the cohomology vector space, $ H^ {p-3} (X, R)
$,  generated  by the equivalence classes of closed differential forms modulo
exact forms,  is used to define the cycles wrapping numbers, or electric and
magnetic RR charges,  in terms of  volume integrals.   
The cycles intersection numbers are defined in terms of topologically
invariant integrals obeying the antisymmetry property, $ [\Pi _\mu ]
\cdot [\Pi _\nu ] =- [\Pi _\nu ] \cdot [\Pi _\mu ] = I_{\mu \nu } .$
The homology basis of fundamental one-cycles $ [\hat a ^I] $ and $
[\hat b ^J] $ for $T^2_I $, have the intersection numbers, $ [\hat a
^I] \cdot [\hat b ^J] =- [\hat b ^I] \cdot [\hat a ^J] = \d ^{IJ}$.
To avoid filling the internal space, one commonly restricts to the
integer quantized homology classes, $H_{p-3} (X, Z)$.  For
convenience, one also usually limits consideration to the subset of
factorizable three-cycles, $ \Pi _\mu = \prod _{I=1}^3 \Pi ^I _\mu ,$
products of one-cycles of the $T^2_I $ tori.  In the lattice bases
generated by the dual bases of one-cycles, $ (\hat a^I ,\ \hat b^I) $,
along the $T^2_I $ tori vielbein vectors, $ (e_1^I,\ e_2^I ) $, these
cycles are parameterized by the three pairs of integer quantized
wrapping numbers, $ (n_\mu ^I, m_\mu^I) $. The orthogonal bases
representations for the factorisable three-cycles and their
orientifold mirror images, $\Pi _{\mu } = \prod _{I=1}^3 \Pi ^I_\mu ,
\ \Pi _{\mu '} = \prod _{I=1}^3 \Pi ^I _ {\mu '} $, denoted by $
(n_\mu ^I, \tilde m_\mu ^I) ,\ (n_ {\mu '} ^I, \tilde m_ {\mu '} ^I) ,
$ are related to the lattice bases representations  as \bea && \Pi ^I
_\mu = n ^I_\mu [\hat a^I] + m^I_\mu [\hat b^I] = n ^I_\mu [a^I] +
\tilde m^I_\mu [b^I] , \cr && \bigg [ [\hat a ^I]= [a ^I] - U_1^I [b
^I],\ [\hat b ^I]= [b ^I] ,\ \tilde m ^I _\mu = m ^I_\mu - U_1^I n
^I_\mu
\bigg ] \cr && [\Pi ^I_{\mu '}] \equiv n^I_ {\mu '} [a^I] + \tilde m
^I _{\mu '} [b^I] = n^I_ {\mu } [a^I] -\tilde m ^I _{\mu } [b^I] =
n^I_ {\mu } [\hat a^I] +(-m ^I_ {\mu }+ 2 U_1^I n^I_ {\mu }) [\hat
b^I] .  \label{eqlatcomp} \eea

The invariant volume of three-cycles, $ V_{\Pi _\mu } $, and the
topologically conserved number of intersection points of three-cycle
pairs, $I_{\mu \nu } $, are given by \bea && V_{\Pi _\mu } \equiv
\prod _{I=1}^3 (2\pi L_\mu ^I) ,\ [L^I _\mu = ( n^{I2} _\mu r_1 ^{I2}
+ \tilde m^{I2} _a r_2 ^{I2} ) ^\ud ] \cr && I_{\mu \nu }= [\Pi _\mu ]
\cdot [\Pi _\nu ] = \prod _{I=1}^3 ( n^I_\mu m^I_\nu - n^I_\nu m^I_\mu
). \eea
The one-cycles in $T^2_I$ can also be described in terms of the
rotation angles relative to the real axis one-cycle, $(1,0)$, wrapped
by the $O6$-planes \bea && \phi _\mu ^I = \pi \t _\mu ^I = \arctan ({
\tilde m_\mu ^I \over n_\mu ^I U_2 ^I }) = \arctan ({ \tilde m_\mu ^I
\chi ^I \over n_\mu ^I }) ,\ [\chi ^I \equiv {1\over U_2^I}= {r_2 ^I
\over r_1 ^I } ] .
\label{sec2.angle} \eea 
We use conventions in which the angles between $Dp$-branes and the
$O6$-plane, $\t ^I _a\equiv {\phi _a ^I \over \pi } $ and the $
Dp_a/Dp_b$-branes, $\t ^I _{ab} = \t ^I _b- \t ^I _a $, vary inside
the range, $ \t ^I _a \in [-1 , 1]$, with the positive angles
associated with counterclockwise rotations.  Our determination of the
$D$-brane-orientifold angle is related to the angle determination
given by the inverse-tangent function, $ \hat \t _\mu ^I \in [-\ud ,
\ud ] $, by the identification: $ \t ^I _\mu = \hat \t ^I _\mu$ for $
n^I _\mu \geq 0 $ and $ \t ^I _\mu ={ \tilde m^I _\mu \over \vert
\tilde m^I _\mu \vert } + \hat \t ^I _\mu$ for $ n^I_\mu \leq 0 $.
Furthermore, our determination of the interbrane angles is related to
the angle differences, $ \hat \t ^I_{ab} = \t ^I_b - \t ^I_a $, by the
identification: $ \t _{ab} = \hat \t _{ab} $, if $ \vert \hat \t _{ab}
\vert \leq 1 $, or $ \t _{ab} = - 2 {\hat \t _{ab} \over \vert \hat \t
_{ab} \vert } + \hat \t _{ab} $, if $ \vert \hat \t _{ab} \vert \geq 1
$.

The Dirac-Born-Infeld world volume action gives the mass of single
$Dp$-branes as the product of the tension parameter $ \tau _p $ by the
wrapped cycle volume, $ M_\mu = \tau _p V _{\pi _\mu } = \tau _p
\prod_{I=1}^3 L ^I _\mu $, using the definitions of $\tau _p$ given in
Eq.~(\ref{eqstccs}) below. This suggests that the construction of
energetically stable configurations of $D$-branes should consider the
cycles of minimal volume.  For the CY complex $d$-folds, equipped with
the metric tensor $ G_{I\bar J} $ preserving the complex structure $
J_I ^J = i \d _{IJ} $, there exist two types of volume minimizing
sub-manifolds, which correspond to the sets of two-cycles and
$d$-cycles~\cite{becker95} calibrated by the K\"ahler and holomorphic
volume forms, $\calj = i G_{I\bar J} d X^I \wedge d \bar X^{\bar J} $
and $ \O _{d,0} ,\ \bar \O _{0,d} $, respectively. Because of the relation linking
these forms, $ {1\over d! }\calj ^d  = (-1)^ {d(d-1) \over 2} i ^d \O _{d,0}
\wedge \bar \O _{0,d}$, and the reality condition on the manifold
volume, $ V_X = {1\over d! }\int _X \calj ^d $, the holomorphic form arise as
the one-parameter family, $ e ^{i\varphi } \O _{d, 0}$, parameterized
by the angle $\varphi .$ One then defines~\cite{joyce01} the
$d$-cycles $\Pi _\mu $ calibrated with respect to $ e ^{i\varphi } \O
_{d, 0}$ as the Lagrangian sub-manifolds (with a vanishing restriction
of the K\"ahler $(1,1)$ form, $ \calj \vert _ {\Pi _\mu } =0 $) on which
the holomorphic $d$-form obeys the reality condition, $\Im (e
^{i\varphi } \O _{3,0} ) \vert _{\Pi _\mu } = 0$.  The calibrated or
special Lagrangian (sLag) sub-manifolds are defined by embedding maps
which obey first order differential equations expressing the
preservation of supersymmetry charges.  These cycles have the property that
their volume integral, $ V_ {\Pi _\mu } = \int_ {\Pi _\mu } \Re (e
^{i\varphi } \O _{d,0} )$, is minimized among the elements belonging
to the same homology class $[\Pi _\mu ] $.  From the action of the
antiholomorphic reflection on the covariantly constant forms, $\calr
:\ \calj \to -\calj , \ \O _{d,0} \to e ^{2i\varphi } \bar \O _{0,d}
$, it follows that the orientifold $Op$-hyperplanes, as fixed point
loci of $\calr $, must wrap the sLag cycles.  In order to construct a
supersymmetry preserving open string sector, one must then consider
setups of $Dp$-branes which wrap the sLag cycles.  For the
factorizable $T^6$ tori, where $ \calj = i dX^I \wedge d \bar X^{I} $
and $ \O _{3,0} = dX ^1 \wedge dX^2 \wedge dX^3 $, the sLag cycles
intersect at angles $ \t ^I$ describing $SU(3)$ rotations, in such a
way such that the brane-orientifold intersection angles, $\t ^I_\mu $,
defined by Eq.~(\ref{sec2.angle}), or the interbrane intersection
angles $\t ^I_{\mu \nu } = \t _\nu ^I - \t _\mu ^I $, obey conditions
of form, $ \sum _I \pm \t ^I = 0 \ \text{mod} \ 2$. 
For completeness, we note that the supersymmetry conditions for  the  
M-theory  intersecting branes have been discussed  in Ref.~\cite{ohtasend97}.

The RR charge cancellation means the absence of RR tadpole divergences
in the open strings one-loop vacuum amplitude. This condition suffices
to guarantee that the world brane effective field theory is anomaly
free.  For general setups consisting of $ K$ stacks of parallel $N_\mu
\ D6 _\mu $-branes and their orientifold mirrors, $(D6_\mu + D6_{\mu
'} ) ,\ [ (\mu , \mu ') =1, \cdots , K = a, b ,\cdots ] $ the RR
tadpole cancellation condition requires that the sum over all the
wrapped three-cycles belongs to the trivial homology class, $ \sum _
{\mu =1}^K N_\mu ([\Pi _\mu ] +[\Pi _{\mu '}] ) + Q_ {O6} [\Pi _ {O6}
] =0 $. Here, the $Op$-plane charge is given by, $Q_ {Op} = \mp
2^{p-4} f_p$, where $f _p = 2 ^{9-p}$ counts the number of distinct
$Op$-planes and the $\mp $ sign is correlated with the (orthogonal or
symplectic gauge group) orientifold projection condition on the
Chan-Paton (CP) matrices and with the sign of the $Op$-plane tension
parameter, $\tau _ {Op} = \mp 2^{p-4} \tau _p $, as we discuss in the
next paragraph.
For toroidal orientifolds, using the relation, $ Q_{O6} [\Pi_{O6} ] =
\mp 32 \prod _I [\hat a^I]$, along with the decomposition in
Eq.~(\ref{eqlatcomp}), translates the RR tadpole cancellation
condition into the four equations for the wrapping numbers \bea &&
\sum _\mu N_\mu n_\mu ^ 1 n_\mu ^ 2 n_\mu ^3 \mp 16 =0, \ \sum _\mu
N_\mu n_\mu ^ I \tilde m_\mu ^ J \tilde m_\mu ^ K =0, \ [I\ne J \ne K]
\eea where, to avoid double counting, one must exclude the orientifold
image branes from the above summations over brane stacks.  For
definiteness, we develop the following discussion in the case of
orientifolds with negative tension and RR charge, $ Q_{Op} < 0$,
corresponding to the $SO$ group type projection.

The open string sectors are associated with the distinct pairs of
$Dp$-branes supporting the end points.  In orientifolds, the diagonal
and non-diagonal sectors include the pairs, $ (a,a),\ (a, a') $ and
$(a,b),\ (a,b') $, with the equivalence relations between mirror
sectors, $ (a, a) \sim (a', a'),\ (a,b) \sim (b',a') , \ (a,b') \sim
(b,a') .$ The gauge or Chan-Paton (CP) factors are time independent
wave functions, described for the diagonal and non-diagonal sectors by
$(2N_a \times 2N_a)$ and $ (2 N_a +2 N_b) \times (2 N_a +2 N_b) $
matrices, $ (\l ^{(a,a)} _A )_{ij} $ and $ (\l ^{(a,b)} _A ) _ {ij} $,
with the labels $ i,\ j$ running over members of the $Dp$-brane stacks
and the labels $A$ running over components of the gauge group
representations.  The matrices $\l ^{(a,a)} _A $ decompose into 4
blocks of size $ N_a\times N_a $ and the matrices $\l ^{(ab)} _A $
into 8 diagonal blocks of size, $ N_a\times N_a,\ N_b\times N_b,$ and
8 non-diagonal blocks of size, $ N_a\times N_b $.  Since the modes in
the conjugate sectors, $(b,a) \sim (a,b) ^\dagger $, have opposite
signs intersection numbers, $ I_{ab} =- I_{ba},$ opposite helicities
and conjugate gauge group representations, these are assigned the
Hermitean conjugate matrices, $\l ^{(b,a)} = \l ^{(a,b) \dagger } .$
We omit writing henceforth the upper suffix labels on the CP  matrices specifying the
sectors.  The normalization and closure sum for the CP matrices are
described in consistent conventions as \bea Trace(\l _A \l_B ^ \dagger
) = \d _{AB} ,\quad \sum _A Trace(O_1 \l _A ) Trace(O_2 \l _A ) =
Trace(O_1 O_2), \eea where the summation extends over the complete set
of states $A$ in the gauge group representation.  For the unitary
group, $U(N_a)$, the completeness sum over the adjoint group
representation uses the identity, $\sum _A (\l _A )_{ij} (\l _A )_{kl}
= \d_ {il} \d_ {jk} $.

The orientifold symmetry $\O \calr $ is embedded in the gauge group
space of a $ Dp$-brane setup through a unitary twist matrix, $ \g _{\O
\calr } $, by imposing the projection condition, $ \l _A = - \g _{\O
\calr } \l ^{ T} _A \g _{\O \calr } ^{-1} $.
One convenient construct for  $ \g _{\O
\calr } $  is given by the direct product, $ \g _{\O
\calr } = \otimes _\mu \g _{\O \calr ,\mu } $.  The anomaly
cancellation constraint commonly imposes the tracelessness condition,
$Trace(\g _{\O \calr } )=0$, along with the symmetry conditions, $\g
_{\O \calr } ^T = \pm \g _{\O \calr } $, in correspondence with the $
SO $ and $ USp $ type projections.  In the special case of a $
D6_a$-brane stack overlapping the $O6$-plane, hence coinciding with
the mirror image $ D6_{a'} $-brane, the gauge symmetry in the $ SO $
and $ USp $ projections is enhanced to the rank $ N_a $ orthogonal or
symplectic groups, $ SO(2N_a) , \ USp(2N_a) $, in correspondence with
the negative and positive signs of the $O6$-plane RR charge $Q_ {O6}$
and tension parameter.  To detail the construction of CP factors, we
consider, for the sake of illustration, the $(a,a)+ (a,a')$ sector.
The $SO$ type projection matrix, $ \g _{\O \calr , a} = \pmatrix{0 & I
_{N_a} \cr I _{N_a} & 0 } $, yields the $SO(2N_a)$ group adjoint
representation, $ \l ^{\bf [Adj] } = \pmatrix{m & a_1 \cr a_2 & - m^T}
$, with the restriction to $ a_1=a_2 =0$ yielding the $U(N_a)$ group
adjoint representation.  We here use conventions where the symbols $
m,\ s $ and $ a$ for block entries designate generic, symmetric and
antisymmetric matrices, respectively.  The antisymmetric
representations of $U(n_a)$ (in the $SO$ type projection) are realized
in the $(a,a') $ sector by the $ 2 N_a \times 2 N_a $ matrix solution,
$ \l ^ { [\bfA ] } = \pmatrix{0 & a \cr 0 & 0} $, and its conjugate, $
\l ^ { [\bfA ] \dagger } $.  The bifundamental modes in the
non-diagonal $(a,b)$ sectors are realized by the $ (2 N_a +2 N_b)
\times (2 N_a +2 N_b) $ matrix solutions \bea && \l ^{ [(N_a, N_b)] }
= \pmatrix{0 & B \cr B' & 0},\ [B=\pmatrix{\a & \b \cr \g & \d },\ B
'=\pmatrix{ -\d ^T & - \b ^T \cr - \g ^T & -\a ^T} ] \eea obtained by
setting successively the $ N_a\times N_b$ block entries, $\a ,\ \b ,\
\g , \ \d $, to non-vanishing values.  We recommend
Refs.~\cite{polchb,johnson99,giveon99} for a further discussion of
$D$-branes.

Proceeding now to the gauge group, we consider first the case of a
single stack of $ N_a $ coincident $ D6_a$-branes and its mirror $
D6_{a'}$-brane stack wrapped around three-cycles at generic angles $\t
^I_a $ and $ \t ^I_{a'} $ relative to the $O6$-plane, hence not
overlapping the $O6$-planes.  The massless states in the diagonal
sectors, $(a,a) \sim (a', a') $, include the gauge bosons of the gauge
group $U(N_a)$, along with adjoint representation matter modes.  For
the pair of intersecting $D6_a/D6_b$-brane stacks and their mirror
image $D6_{a'}/D6_{b'}$-branes carrying the gauge symmetry group,
$U(N_a) \times U(N_b)$, the non-diagonal sectors consist of conjugate
pairs, $ [(a,b) + (b,a)] \sim [ (b',a') + (a',b')] $ and $[(a,b') +
(b',a)] \sim [(b,a') + (a',b)]$, with localized (improperly named
twisted) states carrying the bi-fundamental representations, $ I_{ab}
(N_a,\bar N_b) $ and $ I_{ab'} ( N_a, N_b) $, with multiplicities
given by the wrapped cycles intersection numbers \bea && I_{ab} =\prod
_I (n^I_a \tilde m^I_b - n^I_b \tilde m^I_a) ,\quad I_{ab'} =\prod _I
(n^I_a \tilde m^I_{b'} - n^I_{b'} \tilde m^I_a) = \prod _I - (n^I_a
\tilde m^I_{b} + n^I_{b} \tilde m^I_a) .\eea The sectors $(a,a')$ have
a total number of intersection points, $I_{aa'} = \prod _I (-2 n_a^I
\tilde m_a^I) $, of which the $ I ^{(A)} _{ 1, aa'} = {I_{aa'} \over
\prod _I n ^I_{a} } $ points, symmetric under the reflection $\calr $,
give rise
to modes carrying (in the $SO$ type projection) the antisymmetric
representation of the gauge group $U(N_a), \ \l = - \l ^T $, while the
remaining modes split into pairs of modes carrying the symmetric and
antisymmetric representations with the same multiplicity, $I
^{(S+A)}_{2, aa'} = \ud I_{aa'} (1- {1\over \prod _I n ^I_{a} } )$.
The net multiplicities of the symmetric and antisymmetric
representations are thus given by, $ I_{aa'} ^{S,A} = \ud I_{aa'}
(1\pm {1\over \prod _I n ^I_{a} } )$.

We follow the familiar description of fermion and boson modes in terms
of the basis of left chirality states, $( f_L, \ f_L ^c) $, where the
right chirality states are obtained by applying the complex
conjugation operator exchanging particles with antiparticles, $ f_R
\sim \bar f_L ^c = (f_L ^c) ^ \dagger, \ f ^c_R \sim \bar f_L = f_L ^
\dagger $.  Note that the correspondence relations for the electroweak
$SU(2)$ group doublets include extra signs, with, for instance, the
quark doublet fields given by, $ f_L = (u _L, \ d _L) , \ f_R = (u _R,
\ d _R) , \ f_L ^c = (d_L ^c, \ - u _L ^c) .$ For the spectrum of
modes with the left-right chiral asymmetries, $ I _{ab} f_L $ and $ I
_{cd} f_L ^c , $ the presence of $\D _{ab} $ and $ \D _{cd} $ mirror
vector pairs results in the non-chiral spectrum, $ (I_{ab} + \D _{ab}
) f_L + \D _{ab} f_R ^c , \ (I_{cd} + \D _{cd} ) f_L^c + \D _{cd} f _R
$.  Going from positive to negative intersection numbers entails
changing the sign of the chirality (helicity for massless fermions)
and conjugating the gauge group representations.  For instance, the
massless fermions with negative multiplicities $I_{ab}, \ I_{cd},$
would refer to right chirality fermions (or left chirality
antifermions) carrying the conjugate bi-fundamental representations,
$\vert I_{ab} \vert (\bar N_a ,N_b ),\ \vert I_{cd} \vert (\bar N_c,
\bar N_d)$.

\subsubsection{First quantization formalism}

We only discuss here the non-diagonal open string sectors $(a,b)$.
The coordinate and spinor field components along the flat $M_4$
space-time dimensions obey the N conditions, $\dh _\s X ^ {\mu } =0,\
\psi ^{\mu } \mp {\tilde \psi } ^{\mu } = 0 ,$ at both end points, $
\s = (0, \pi ) $, where the upper and lower signs apply to the R and
NS sectors.  For $D6_a/D6_b$-brane pairs wrapped at the angles $\phi
_{a, b}^I = \pi \t _{a, b}^I$ in $T^2_I$, the rotated complexified
coordinate components, $ e ^{-i \phi _{a,b} ^ I } X^I ,\ e ^{-i \phi
_{a,b} ^ I } (\psi \mp \tilde \psi ) $, split into real and imaginary
parts, longitudinal and transverse to the branes, hence obeying the N
and D boundary conditions, $ \dh _\s \Re ( e ^{-i \phi _{a,b} ^ I }
X^I )=0 $ and $ \dh _t \Im ( e ^{-i \phi _{a,b} ^ I } X^I )=0.$ The
corresponding conditions for the rotated spinor field components read,
$ \Re ( e ^{-i \phi _{a,b} ^ I } (\psi ^ I\mp \tilde \psi ^ I ) ) =0 $
and $\Im ( e ^{-i \phi _{a,b} ^ I } (\psi ^ I\pm \tilde \psi ^ I) ) =0
$, with the upper and lower signs referring to the R and NS sectors.
In terms of the complex plane $z,\ \bar z $ variables, the N and D
boundary conditions along the real axis for the pair of $\phi _{a,
b}^I $ rotated $D$-branes read in full as \bea && (\dh -\bar \dh )
[e^{-i \phi _{a,b}^ I } X^I + e^{+i \phi _{a,b}^I } \bar X^I ] = 0, \
e^{-i \phi _{a,b} ^I } (\psi ^I \mp {\tilde \psi ^{ I}} ) + e^{+ i
\phi _{a,b} ^I } (\bar \psi ^{I} \mp \bar {\tilde \psi ^{I} } ) = 0 ,
\cr && (\dh + \bar \dh ) [e^{-i \phi ^{a,b}_I } \bar \dh X^I - e^{i
\phi _{a,b}^ I } \dh \bar X ^I ] = 0 ,\ e^{- i \phi _{a,b} ^I } (\psi
^I \pm {\tilde \psi ^{ I} } ) - e^{i \phi _{a,b} ^I } (\bar \psi ^{I}
\pm \bar {\tilde \psi ^{I} } ) = 0,
\label{eqbcsxpsi} \eea where the labels $ [a, \ b]$ correspond to the
open string end points, $\s = [0, \pi ] $.  For convenience, we extend
the notation for the interbrane angles to, $\t ^a _{ab} = (\t ^I _{ab}
, \t ^4 _{ab} ) $, with the understanding that $ \t ^4 _{ab} =0 $ in
our  present discussions.  In
the light cone gauge of the 2-d world sheet superconformal field
theory, the open string states include the quantized oscillator modes
described by the number operators, $N_X,\ N_\psi $, and the zero modes
described by the momentum and winding quantum numbers, $ p ^I _{ab} ,
\ s ^I _{ab} \in Z $, and by the interbrane transverse distances, $Y
^I_{ab} $, for the complex directions along which the branes are
parallel.  The string oscillation frequencies along the complex
dimensions have integral modings shifted by the $D$-brane angles, $n^
a_\pm = n ^a \pm \t ^a _{ab} ,\ [n ^a\in Z, \ a=1,2,3,4 ] $.  In the
boson representation of spinor fields, the oscillator number
operators, $N_ {\psi ^a} $, are replaced by the $H^a$ fields momentum
vectors, $r^a = (r^I, r^4) , \ [I=1,2,3]$ corresponding to the weight
vectors for the Lorentz group $SO(8)\sim Spin(8)$ Cartan torus
lattice.  The GSO projection for the world sheet fermion number parity
symmetry, $(-1)^F$, restricts the weight vectors to the sub-lattice,
$r^ a \in Z + \ud + \nu ,\ [ \sum _{a=1}^4 r^a \in 2Z +1 ] $ where the
boson and fermion (NS and R) sector modes with $\nu = \ud ,\ 0 $, are
assigned vector and spinor type weight vectors.  The fourth component
of the $SO(8)$ weight vectors, $ r^4 =0,\ \pm \ud , \ \pm 1 , \cdots
$, describes the chirality (or helicity quantum number for massless
fermions) for the $ SO(2)$ little group of the flat space-time Lorentz
group. The remaining three components, $ r^I$, describe the helicity
quantum numbers for the three $SO(2)_I$ subgroup factors of $SO(8)$.
The quantized string mass shell condition for the $(a,b)$ sector is
expressed by the general formula for the string squared mass spectrum
\bea && \a ' M^2 _{ab} = \sum_{I=1}^3 \a 'M^{(0) 2 } _{I, ab} + \sum _
{a=1} ^4 N_{X^a} (\t ^a_{ab}) + \sum _{a =1} ^{4} {(r^a + \vert \t ^a
_{ab} \vert )^2 \over 2} - \ud + \sum _{I=1}^3 \ud \vert \t ^I _{ab}
\vert (1- \vert \t ^I _{ab} \vert ) , \cr && \bigg [ \a ' M^{(0) 2 }
_{I, ab} = \d ^I _{ab}\sum _{p_{ab} , s_{ab} \in Z} {\vert p_{ab} + s
_{ab} T ^I \vert ^2 \over T ^I_2} { U ^I _2 \over \vert m _a ^I- n_a
^IU ^I\vert ^2} + {Y ^{I2} _{ab} \over 4 \pi ^2 \a '} \bigg ]
\label{sec2.mass}  \eea
where we continue using the conventional range for the brane
intersection angles, $ -1 < \t ^I _{ab} \leq +1$. The last two terms
in the squared mass, $\a ' M^2 _{ab} $, describe the string zero
energy vacuum contributions, while the first term, $M^{(0) 2 } _{I,
ab}$, given explicitly in the second line entry, includes the
contribution from the momentum and winding string modes and from the
transverse separation distance $Y ^I _{ab}$ of the $D6_a/D6_b $-branes
along the $T^2_I$ where they are parallel. The latter point is
reminded by the symbol $\d ^I _{ab} $ which is non-vanishing whenever
the displaced $D6_a/D6_bb$-branes are parallel along some complex
plane $I$ so that $\t ^I _{ab} =0$.

We now explicitly describe the low lying string modes.  The solutions
for massless spin-half fermion modes select the unique conjugate pair
of spinor weights, $ r^a \equiv (r^I, r^4)= \pm ( - \ud , - \ud ,- \ud
, \ud ) $, with the overall $\pm $ signs corresponding to the two
possible spatial helicities.  The abbreviated notation for the spinor
weights illustrated by, $ \pm ( - \ud , - \ud ,- \ud , \ud ) \equiv
\pm (---,+)$, will be adopted for convenience.  The scalar modes of
smallest squared mass select the four solutions for the vector
weights, $ r^a= \pm (\underline{-1, 0,0}, 0) , \ \pm (-1,-1,-1,0) $,
with the underline symbol standing for the three distinct permutations
of the entries. The resulting four solutions enter with the squared
masses \bea && \a ' M^2 _{ab} =\bigg [\ud (- \vert \t _{ab} ^1 \vert +
\vert \t _{ab} ^2 \vert + \vert \t _{ab} ^3 \vert ) ,\ \ud (+ \vert \t
_{ab} ^1 \vert - \vert \t _{ab} ^2 \vert + \vert \t _{ab} ^3 \vert
),\cr && \ud ( \vert \t _{ab} ^1 \vert + \vert \t _{ab} ^2 \vert -
\vert \t _{ab} ^3 \vert ),\ 1 -\ud ( \vert \t _{ab} ^1 \vert + \vert
\t _{ab} ^2 \vert + \vert \t _{ab} ^3 \vert ) \bigg ].
\label{sec2.tetra} \eea 
The lowest lying vector boson mode arises from the vector weight, $
\pm ( 0, 0,0, 1 ) $, with squared mass, $ M^2 _{ab} = \ud \vert \t
_{ab} ^I \vert $.  For completeness, we note that the towers of
so-called Regge resonance gonion modes~\cite{aldaisb01} of scalar and
vector boson types correspond to the oscillator excited states, $\psi
^{[I, \mu ]} _\ud (\psi ^I _{-r_+} \a ^I _{-n _+} )^ {m ^I} \vert 0>
_{NS} $, with mass squared, $ M^2 _{ab} = (m ^I \mp \ud ) \vert \t
_{ab} ^I \vert ,\ [m ^I \in Z]$.  We recommend
Ref.~\cite{lustepple03,epple04} for a further discussion of the mass
spectrum in intersecting brane models.

In parallel with the closed string geometric moduli, there arise open
string sector moduli which correspond to order parameters of the world
brane gauge field theory associated with the $D$-branes positions and
orientations.  Thus, the transverse coordinates of a $D6_a$-brane
stack are moduli fields in the adjoint representation of the $U(N_a)$
gauge theory which parameterize its Coulomb branch deformation.  The
recombination of a pair of intersecting branes into a single brane, $
a+ b\to e$, or the reconnection of two branes, $ a+ b\to c+d $, are
described in terms of the moduli fields in bi-fundamental
representations of the $(a,b)$ sector which parameterize the Higgs
branch of the $U(N_a)\times U(N_b) $ gauge theory.  The brane
splitting fixes the VEVs of open string moduli while the brane
recombination redefines the VEVs of open string localized moduli
needed to avoid the vacuum instability from tachyon modes, in analogy
with the Higgs gauge symmetry breaking mechanism.  The splitting and
recombination processes are accompanied by mass generation mechanisms
which decouple pairs of fermion modes in vector and chiral
representations.  Representative examples of these deformations are
the Higgs mechanisms for the unified and the electroweak gauge
symmetries. The consistent description of $D$-brane recombination
using non-factorisable cycles~\cite{rab01,iban1,Cvetic:2005lll} does
indeed lead to a reduction of the wrapped cycles volume and of the
fermion spectrum chiral asymmetry, in agreement with the Higgs
mechanism.  In spite of the poor information on string
non-perturbative dynamics, interesting results have been established
concerning the existence of bound states for $ Dp/D (p +4) $-brane
pairs and for $ Dp/D (p +2) $-brane pairs in backgrounds involving
NSNS or magnetic field fluxes~\cite{gava97,mihail01,witten98} or the T-dual
backgrounds of $Dp_a/Dp_b$-brane pairs wrapping intersecting
cycles~\cite{bluming0}.  We also note that the recombination process
can be partially formulated in the context of branes realized as gauge
theory solitons~\cite{lustepple03,epple04,ohtazhou98}.

\subsubsection{Conformal   field theory formalism}

The conformal field theory provides a powerful approach to calculate
the on-shell string S-matrix in perturbation theory.  The open string
amplitudes are obtained by integrating the vacuum correlation
functions of the modes vertex operators inserted on the world sheet
boundary.  We focus here on the tree level amplitudes of the $(a,b)$
non-diagonal sectors of the $D$-brane pair, $D6_a/D6_b $, intersecting
at the angles, $\phi ^I _{ab} = \phi ^I _{b} - \phi ^I _{a} $.  With
the field doubling prescription, the world sheet field propagators are
simply given by \bea && <X^M(z_1) X ^N(z_2) > =- {\a ' \over 2} G
^{MN} \ln (z_{12}), \ <\psi ^M(z_1) \psi ^N(z_2) > = { G ^{MN} \over
z_{12}}, \cr && \varphi (z_1) \varphi (z_2) > = - \ln (z_{12}) ,\
<H^A(z_1) H ^B (z_2) > =- \d ^{AB} \ln (z_{12}) , \eea where $ z_{12}
= z_1 - z_2$.  Since the coordinate and spinor field components of
$M_4$ obey the N boundary conditions, $ (\dh - \bar \dh ) X ^{\mu }
=0, $
one can formally replace the Minkowski space coordinate field
components along the complex plane real axis boundary as, $ [X ^\mu
(z)+\tilde X ^\mu (\bar z)] \to 2 X ^\mu (x) $.  The insertion of the
open string mode $(a,b)$ at the real axis point, $x_i =\Re (z_i) $,
modifies the boundary conditions on the right hand half axis, $x>x_i
$, in such a way that the two orthogonal linear combinations,
associated with the real and imaginary parts of the rotated
complexified coordinate fields, obey the N and D boundary conditions:
$\Re (e ^{-i\phi _{ab} } \dh _\s X^I) =0,\ \Re (e ^{-i\phi _{ab}^I }
(\psi ^I \mp \tilde \psi ^I) ) =0, $ and $\Im (e ^{-i\phi _{ab} } \dh
_t X ) = 0,\ \Im (e ^{-i\phi _{ab} } (\psi ^I \pm \tilde \psi ^I) ) =0
$, where the upper and lower signs refer to the R and NS sectors.  The
left and right half lines, $x\in [-\infty , x_i] $ and $x\in [x_i,
\infty ] $, are mapped to the $D6_a$- and $D6_b$-branes with the
boundary conditions given by Eq.~(\ref{eqbcsxpsi}).  Since only the
interbrane angle really matters, the boundary conditions on the
coordinate and spinor field combinations along $T^2_I$ can be
expressed by the same formulas as in Eq.~(\ref{eqbcsxpsi}) with $\phi
_ {a,b} ^I \to \phi _ {ab} ^I = \phi _ {b} ^I - \phi _ {a} ^I $, along
the half line $x\in [x_i, \infty ] $, and $ \phi ^I _{a,b} \to 0 $
along the half line $x\in [-\infty , x_i] $.
Taking the sum and difference of the two relations yields the
equivalent form of the boundary conditions \bea && \dh X ^I - e^{2i
\phi _{ab} ^I } \bar \dh \bar X ^I =0, \ \bar \dh X ^I - e^{2i \phi
_{ab} ^I } \dh \bar X ^I =0 ,\cr && \psi ^I \mp e^{2i \phi _{ab} ^I }
\bar {\tilde \psi } ^I =0,\ \tilde \psi ^I \pm e^{2i \phi _{ab} ^I }
\bar \psi ^I =0. \eea Note that our sign convention for the brane
angles is opposite to that used in Refs.~\cite{polchb,Cvetic:2003ch}
and that we differ from Ref.~\cite{Cvetic:2003ch} in certain relative
signs.

We now discuss the covariant conformal gauge formalism of the world
sheet theory.  Each open string state of the non-diagonal sector, $
C\in (a,b)$, is assigned a primary vertex operator of ghost charge $q$
and unit conformal weight, $ V _{C , (a,b)} ^{(q)} (z_i , k_C, \l _C )
$, with $ k_C $ denoting the incoming four momentum and $\l _C$ the
gauge wave function factor. The building blocks in constructing the
vertex operators are the coordinate fields, $ X^{A,\bar A} (z) $,
their derivatives, $\dh X^{A,\bar A} (z) $, and exponential maps, $e^
{i k _C \cdot X (z,\bar z ) } $; the spinor fields, $ \psi ^{A,\bar A}
(z) = e ^{iH^{A,\bar A} (z) } $; the superconformal ghost scalar field
$\varphi (z)$ exponential map, $ e ^{q \varphi (z) } $, of ghost
charge $q$; the spin and twist operators for spinor fields along the
flat space-time and internal space directions, $ S _{ r ^\a } (z) = e
^{i r^\a (z) H^\a (z) } $ and $ s ^{r^I} _{\pm \t ^I} (z) ,\ [r^\a = (
r^4, r^0 ) ,\ r^I=(r^1, r^2,r^3) ]$; and the twist operators for
coordinate fields along the internal space directions, $ \s _{\pm \t
^I} (z) $.  The weight vectors, $ r^A = (r^I, r^a )$, denote the
momentum vectors of the complex scalar fields, $ H ^A (z) = [ H^{\a }
(z),\ H^{I } (z) ] $, belonging to the Cartan torus lattice of the
$Spin(10) $ Lorentz group. The twist and spin operator factors are
needed to produce the requisite branch point singularities at the
modes insertion points.  These operators create the ground states of
the twisted sectors upon acting on the $ SL(2,R)$ invariant ground
states of the NS and R sectors.  For the low lying non-diagonal sector
modes with excited coordinate oscillator states along the internal
space directions, alongside with the ground state twist field, $ \s^
{\a } _{\pm \t ^I} (z) $, one needs to introduce the excited twist
field operators, $\tau _{\pm \t ^I} (z) ,\ \tilde \tau _{\pm \t ^I}
(z) $.  The spinor field ground state and low lying excited twist
field twist operators, $s^ {r } _{\pm \t ^I} (z) ,\ t^ {r } _{\pm \t
^I} (z),\ \tilde t^ {r } _{\pm \t ^I} (z)$, are explicitly realized by
the free field vertex operators, \bea && s^ {r ^I } _{\pm \t ^I} (z) =
e^{\pm i (\t ^I +r ^I ) H_I (z) } ,\ t^ {r^I } _{\pm \t ^I} (z) =
e^{\pm i (\t ^I + r ^I +1 ) H_I (z) } ,\ \tilde t^ {r ^I } _{\pm \t
^I} (z) = e^{\pm i (\t ^I +r ^I - 1 ) H_I (z) } , \eea labeled by the
angle $\t ^I$ and the $SO(6)$ Lorentz group weight vector, $r ^I $.
The GSO projection for the world sheet fermion number parity symmetry,
$ (-1)^F$, correlates the weight vectors for the flat space-time
$SO(1,3) \sim SO(4) \sim SO(2)\times SO(2)$ helicity, $r^\a
=(r^4,r^0)$, with those of the internal $SO(6)$ helicity, $ r^I =
(r^1,r^2,r^3) $.  For the R sector fermions, this requires the number
of $-\ud $ entries in the five-component spinor weight vectors, $ r^A
= (r^I, r ^\a ) $, to have a fixed parity (odd in our conventions).
The left and right helicity (chirality) fermions are thus described by
the spin operators, $S_{r ^\a } =e^{ir^4 H_4 +ir^0 H_0 }$, with
weights: $ r^4= r^0 =\pm \ud $ and $ r^4=- r^0 =\pm \ud $,
respectively.  Note that the $SO(10)$ group weight vectors, $ r ^A =
(r^I, r ^\a ) = (r^a, r^0)$, reduce in the light cone gauge to the
$SO(8)$ group weight vectors, $ r ^a = (r^I, r ^4 ) $.  The same
description applies to the $(a,a)$ diagonal open string sectors upon
introducing the spin fields and the spinor twist fields, $ s ^{r^I}
(z) = e ^{i r^I H^I (z) } $, with vanishing angles.  To develop a
unified formalism for both the diagonal and non-diagonal sectors, we
adopt the self-explanatory notation for the twist operators, $ s
^{r^A} _{ \pm \t ^A } (z) = e ^{\pm i (\t ^A + r^A ) H^A(z) } $,
encompassing the case, $ \t ^A =0$.

Unlike the spinor field twist operators, the coordinate field twist
operators do not have a free field representation.  An implicit
definition can still be obtained by specifying the leading branch
point singularities in the operator product expansions of these
operators with the primary operators constructed from the coordinate
fields \bea && \dh X ^I (z_1 ) \s _{\t ^I } (z_2) \sim z_{12} ^{(\t ^I
-1)} \tau _{\t ^I } (z_2) + \cdots , \ \dh \bar X ^{ I} (z_1) \s _{-\t
^I } (z_2) \sim z_{12} ^ {(\t ^I -1)} \tau _{-\t ^I } (z_2) + \cdots
,\cr && \dh X ^I (z_1 ) \s _{-\t ^I } (z_2) \sim z_{12} ^{-\t ^I }
\tilde \tau _{-\t ^I } (z_2) + \cdots , \ \dh \bar X ^{ I} (z_1) \s
_{+\t ^I } (z_2) \sim z_{12} ^ {-\t ^I}\tilde \tau _{+\t ^I } (z_2) +
\cdots ,
\label{app.XOPE} \eea 
where $ \tau _{\pm \t } , \ \tilde \tau _{\pm \t } $ are the excited
state twist field operators introduced earlier, and the dots denote
contributions operators which are regular in the limit $ z_{12} \equiv
z_1-z_2 \to 0 $.  We have used here the abbreviated notation for the
brane angles, $\t _{ab} ^I \to \pm \t ^I ,\ [\t ^I \in [0, 1] ]$ with
the sign made explicit in such a way that the results for the negative
and positive brane intersection angles, $ \mp \t ^I $, are related by
the substitution, $ \t ^I \to 1-\t ^I $.  For completeness, we also
quote the operator product expansions for the spinor twist field
operators in terms of the excited twist operators introduced above,
\bea && \psi ^I (z_1) s ^ {r ^I }_{\t ^I } (z_2) \sim z_{12} ^{(\t ^I
+ r ^I) } t ^ {r ^I }_{ \t ^I } (z_2) + \cdots , \ \bar \psi ^{I}
(z_1) s ^ {r ^I} _{- \t ^I } (z_2) \sim z_{12} ^{(\t ^I + r ^I) }t ^
{r ^I} _{-\t ^I } (z_2) + \cdots , \cr && \psi ^I (z_1) s ^r _{-\t ^I
} (z_2) \sim z_{12} ^{-(\t ^I + r ^I) }\tilde t ^ {r ^I} _{- \t ^I }
(z_2) + \cdots , \ \bar \psi ^{I} (z_1) s ^ {r ^I} _{\t ^I } (z_2)
\sim z_{12} ^{-(\t ^I +r ^I) }\tilde t ^{r ^I}_{\t ^I } (z_2) + \cdots
\label{app.POPE} \eea It is of interest to note that the singular
dependence on the brane angles cancels out in the operator product
expansions of the coordinate and spinor twist fields with the
energy-momentum and supersymmetry generators, \bea && T (z) = - G_{MN}
({1\over \a '} \dh X ^M \dh X ^N + \ud \psi ^M \dh \psi ^N ), \cr &&
T_F (z) = i \sqrt {2\over \a ' } G_{MN} \psi ^M \dh X^N (z) = i \sqrt
{2\over \a ' }( \psi ^\mu \dh X_\mu + \psi ^I \dh \bar X^{ I} + \bar
\psi ^{ I} \dh X^I ) . \eea For instance, $ T_F (z_1) (s ^r _ {\t ^I}
\s_ {\t ^I} )(z_2) \sim i \sqrt {2 \over \a ' } ( z_{12} ^{-1-r ^I}
\tilde t ^r _ {\t ^I} \tau _ {\t ^I} + z_{12} ^{r ^I} t ^r _ {\t ^I}
\tilde \tau _ {\t ^I} ).$

The following formulas are of use in evaluating the conformal weights
of various operator factors, \bea && h (e^ {q\varphi (z) }) =
-{q(q+2)\over 2 } ,\ h (e^ {\pm i r ^A H_A (z) }) = {r^{A2} \over 2}
,\cr && h (\s _{ \pm \t ^I} )= \ud \t ^I (1-\t ^I) ,\ h(\tau _ { \pm
\t ^I} ) = \ud \t ^I (3-\t ^I) , \ h(\tilde \tau _ {\pm \t ^I} ) = \ud
(1-\t ^I) (2+\t ^I) .\eea The mass shell condition for a mode of mass
squared, $M_C ^2 $, is then determined by requiring that the total
conformal weight of the mode $C$ vertex operator, $ V _C (z) = V_ C
(z; k_C, \l _C ) $, amounts to unity, $ 1= h (V_C ) \equiv k_C ^2 +
\cdots = - M_C ^2 + \cdots $.

The vertex operators take different forms depending on the
superconformal ghost charge~\cite{polchb}, $q \in Z +\nu +\ud $,
carried by the scalar ghost field exponential, $e^ {q\varphi (z) } $,
with $ \nu = \ud , 0 $ in the NS and R sectors.  The canonical
pictures (unintegrated with respect to the superspace variable, $\t $)
involve the superconformal scalar ghost field factors, $ V^{(-1)} (z)
\sim e^{-\varphi (z) } O ^{ (-1 )} (z) $ and $ V^{(-\ud )} (z) \sim
e^{- {\varphi (z) \over 2 } } O ^{ ( -\ud )} (z) $, whereas the
integrated (with respect to $\t $) vertex operators, of higher
superconformal ghost charges, are obtained by acting on the canonical
operators with the picture changing operator, $
G_{-\ud } = e^{\varphi (z)} T_F (z) + \cdots $, where the dots refer
to ghost field terms.  The isomorphic representations of the vertex
operators of increasing ghost charges are obtained by the stepwise
incrementation, $V ^{ (q +1) } (z) = \lim _{w\to z} \calp (z,w) V
^{(q)} (w) = \lim _{w\to z} T_F (z) e^{\varphi (z)} V^{(q)} (w)
+\cdots $.  Since the vacuum for the world sheet surface of genus $g$
carries the defect ghost charge, $( 2g-2)$, in order to conserve the
ghost charge in the vacuum correlator involving $n_F$ and $n_B$
fermion and boson vertex operators carrying the natural charges, $
-\ud ,\ -1 $, one must apply the picture changing operator (PCO) on
the number of vertex operator factors, $N_{PCO}= n_B + {n_F\over 2} +
2g-2 $.
For instance, the four point vacuum correlators on the disk surface
require, $ N_{PCO} = n_B + {n_F\over 2} -2 $, while those on the
annulus surface require, $ N_{PCO} = n_B + {n_F\over 2} $.
   
We are now ready to complete the construction of vertex operators.
The matter and gauge boson modes are described, in the diagonal
sectors of parallel $D6$-branes, by the following vertex operators in
the canonical and once-derived ghost pictures, with charges $ q=-1, 0$
for bosons and $ q= -\ud , +\ud $ for fermions, \bea && \bullet \
V_{C^I} ^{(-1)} (z) = \l_{C^I} e ^{- \varphi }
\psi ^I e ^{i k \cdot X } , \ V _{C^I} ^{(0)}(z) = i \sqrt {2\over \a
' } \l _{C^I} [\dh X^I - i \a ' (k \cdot \psi ) \psi ^I ] e ^{i k
\cdot X } , \cr && \bullet \ V_{C^I} ^{(- \ud )} (z) = \l_{C^I} e ^{-
{\varphi \over 2} } u ^\a (k) S_\a e^{i r ^I_ {s} H_I (z)} e ^{i k
\cdot X } , \ V_{C^I} ^{(+ \ud )} (z) = i \sqrt {2\over \a ' } \l
_{C^I} e ^{+{\varphi \over 2} } u _\a (k) S_\b e^{i r ^I _{s} H_I }
\cr && \times [\sum _{J=1}^3 (e ^{-i H_J} \dh X^J \d _{r_s ^J, \ud } +
e ^{i H_J} \dh \bar X^J \d _{r_s ^J, -\ud } ) \d ^{\a \b } +{1\over
\sqrt 2} (\g _\mu )^{\a \b } \dh X^\mu ] e ^{i k \cdot X } , \cr &&
\bullet \ V _{A^a _\mu } ^{(-1)}(z) = \l _{A^a _\mu } e ^{- \varphi }
\e _{a \mu } (k) \psi ^\mu e ^{i k \cdot X } , \ V _{A^a _\mu }
^{(0)}(z) = i \sqrt {2\over \a ' } \l _{A^a _\mu } \e _{\mu } (k) [\dh
X ^\mu - i \a ' (k \cdot \psi ) \psi ^\mu ] e ^{i k \cdot X } , \cr &&
\eea where, $\l _ { C^I},\ \l _ { A_\mu ^a }$, denote the CP factors,
$ k$ the 4-d momenta, $u (k), \ \e _\mu (k)$ the Dirac spinor and
polarization vector wave functions for spin $\ud , \ 1 $ particles,
and the suffix labels $ v,\ s$ in $r^ A_{v, s} $ are used to remind
ourselves that boson and fermion modes carry vector and spinor
$SO(10)$ group weight vectors.  In the non-diagonal (`twisted')
sectors, $(a,b)$, the boson and fermion mode vertex operators in the
canonical and once-derived ghost pictures are given by the following
formulas: \bea && \bullet \ V _{C_{\t ^I} } ^{(-1)} (z) = \l _{C_{\t
^I} } e ^{- \varphi } \prod _I (s ^{r _v} _{\t ^I} \s _{\t ^I} ) e ^{i
k \cdot X } , \cr && V _{C_ {\t ^I} } ^{(0)} (z) = i \sqrt {2\over \a
' } \l _{C_{\t ^I} } \bigg [ \sum _J t ^ {r _v} _{\t ^J} \tilde \tau
_{\t ^J} \prod _{I \ne J} ( s ^{r _v}_{\t ^I} \s _{\t ^I} ) - i \a '
(k \cdot \psi ) \prod _{I} ( s ^ {r _v} _{\t ^I} \s _{\t ^I} ) \bigg ]
e ^{i k \cdot X } , \cr && \bullet \ V _{C_{\t ^I} } ^{(-\ud )} (z) =
\l _{C_{\t ^I} } e ^{- {\varphi \over 2 } } u^\a S_\a ( \prod _I s ^{r
_ s} _{\t ^I} \s _{\t ^I} ) e ^{i k \cdot X } , \cr && V _{C_{\t ^I} }
^{(\ud )} (z) = i \sqrt {2\over \a '} \l _{C_{\t ^I} } e ^{\varphi
\over 2} u^\a S_ \a \bigg [ \sum _J ( t ^{r _s} _{\t ^J}\tilde \tau
_{\t ^J} + \tilde t ^{\a _s} _{\t ^J} \tau _{\t ^J}) \prod _{I\ne J}
(s ^{\a_ s} _{\t ^I} \s _{\t ^I} ) + \cdots \bigg ] e ^{i k \cdot X }
,
\label{eqvertx2}\eea where the dots in $V
_{C_{\t ^I} } ^{(\ud )} (z) $ represent $O(k)$ terms of complicated
form that we shall not need in the sequel.  We recommend
Refs.~\cite{kost86,kost87} for further discussions of the vertex
operator construction.

The processes of interest to us in this work involve four massless
fermions belonging to two same or distinct pairs of conjugate modes, $
f=f', \ \t =\t '$ and $ f\ne f' , \ \t \ne \t '$.  In the Polyakov
functional integral formalism for the string world sheet, the
$n$-point open string tree amplitudes are represented by the disk
surface punctured by $n$ points $x_i$ inserted on the boundary with
ghost charge obeying the selection rule, $ \sum _{i =1} ^n q_i =-2 $.
Since the unpunctured disk surface has no moduli, the integration over
the moduli space consists of integrals over the ordered real
variables, $ x_i \in R $, summed over their cyclically inequivalent
permutations and divided by the M\"obius symmetry group, $SL(2,R)$,
generated by the homography transformations of the disk boundary.
Following the familiar Faddeev-Popov procedure of gauge fixing and
division by the volume of the conformal Killing vectors (CKV) group,
one can write the four-point tree string amplitude, $ \cala _4 = \cala
( f (k_1) f ^\dagger (k_2) f '(k_3) f ^ {'\dagger }(k_4) ) $ as the
integral of the vertex operators vacuum correlator, \bea && \cala _4 =
\sum _{\s } \int { \prod _{i=1}^4 dx_i \over V_{CKG} } < V ^{(q_1 )}_{
-\t , (D,A) }( k_1, x_ {\s _1}) V ^{(q_2 )}_{ +\t , (A,B)}( k_2,x_ {\s
_2}) V ^{(q_3 )}_{ -\t ' , (B,C)}( k_3, x_{\s _3} ) V ^{(q_4 )}_{ +\t
' , (C,D)}( k_4, x_{\s _4}) > ,\cr && \label{eqcorr}\eea where we
follow the familiar convention in which all of the particle quantum
numbers are incoming.  The elements of the permutation group quotient,
$ \s \in S_4/C_4$, consist of the three pairs of direct and reverse
orientation permutations, and the integrations are carried over the
ordered sequences of the $x_i$.  The invariance under the $SL(2,R)$
subgroup of the conformal group is used to fix three of the insertion
points, say, at the values, $ x_1=0,\ x_3=1,\ x_4= X \to \infty $,
with the free variable varying inside the interval, $ x_2 \equiv x \in
[-\infty , \infty ] $, so as to cover the three pairs of cyclically
inequivalent permutations, and $ V_{CKG} = 1/X^2$.  We have labelled
the open string vertex operators in Eq.~(\ref{eqcorr}) by the pairs of
associated branes, such that the disk surface is mapped in the $T^2_I$
complex planes on closed four-polygons whose sides are parameterized
by the linear combinations of N coordinates tracing the equations for
the adjacent branes, $D, A,B,C $.  This map is illustrated in
Figure~\ref{mapbr} of Appendix~\ref{apptwist}.  With the world sheet
boundary represented by the complex plane real axis, the reference
ordering of insertion points for the trivial permutation, $\s =1$,
determines the four segments, $ (-\infty , x_1), \ (x_1, x_2), \ (x_2,
x_3), \ (x_3 , x_4 \to + \infty ) ,$ on which the orthogonal linear
combinations of internal coordinate fields, $ \Re (e ^{-i\phi ^I _{ab}
} X ^I) $ and $ \Im (e ^{-i\phi ^I_{ab} } X ^I ) $, obey N and D
boundary conditions, with $\phi ^I_{ab} $ denoting the fixed
interbranes angles at the corresponding insertion points, as displayed
in Eq.~(\ref{eqappbcs}).
The four point string amplitude may thus be represented by the sum of
three reduced amplitudes, \bea && \cala (1234)= {\sum } ' _{\s \in
S_4/C_4} [Trace(\l _{\s _1} \l _{\s _2}\l _{\s _3}\l _{\s _4})
+Trace(\l _{\s _4} \l _{\s _3}\l _{\s _2}\l _{\s _1}) ] \cr && \times
X^2 \int _{-\infty }^{ + \infty } dx < \hat V ^{(-\ud )}_{ -\t }(x_
{\s _1})\hat V ^{(-\ud )}_{ +\t }(x_ {\s _2}) \hat V ^{(-\ud )}_{ -\t
' }(x_{\s _3} )\hat V ^{(-\ud )}_{ +\t '} (x_{\s _4}) > \cr && =
[A(1234) +(2 \leftrightarrow 4)] + [2\leftrightarrow 3]
+[1\leftrightarrow 2] \cr && = [A(1234) + A(4321)] + [A(1324) +
A(4231)] + [A(1342) + A(2431)] , \label{eqconf4} \eea where we have
denoted, $ -\t = \t _A -\t _D = \t _A -\t _B ,\ -\t ' = \t _C -\t _B =
\t _C -\t _D ,$ and introduced the hat symbol to denote the vertex
operators with the CP matrix factor removed, $ V ^{( q )} (x_i) = \hat
V ^{( q )} (x_i) \l _i $. The factor $X^2$ from the gauge fixing
cancels out with the $X$-dependent contributions from the correlator.
With the incoming flat space-time four-momenta denoted by $ k_i$,
obeying the conservation law, $ k_1 + k_2 + k_3+ k_4 = 0$, one can
express the Lorentz invariant Mandelstam kinematic variables as, $ s
=- (k_1 +k_2) ^2 , \ t = - (k_2 +k_3) ^2, \ u =- (k_1 +k_3) ^2 $.  A
compact representation of the reduced amplitudes may be obtained by
considering the definition of the correlator with the dependence on
the kinematic invariants extracted out, \bea < \hat V ^{(-\ud )}_{ -\t
}( 0 )\hat V ^{(-\ud )}_{ +\t }(x )\hat V ^{(-\ud )}_{ -\t ' }(1 )\hat
V ^{(-\ud )}_{ +\t '}(X ) > = x^{-s-1} (1-x)^{-t -1} \calc _{1234} (x)
, \eea while rewriting the second and third reduced amplitudes,
$A(1324)$ and $ A(1342) $, in terms of the first reduced amplitude,
$A(1234)$, through the change of integration variables, $ x \in [1,
\infty ]\to x'= {x-1\over x} \in [0,1] $ and $ x\in [-\infty ,0]\to
x'= {1\over 1- x}\in [0,1] $.  These steps lead to the compact
representation of the disk level string amplitude \bea && \cala
(1234)= \bigg [ [Trace(\l _{1} \l _{2}\l _{3}\l _{4}) +Trace(\l _{4}
\l _{3}\l _{2}\l _{1}) ] \int _{0 }^{1} dx x^{-s-1} (1-x)^{-t -1}
\calc _{1234} (x) \cr && + [Trace(\l_1 \l_3 \l_2 \l _4) + Trace(\l_4
\l_2 \l_3 \l _1) ] \int _{0}^{1 } dx x^{-t-1} (1-x)^{-u} \calc _{1324}
({1\over 1-x}) \cr && + [Trace(\l_2\l_1 \l_3 \l _4) + Trace(\l_4 \l_3
\l_1 \l _2) ]\int _{0}^{1} dx x^{-u} (1-x)^{-s-1} \calc _{1342}
({x-1\over x}) \bigg ] .  \eea

\subsection{String   amplitudes  from world sheet correlators}
\label{sub12}

We discuss here some practical details of use in evaluating the open
string amplitudes for the configuration of non-diagonal sector modes
involved in Eq.~(\ref{eqcorr}) for the amplitude $ \cala (f _1 f
_2^\dagger f '_3 f _4^{'\dagger } )$.  Since the ordering of adjacent
$D6$-branes is determined by that of the vertex operator insertion
points, $ x_i$, we deduce by simple inspection that only the direct
and reverse orientation permutation terms for the first reduced
amplitude, $ A(1234) + A(4321) $, is allowed, while the other two
pairs of reduced amplitudes are forbidden.  Only for symmetric
configurations involving subsets of identical $D$-branes, do
exceptions to this rule occur.

The correlators receive contributions from three sources. There are
first the quantum or oscillator terms coming from the Wick pair
contractions of free field operators, which are determined by the
world sheet field propagators.  The second source is associated with
the CP factors which are grouped inside traces of ordered products.
The third source is associated with the classical action factor in the
functional integral which accounts for the string momentum and winding
zero modes for the coordinate field components along the compact
directions. The heaviest calculational task resides in the coordinate
twist field correlator factor.  The correlation function, $Z _I(x_i)=
<\s _{-\t ^I} (x_1) \s _{\t ^I} (x_2) \s _{-\t ^{'I} } (x_3) \s _{\t
^{'I}} (x_4) > $, is evaluated by making use of the stress energy
source approach initiated by Dixon et al.,~\cite{dixon87} and
Bershadsky and Radul~\cite{bersh87}.  One expresses the constraints
from operator product expansions, holomorphy and boundary conditions
on the two correlators, $ g(z,w; x_i) ,\ h( \bar z,w; x_i) $, obtained
from $ Z ^{I}(x_i) $ by inserting the bilocal operators, $\dh _z X
^{I}\dh _ w \bar X ^{I}$ and $\dh _{\bar z} X ^{I}\dh _ w \bar X ^{I}
$.  The resulting formula for $ Z ^I (x_i)$ comprises two factors
including the contributions from quantum (oscillator) and classical
(zero mode) terms, $ Z ^I(x_i)= Z ^I_{qu}(x_i) \sum _{cl} Z ^I_{cl}
(x_i) $, where the classical summation is over the lattice generated
by the closed 4-polygons with sides along the branes, $A,B,C,D$.  We
have found it useful to provide in Appendix~\ref{apptwist} a
comprehensive discussion of the correlators of open string modes
involving distinct angles, $ \t \ne \t ' $, since this application has
not been addressed in great detail in the literature.  Our
presentation there closely parallels that of B\"urwick et
al.,~\cite{burwick} for the closed string orbifolds.

Two important constraints follow upon requiring that the world sheet
boundary is embedded on closed polygons in the $T^2_I$ planes.  For
the coordinate twist field correlator, $ <\s _{\t _1 , f_{ 1} } (x_1)
\s _{\t _2 , f_{ 2} } (x_2)\s _{\t_3, f_{ 3} } (x_3) \s _{\t _4, f_{
4} } (x_4)>$, the closed four-polygons have edges along the N
directions of the $D$-branes, with vertices $\hat f_i ^\a $ and angles
$\hat \t _i\in [0, 1]$ identified to the intersection points and
angles $ f ^\a _i $ and $\t _i$ of the adjacent branes.  We use here the
index $\a $ to label the intersection points and the notational
convention for the angles, $\hat \t _i = [\t _i , 1-\t _i ] $ for $
\pm \t _i $.  The first condition expresses the obvious geometric
property of the angles, $\sum _{i =1}^4 \hat \t _i ^I = 2 $.  The
second condition is related to the consistent configuration  for the
intersections of the various branes pairs.  Following the initial
discussion for $3$-point couplings by Cremades et al.,~\cite{ibanyuk},
a general comprehensive discussion of this problem was provided by
Higaki et al.,~\cite{higaki05}, whose presentation is closely followed
here. We start by observing that each  pair of branes
$ \a = A,\ B$ intersect at $ I_{AB} = n_A m_B - n_B m_A$ points lying
along the branes $ A, \ B$ with coordinates, $ X_\a = {L_\a k_\a \over
I_{AB} } $, labelled by the integers $ k_A,\ k_B \in [ 0, 1,\cdots ,
I_{AB} -1] $, such that each intersection point is associated with a
unique   choice  for the pair of
integers $ k_A,\ k_B$.  In the case of branes intersecting at the
origin,    solving the complex linear equation, $ X _A = X_B ,\ [ X_\a = \xi _\a L_\a + q_\a
e_1 + p_\a e_2 , \ L_\a = n_\a e_1 + m _\a e_2 , \ \xi _\a \in R, \
(p_\a , q_\a ) \in Z ] $  yields the  explicit representation for the
integer parameters, $ k_\a = n_\a p_{AB} - m_\a q_{AB} ,\ [p_{AB}
=p_{A} - p_{B} ,\ q_{AB} =q_{A} - q_{B} ].$ Since the intersection points are defined
modulo the grand lattice, $\L _{AB} $, generated by $ L_A,\ L_B$, they
form equivalence classes defined modulo the addition of vectors
of $\L _{AB} $.  A convenient way to characterize these $ I_{AB} $
classes is in terms of the shift vectors, $ w_{AB} = X_A -X_B = {L_A
k_A \over I_{AB} } - {L_B k_B \over I_{AB} }$, associated to the
choices of integers $ k_A, \ k_B $ appropriate to the various intersection points.
Since the  vectors $ w_{AB} $ belong to the torus lattice $\L $, generated by
the cycles $ e_1 , \ e_2$, and are defined modulo $\L _{AB}$, they
arise as the independent elements of the lattice coset, $ \L / \L
_{AB} $.  One can also interpret the shift vectors as the $\L $
lattice translations which bring the intersection points on branes $A,\ B$ in
coincidence, or equivalently, as the segments linking the open string
end points located on the branes $A,\ B$.  For the $4$-point
correlator with the configuration of adjacent branes, $ABCD$, the
condition that the $4$-polygon closes, may now be expressed by the
selection rule involving the shift vectors associated to the four
adjacent brane pairs, $ w _{ DA } + w _{AB} + w _{BC} + w _{CD} = 0 $
modulo $\L $.  While the  $ I_{AB}$ independent  classes of shift
vectors  are in one-to-one correspondence with the intersection points, they do not specify the
coordinates of these points which must be calculated independently.
Higaki et al.,~\cite{higaki05} have given a simple useful procedure to
explicitly evaluate the shift vectors.  One  starts by
testing whether the winding numbers of the $A, \ B$ brane pair along
the two lattice cycles are relative primes, by considering their
greatest common divisors (gcd) defined by, $ \text{gcd} (n_A, n_B) =
N_{AB} $ and $\text{gcd} (m_A, m_B) = M_{AB} $. The independent set of
shift vectors is then given by $ w_{AB} = p_{AB} e_2 ,\ [p_{AB} =0, 1,
\cdots , I_{AB} -1] $ if $ N_{AB}=1 $, or by $ w_{AB} = q_{AB} e_1 , \
[q_{AB} =0, 1, \cdots , I_{AB} -1] $ if $ M_{AB}=1 $.  Otherwise, 
for $ N_{AB}\ne 1 ,\ M_{AB}\ne 1 $, the independent set of shift vectors 
can be chosen as, $ w_{AB} = q_{AB} e_1 + p_{AB} e_2 ,\ [p_{AB} =0, 1,
\ \cdots , M_{AB} -1 ,\ q_{AB} =0, 1, \cdots , {I_{AB} \over M_{AB} }
-1 ] $. The above rules readily generalize to the case of $n$-point
correlator, $<\prod _{i=1}^n \s _{\t ^I_i , (A_ i A _{i+1} ) , f ^ {\a
_i} } (x_i) > ,\ [A_{n+1} = A_1]$ where the requirement that the
embedding $n-$polygons, $ A_1 A_2 \cdots A_n$, close in  each $ T^2_I$, is
expressed by the selection rules on the angles and the shift
vectors~\cite{higaki05}, \bea && \sum _{i=1}^n \hat \t _i ^I = (n-2)
,\quad w ^I_{A_ 1 A _{2} } + w ^I_{A_ 2 A _{3} } + \cdots + w ^I_{A_ n A _{1} }
=0. \eea These results hold irrespective of whether the branes $ A_i $
intersect at a common point, chosen  above as the origin of the
coordinate system.  Finally, we  observe that 
there exist a close formal similarity with the shift vectors
and fixed  points, $ w _h,\ f ^\a $, introduced in  $T^2/Z_N$ orbifolds with lattice
$\L $ symmetric under the point group rotations, $\vt ^h , \ [h=0, 1,
\cdots , N] $ by using the definition, $ (1-\vt ^h) (f ^\a +\L ) = w
_h ^{f ^\a } $. The shift vectors described by the lattice
translations which bring the corresponding fixed  points in coincidence with
themselves after applying the rotation $\vt ^h $, arise here as the
representative elements of the lattice coset, $ \L / ( 1-\vt ^h ) \L
$.   However, it is  important to  realize  that    for the open
strings in intersecting  brane  models, in contrast to the  
closed  strings in   orbifold  models, the   selection rules have  nothing  to 
do  with   the point  and space  group symmetries  of the torus
lattice.  

For the four-point string amplitude in the equal angle case, $\t = \t
'$, the embedding 4-polygon is a parallelogram, so that the selection
rule takes the simple geometric form, $ f_2 - f_1 + f_4 -f_3 =0$, in
terms of the intersection points $ f_1,\ \cdots , f_4$.
Since the intersection points are naturally associated with the
generation (flavor) quantum numbers of matter modes, one might wonder
whether generation non-diagonal four fermion processes may be allowed
at the tree level, only subject to suppression from the classical
action factor.  However, the combined constraints from 
gauge symmetries and  the above tree level selection rules on
angles and intersection points, are seen to conserve flavor and hence
disallow the flavor changing neutral current processes.  Assuming for
the sake of illustration that the intersection points label the quark
flavors, $ f_i (q)$, one indeed finds that the $\D S = 1,\ 2$
strangeness changing processes, $s^\dagger _1 d_2 d ^\dagger _3 d_4 $
and $ s^\dagger _1 d_2 s^\dagger _3 d_4 ,$ require the relations, $
f_2(d) - f_1 (s) =0 $ and $ f_2(d) - f_1 (s) + f_4(d) - f_3 (s) =0$,
which cannot be satisfied unless the intersection points for $d, \ s $
quarks are coincident.  From these observations it follows that the
matter fermions trilinear effective Lagrangian couplings with the
non-localized massless or massive boson modes are necessarily flavor
diagonal.  The quark and lepton flavor symmetries are broken only by
the fermions Yukawa couplings with the electroweak Higgs bosons with
the flavor mixing arising in the familiar way through the fermion mass
generation mechanism.

The 4-d space-time structure of amplitudes is strongly restricted by
the symmetry constraints.  The GSO projection correlates the
helicities $r_A (i) $ along the internal and non-compact space
directions (odd number of $ -\ud $ for fermions), as already noted,
while the selection rules from the $SO(10)$ Lorentz symmetry group
imposes the $ H_A$ momentum conservation, $ \sum _{i=1}^n r_A (i) =0$,
summed over the $n$ modes of the correlator. 
The $H_I$ momentum conservation,  following from the symmetry under the
$SO(6)$ space group,  imposes the conditions on the branes intersection
angles, $ \sum _{i=1} ^n \t _i ^I =0 ,\ [ -1 < \t ^I _i <1 ,\ I=1,
2,3]$ which identify with the previously quoted selection rule.
 These conditions often
suffice to uniquely determine the Lorentz group covariant structure of
matrix elements with respect to the Dirac spinors.  For the
configuration, $f_1 (-\t ) ,\ f_2 ^\dagger ( \t ) ,\ f_3 (-\t ' ) ,\
f_4 ^\dagger ( \t ') $, the restrictions on the spinor weight
solutions for the massless localized fermions entail that only the
configurations involving pairs of conjugate modes with same or
distinct angles, $\pm \t ^I $ are allowed, so that only the reduced
amplitude $A(1234)$ is non vanishing.  Since the fermions in the two
pairs of conjugate states with opposite space-time chiralities require
setting the weight vectors as \bea && -r^A (1)= (---,+-) ,\ r^A (2)=
(---,--) ,\cr && -r^A (3)= (---,-+) ,\ r^A (4)= (---,++), \eea
the Dirac spinors can only be contracted via the 10-d vectorial
coupling, $ (u ^T _1 C \G _ M u _2 ) (u ^T _3 C \G ^ M u _4 ) $, which
reduces in 4-d to the matrix element with vector contraction, $(\bar u
_1 \g _ \mu u _2)(\bar u _3 \g ^\mu u _4) .$ This unique structure, up
to Fierz reordering of Dirac spinors, is antisymmetric under all of
pair permutations of the (commuting $c$-number) spinor factors, as it
should be.  Note that the scalar coupling of Dirac spinors would
appear upon considering configurations mixing the fermion and
antifermion modes, $f $ and $ f ^c$.  Translating between different
structure of the Dirac spinors matrix elements is conveniently
performed by making use of the 4-d Fierz-Michel identities, given for
the $c$-number Dirac spinors by \bea && \bar u_{1L} ^c \g ^\mu u_{2L}
^c =\bar u_{2R} \g ^\mu u_{1R} ,\ (\bar u_{1H} \g ^\mu u_{1H} )(\bar
u_{3H} \g _\mu u_{4H}) =- (\bar u_{1H} \g ^\mu u_{4H})( \bar u_{3H} \g
_\mu u_{2H} ) , \ [H=L, R] \cr && (\bar u_{1L} \g ^\mu u_{1L })(\bar
u_{3R} \g _\mu u_{4R}) =2 (\bar u_{1L} u_{4R})( \bar u_{3R}
u_{2L}). \eea

\subsection{Orbifold compactification} 
\label{sub13}

The covering space formalism of orbifold compactification is developed
by including all the states produced by the orbifold group action
prior to projecting on the physical states invariant under the
orbifold symmetry.  We restrict consideration to the subset of Abelian
orbifolds, $T^6 /Z_N $, with the cyclic groups generated by the order
$N$ unitary matrices, $\T \in SU(3) ,\ [\T ^N =1]$ yielding $\caln =2
$ supersymmetry in the closed string sector.  The complexified bases
of coordinate and spinor fields of the symmetric 6-d factorisable
tori, $T^6 =\prod _I T^2_I$, transform by the diagonal unitary matrix
transformations \bea && X ^I_{L,R} (z) \to \T ^g X ^I_{L,R} (z) , \
\psi ^I_{L,R}(z) \to \T ^g \psi ^I_{L,R} (z) , \cr && [\T ^g =
\text{diag} \ (e ^{2 i \pi g v ^1 } ,\ e ^{2 i \pi g v ^2 } ,\ e ^{2 i
\pi g v ^3 } ) , \ g=0, 1, \cdots , N-1] \eea where the generator $\T
$ is represented by the twist vector, $ v = [v^I] $, satisfying the
conditions, $\sum _I v^I =0,\ N v^I = 0 \ \text{mod} \ N .$
For the compactification on $ M^4 \times T^6/ (Z_N + Z_N \O \calr )$,
with the orientifold point symmetry group including the elements, $ \O
\calr \T ^g, \ [g=0, \cdots , N-1]$ one must require that the
generator $\O \calr $ acts crystallographically on the 6-d torus $ T^6
$.  This introduces conditions on the torus moduli which transform
certain continuous vacuum degeneracies into discrete
ones~\cite{Blumenhagen:1999md,Blumenhagen:1999ev,angela99}.  Thus, for
$T^2$ tori, the reflection $\O \calr $ has only two inequivalent
actions up to coordinate rescalings. The first corresponds to the
diagonal reflection about one of the two torus cycles, say, $ e_1$
(Case $ \bfA $) and the second to the reflection about the diagonal
sum of cycles, say, $ e_1+e_2$ (Case $ \bfB $). An equivalent action
to Case $ \bfB $ corresponds to the diagonal coordinate reflection
about the single cycle, $ e_1$, followed by a complex rotation, $ X ^I
\to e ^{i\g ^I} \bar X^ { I}$.  Explicit solutions for the allowed
$\bfA ,\ \bfB $ lattices have been obtained in the various Abelian
orbifolds~\cite{blumz402,honec04}.  Extensions to non-factorisable
tori~\cite{blumlonsu04} and to exceptional cycles in orbifolds and
smooth Calabi-Yau manifolds~\cite{blumlust02} have also been discussed
in the literature.

The invariance under the orientifold group, $ Z_N \O \calr $, produces
$N$ distinct orientifold planes, $ O6_g$, defined as the fixed loci of
$ \O \calr \T ^g , \ [g=0, 1, \cdots , N-1]$.  Interpreting the
operator identity, $ \T ^{ -g/2} \O \T ^{g/2} =\O \calr \T ^g $, as a
similarity transformation by the generator $\T ^{-g/2} $, shows that
the $ O6_g$-planes trace in $ T^2 _I$ a sequence of $N$ lines related
by the half-rotation angles, $\T ^{ -\ud } $, with $ O6_g = \T ^ { -
{g \over 2} } O6_0$.  The orbifold symmetry also imposes conditions on
the individual $ D6$-brane stacks and the open string sectors.  For
the case of pairs of brane stacks $\mu , \ \nu $ passing by orbifold
fixed points, the invariance under $ \T ^g , \ \O \calr \T ^g $ is
ensured by requiring that the CP gauge factors of open string sectors,
$ (\mu ,\nu ) $, realize projective representations for the gauge
embedding unitary matrices, $ \g _{ \T ^g } , \ \g _{ \O \calr \T ^g }
$, obeying the projection conditions \bea && \l _A = \g _{ \T ^g , \mu
} \l _A \g _{ \T ^g , \nu } ^{-1} , \ \l _A = - \g _{\O \calr \T ^g ,
\nu } \l _A ^T \g _{\O \calr \T ^g , \mu } ^{-1} , \eea holding for
$\mu =\nu $ or $\mu \ne \nu $, where one must allow for mode dependent
complex phase factors determined by the quantum numbers of the modes.
The RR tadpole cancellation conditions generally admit the simple
solution involving traceless twist gauge embedding matrices for the
orbifold group elements, $ Trace (\g _{\T , \mu } )= 0 $.

 For the case of branes intersecting the orientifold planes at generic
angles, $ \t _a ^I$, hence not traversing the orbifold fixed points,
the rotations $\T ^g $ act non-trivially on brane stacks, so that the
CP factors are not constrained.  To ensure that the $D$-brane setup is
orbifold group invariant, one must include for each stack of $ D6_\mu
$-branes its $N-1$ images $ D6_{\mu _g}$ under the rotations $\T ^g$,
and similarly for the mirror images, $ D6_{\mu '_g}, \ [g=0, 1, \cdots
, N-1]$.  Thus, the intersecting $D6 _\mu $-brane stack wrapped around
the three-cycles $ [\Pi _\mu ] $ is made symmetric under the $\T ^g $
identification of $T^6$ by introducing the image $D6 _{\mu _g}$-brane
stack wrapped around the image three-cycles, $\Pi _{\T ^g \mu } $.
Each $D6_\mu $-brane stack at generic angles is then described by the
equivalence class (orbit) of $N-1 $ branes, rotated images of the
reference brane, $ \mu _g = \T ^g \mu $, accompanied by the orbit of $
N $ rotated mirror image branes, $ \T ^g \mu '= \O \calr \T ^g \mu $.
The non-diagonal open string sectors are described by the orbits, $
(\mu , \nu _{g } ) \sim (\nu '_{g }, \mu ' ) ,\ ( \mu , \nu '_{g } )
\sim (\nu _{g }, \mu ') $.  Note that the reverse orientation pairs, $
(\mu , \nu _g) \sim ( \nu _g,\mu ) ^\dagger $, are related by
conjugation, as discussed previously, and that being identical, the
modes $ (\mu , \nu _g)$ and $ (\mu _k , \nu _{{g+k}} ) $ need not be
included simultaneously.  The $Z_N+ Z_N \O \calr $ group elements are
represented on the 2-d vector space of wrapping numbers by the matrix
transformations \bea && \T ^g \cdot {n_\mu \choose m_\mu } = \T ^g
{n_\mu \choose m_\mu } = {n_{\mu _g} \choose m_{\mu _g} } ,\ \O \calr
\T ^g \cdot {n_\mu \choose m_\mu } = \T ^ {'g} {n_\mu \choose m_\mu }
= {n_{\mu '_g} \choose m_{\mu '_g} } , \eea such that the column
vectors of lattice cycles, $ {e_1 \choose e_2} $, transform by the
transposed matrices, $\T ^g $ and $ \O \calr \T ^g .$
The 3-cycle volume is a function of the cycles equivalence classes
given by the product of three one-cycle lengths, \bea && V(Q_\mu)=
\prod _{I=1}^3 (2\pi L_\mu ^I ) , \ [L_\mu^ {I2} = (n_\mu ^Ir^ I_1 \
m_\mu ^I r^ I_2) \cdot g ^I \cdot {n_\mu ^I r^ I_1 \choose m_\mu ^I r^
I_2} ] \eea with $ g^I$ denoting the $2\times 2 $ matrix for the
metric tensor in the lattice coordinate basis.

The diagonal orbit for the $N_a$ mirror stacks, $ (D6_{[a]}, \
D6_{[a']}) = (D6_{a_g}, \ D6_{a'_g}) $, generates the gauge symmetry
group, $ U(N_a)$, as the diagonal subgroup of the $ 2N$ direct product
of group factors.  The $N^2$ sectors $ ([a], [b] ) = (a_{g_1}, b
_{g_2})$ for the pair of $ N_a, \ N_b$ stacks of $ D6_a/D6_b$-branes,
carrying the gauge group $ U(N_a) \times U(N_b)$, consist of $N$
distinct subsectors, $ (a , b_g) $, labelled by the relative
rotations, $ g = g_2 g_1 ^{-1} $.  The $N$ distinct mirror subsectors,
$ (a ', b_g) $, are similarly defined.  The off-diagonal open string
sectors, $([a],[b]) \sim (a, b_g) $ and $([a],[b']) \sim (a,b '_g) $
carry bifundamental representations for $U(N_a) \times U(N_b) $ whose
multiplicities must be combined algebraically.  Recalling that
opposite signs are assigned to the complex conjugate modes of opposite
helicities, one can express the net chiral multiplicities of the
bifundamental modes as \bea && (N_a, \bar N_b):\ I ^{\chi } _{a b } =
\sum _{g=0}^{N-1} I_{ a b_{g } } ,\ ( N_a, N_b ):\ I ^{\chi } _{ a b'
} = \sum _{g=0}^{N-1} I_{ a b ' _{g } } ,\cr && [ I_{ab_g} =\prod _I (
n^I _a \tilde m^I _{b_g} - n^I _{b_g} \tilde m^I _a ),\ I_{ab'_g} =
\prod _I - (n^I _a \tilde m^I _{b_g} + n^I _{b_g} \tilde m^I _a) ]
\eea where the summations extend over the $N$ distinct subsectors in a
given equivalence class belonging to the same gauge group
representations.  The non-chiral spectrum may thus be expressed as, $
(I_{ab} + \D _{ab} ) (N_a ,\bar N_b) + \D _{ab} (\bar N_a , N_b) $ and
$ (I_{ab'} + \D _{ab'} ) (N_a ,N_b) + \D _{ab'} (\bar N_a ,\bar N_b)$,
with the model dependent integer numbers of vector pairs denoted by, $
\D _{ab}$ and $ \D _{ab '}$.

The diagonal sectors, $([a], [a]) $ and $([a], [a ']) $, include the
$N$ distinct subsectors, $ (a, a_g)$, carrying the adjoint
representation ${\bf Adj _a} $ of $U(N_a)$, and the $N$ distinct
subsectors, $ (a, a '_g)$, carrying the antisymmetric and symmetric
tensor representations, ${\bf A_a,\ S _a } $, of $U(N_a)$. The net
chiral multiplicities for these modes are given by \bea && {\bf Adj
_a} :\ I ^{Adj} _{a}= \ud \sum _g I_{aa_g} ,\
    [ {\bf A }_a ,\ {\bf S }_a ] :\ I ^{\chi [A_a, S_a ]} _{aa'} =
\sum _g \ud I_{aa' _g} (1 \pm {1 \over \prod _I n ^I_{a_g} } ) , \eea
where the multiplicity for the adjoint (real) representation modes is
halved in order not to double count the equivalent charge conjugate
pairs, $(a, a_g)$ and $(a, a_{N-g} ) $.  Since the rotation angles in
the supersymmetric type $Z_N$ orbifolds are given by $SU(3)$ unitary
matrices, the adjoint representation modes, $ ([a], [a]) $, are
localized at branes with angles obeying the supersymmetric cycle
conditions, $ \sum _I \t ^I _{ab} = g \sum _I \T ^I = 0 \ \text{mod}
(1) $. Hence, they form chiral supermultiplets of $\caln = 1$
supersymmetry.  By contrast, for branes intersecting at generic
angles, none of the non-diagonal sector modes form chiral
supermultiplets.  We also note that the extra vector mode pairs, $ \D
_{ab} $, in bifundamental representations are expected to decouple
through the tree level Yukawa couplings, $ f _{ab} f ^\dagger _{ab} K
_{a} ,\ f ^\dagger _{ab} f_{ab} K _{b} $, involving the singlet
components of the adjoint representation scalar modes, $K _a $ and $K
_b $ of $U(N_a) $ and $U(N_b) $.

\section{Tree level string amplitudes for baryon number violating processes}
\label{sect2}

We discuss in the present section the string amplitudes for the baryon
number violating tree level processes taking place in the gauge
unified models with intersecting branes.  For the familiar two-body
nucleon decay channels into meson-lepton pairs, the dominant
contributions arise from the subprocesses involving four matter
fermion fields and, in supersymmetry models, from the subprocesses
involving two pairs of matter fermions and sfermions, which must be
subsequently dressed by one-loop gaugino or higgsino exchange
mechanisms.  The low energy limit of these amplitudes is represented
by baryon number violating local operators of dimension $6$ and $5$,
obeying the selection rules, $ \D B = \D L = -1$.  Other exotic
nucleon decay channels can also arise from higher order subprocesses
involving either four fermions interacting with a gauge or scalar
boson or six fermions~\cite{weinbergs}. These are represented by
dimension $7$ and $9$ operators obeying the selection rules, $ \D B =
- \D L = -1$ and $ \D B = -2, \ \D L = 0$. We shall present here a
detailed treatment for the former processes but only a qualitative
treatment for the latter.

\subsection{Four fermion processes} 
\label{sub21}

\subsubsection{General structure of amplitude} 

The string amplitude for processes involving two distinct conjugate
pairs of incoming massless fermions, $ \cala '\equiv \cala ' (f _1 f
_2^\dagger f'_3 f _4^{ ' \dagger } ) $, is obtained from the general
formula in Eq.~(\ref{eqcorr}) in the simplified form
\bea && \cala ' = (\calT _1 + \calT _2) X ^2 \int _0^1 dx <\hat V^{
  -\ud } _{- \t , (D,A)} (0) \hat V^{-\ud } _{ \t , (A,B)} (x) \hat
V^{-\ud } _{- \t ' , (B,C) } (1 ) \hat V^{-\ud } _{ \t ' , (C,D) } (X)
> , \cr && [ \calT _1 = Tr( \l _1 \l _2 \l _3 \l _4),\ \calT _2 =
  Tr(\l _4 \l _3 \l _2 \l _1) ] \eea involving only the reduced
amplitude $A(1234)$ and its reverse orientation counterpart, since the
other two permutations refer to forbidden target space embeddings.  We
find it convenient in the present work to introduce the primed
amplitude, $ \cala ' $, obtained by removing the space-time momentum
conservation factor, $ \cala ' \equiv \cala / [i (2\pi ) ^4 \d ^{(4)}
  (k_1 + k_2 + k_3 + k_4)] $.  Making use of the useful formulas \bea
&& < e ^{- {\varphi (x_1) \over 2 } } \cdots e ^{- {\varphi (x_4)
    \over 2 } } > = [x (1-x) ] ^{-1/4 } , \ < e^{ik_1 \cdot X(x_1) }
\cdots e^{ik_4 \cdot X(x_4) } > = x^{-s} (1-x)^{-t} , \cr && (u _1^
       {\a _1} \cdots u _4^ {\a _4} ) < S _{\a _1 } (x_1) \cdots S
       _{\a _4 }(x_4) > = x ^{-\ud } (1-x) ^{-\ud } (\bar u_1 \g ^\mu
       u_3 ) (\bar u_2 \g ^\mu u_4) , \cr && <s _ {-\t ^I } (x_1)
       \cdots s _ {+ \t ^{'I} } (x_4) > = x ^{- (\t ^I -\ud )^2 }
       (1-x) ^{- (\t ^I -\ud ) (\t ^{'I} -\ud )}, \cr && \hat Z ^ I
       (x) = Z _{qu } ^I (x) Z _{cl} ^I (x) = <\s _ {-\t ^I } (x_1)
       \cdots \s _ {+ \t ^{'I} } (x_4) > ,\cr && [ Z _{qu} ^I (x) = C
	 _\s x ^{ -\ud \t ^I(1-\t ^I) } (1-x) ^{ -\ud \t ^I(1-\t ^{'I}
	   ) - \ud \t ^{'I} (1-\t ^I) } I _I ^{-\ud } (x) ] \eea one
       finds that the dependence on intersection angles in the power
       exponents of $x $ and $(1-x)$ cancels out upon combining the
       various correlator factors.  The coordinate twist field
       correlator consists of quantum and classical partition function
       factors, $\hat Z _{qu } ^I (x) $ and $ Z _{cl } ^I (x) $, which
       are evaluated in Appendix~\ref{apptwist}.  The combined
       contributions from the trace over CP factors, the Wick
       contractions of superconformal ghost fields and of spinor twist
       fields and the coordinate twist field correlator lead to the
       following final formula for the string amplitude \bea && \cala
       '
= C' \int _0^1 dx \bigg [\cals _1 \calT _1 x^{-s-1} (1-x)^{-t-1}
I^{-\ud } (x) \sum _{cl} e ^{-S ^{(1)}_{cl} (x) } \cr && - \cals _2
\calT _2 x^{-t-1} (1-x)^{-s-1} I^{' -\ud } (x) \sum _{cl} e ^{-S
^{(2)}_{cl} (x) } \bigg ] \cr && = C (\calT _1+ \calT _2 ) \cals _1
\int _0^1 dx x^{-s-1} (1-x)^{-t-1} \prod _I \bigg ( 2\sin (\pi \t ^I )
I _I^{-\ud } (x) \sum _{cl} e ^{-S ^{I}_{cl} (x) } \bigg ) , \cr &&
[\cals _1 =(\bar u_1 \g ^\mu u_2) (\bar u_3 \g _\mu u_4),\
\cals _2 =(\bar u_1 \g ^\mu u_4) (\bar u_3 \g _\mu u_2), \cr &&
I(x) = \prod _I I_I (x),\quad I_I (x) = {\sin (\pi \t ^I) \over \pi }
(B_2 \bar G_1 H_2 + B_1 \bar G_2 H_1 ) , \cr && C' = C \prod _I 2\sin
(\pi \t ^I),\ C = 2 \pi g_s ] \cr && \label{eqampstr}\eea where the
coefficients, $ B _1, B_2 $, and the functions, $ G_1 (x) , G_2(x) ,
H_1(x) , H_2(x) $ and $I_I (x)$, are defined by the formulas quoted in
Eq.~(\ref{app.ampqu}), while the classical action factors, $ S
^{(1)}_{cl} (x) ,\ S ^{(2)}_{cl} (x) $, are defined in
Eqs.~(\ref{eqclassact}) of Appendix~\ref{apptwist}, and will be
discussed in more detail in the next subsection.  The equality of the
direct and reverse orientation reduced amplitudes, which is used in
obtaining the second form of the amplitude with the dependence on CP
factors factored out, follows in a non-trivial way from the selection
rules on the branes intersection points and angles, the relation
between the Dirac spinor matrix elements, $ \cals _1 = - \cals _2$,
differing by the permutation $ 2\leftrightarrow 4$, and the
transformation properties of the function $ I _I(x)$ and of the
classical action under the $x \to 1-x$ change of integration variable,
$ I ' _I (x) \to I _I (1-x),\ S _{cl} ^{(2)} (x) \to S_{cl} ^{(1)}
(1-x) $.  We shall keep track in intermediate results of the direct
and reverse orientation reduced amplitudes, $ A(1234) + A(4321) $,
despite the fact that these are equal in orientifold models.

The overall normalization factor $C = 2\pi g_s $ in
Eq.~(\ref{eqampstr}) is determined from the factorization of the low
energy amplitude with respect to the massless gauge boson $s$- and
$t$-channel exchange poles. These are asssociated with the
contributions from the regions $ x \to 0$ and $ x \to 1$, as will be
discussed in detail below.
For a shortcut derivation at this stage, we consider the large radius
limit, $r \to \infty $, where the classical action factor reduces to
unity, $ e ^{ -S _{cl } ^I }\to 1$,
and the $ x \to 0$ limit of the amplitude reproduces the $s$-channel
massless gauge boson exchange pole of the 7-d gauge theory on the $D6$
world brane with gauge coupling constant, $ g^2 _{7} = g^2 _{D6} =
2\pi g_s \a ' (2\pi \sqrt {\a '} ) ^3 $.  Matching the leading
massless pole term in the string amplitude $ \cala ' _{st} $ to the
massless gauge boson exchange term in the field theory amplitude,
$\cala ' _{ft} $, suitably transformed by applying the closure formula
for the trace over four CP factors, \bea && \bigg [\cala ' _{st}
\simeq C (\calT _1 + \calT _2) \cals _1 \pi ^{3\over 2} \int _0 ^1 dx
x^{-s-1} (\ln (1/x) )^{-{3\over 2 } }
\bigg ] \cr && = \bigg [\cala ' _{ft} = g^2 _{D6} (\calT _1 + \calT
_2) \cals _1 \int {d^3 \vec q \over (2\pi )^3} {1\over \vec q ^2 -s }
\bigg ] , \eea yields the previously quoted result~\cite{KW03}, $ C =
{ g^2 _{D6} \over (2\pi \sqrt {\a '} ) ^3 } = 2\pi g_s \a ' $.

\subsubsection{Classical  action factor} 

The tree level contributions to the classical partition function from
zero modes , $ Z_{cl} (x) = \sum _{cl} e ^{ -S_{cl } } \sim \sum _{cl}
e ^{ - {\text{Area} \over (2 \pi \a ') } } $, are represented by the
sum over the $T^6$ embedding of the disk on the lattice of
four-polygons weighted by the exponential of the classical action
which identifies with the polygons area. These correspond formally to
the holomorphic and antiholomorphic instanton embeddings of the disk
on $T^6$.  The classical action term in the reduced amplitude $ \cala
(1234)$ contains a factor for each $T^2_I$ given by a double series
sum over the lattice of the large 2-d tori with cycles given by the
$D6_A/D6_B$-brane segments, $ (L ^I_A, \ L ^I_B )$, images of the disc
intervals, $ (x_1, x_2),\ (x_2, x_3) $, as illustrated by
Figure~\ref{mapbr} in Appendix~\ref{apptwist}.

Using the complex number notation for the $T^2$ plane coordinates, we
denote the equations for the branes $A,\ B$ as, $ X_A = \xi _A L_A, \ X_B=
\xi _BL_B ,\ [ \xi _A ,\ \xi _ B \in R]$; the position of intersection
points as, $ f_1 = X(x_1) , \ f_2 = X(x_2),\ f_3 = X(x_3) $; the
straight line distances between them as, $ \d ^A_{12} = f^A_2 - f^A_1
,\ \d ^B_{23} = f^B_3 - f^B_2 $; and the winding numbers around the
large tori as, $ v_A , \ v_B $.  For notational simplicity, we
suppress here the index $I$ of the $T^2 _I$ complex planes.  While
superfluous, the suffices $A,\ B$ on $ f_i $ and $\d _{ij} $ are
retained for the sake of mnemonics.  Upon circling the cycles
$\calc_1, \ \calc_2$ surrounding the segments $(x_1,x_2),\ (x_2,x_3)$
the classical part of the coordinate fields along the branes $ A,\ B$
transform by elements in the lattice $(L_A, L_B)$ shifted by the
distance of intersection points.  The extended boundary conditions are
expressed by the monodromies \bea && \sqrt 2 \D _{\calc _1} X = 2 \pi
v_A = 2\pi (1 - e ^{2 i \pi \t } ) (\d ^A _{12} + p_A L_A ) , \ [\d
^A_{12} = f^A_2 -f^A_1 , \ L_A= n_A e_1 + m_A e_2 ,\ p_A \in Z ] \cr
&& \sqrt 2 \D _{\calc _2} X = 2 \pi v_B = 2\pi (1 - e ^{2 i \pi \t } )
(\d ^B _{23} + p_B L_B ), \ [\d ^B_{23} = f^B_3-f^B_2 ,\ L_B= n_B e_1
+ m_B e_2 , \ p _B \in Z ]\cr && \label{eqmon} \eea using the
definition in Eq.~(\ref{eqglomo}), where the integers $ p_A, \ p_B$
label the winding numbers of classical solutions and the factors $2\pi
$ represent the $T^2 _I $ tori periodicities. The factors depending
explicitly on the open string sector intersection angles, $\t _{DA}
\equiv - \t $, reflect our use of closed contours, $\calc _1 = C_1 -
C'_1$ and $\calc _2 = C_2 - C'_2$, composed of the mirror contours
around $ (x_1, x_2) +(x_2, x_1) $ and $ (x_2, x_3) + (x_3, x_2) $,
while noting that the coordinates along the lower and upper paths are
related by the complex rotation of angle, $ 2 \pi \t _{DA} = - 2 \pi
\t $.

A convenient parameterization for the $D$-brane equation is obtained
by introducing the longitudinal $(L)$ and transverse $(t)$ vector
directions with respect to the wrapped cycle, $ (n, m)$, \bea && L
\equiv L _1+iL_2 = n e_1 +m e_2 = (m + \tau ' n) e_2 = (- n \tau ' _2
+ i\tilde m ) , \cr && [\tilde m= m +n \tau '_1 , \ \tau ' \equiv {e_1
\over e_2} \equiv \tau '_1 + i\tau '_2 = -U, \ \vert L \vert = (\tilde
m^2 + n ^2 \tau _2 ^{'2} ) ^\ud  =  (L ^\star L ) ^\ud  ] \cr && t\equiv t^1+it^2 ={1\over
\vert L \vert } (\tilde m + i n \tau ' _2 ) ,\ [\vec t \cdot \vec L =0
= \Re (t ^\star L )=0, \ \vec t ^2 = \Re (t ^\star t )=1] \eea
where we continue using the  case  of  up-tilted torus,  while 
setting  the length scale  to unity,  $ e_2 =i$,  for simplicity. 
The c-numbers $ L$ and $ t$ may also be represented geometrically
by the 2-d orthogonal vectors $\vec L $ and $\vec t $ with Cartesian
components given by the real and imaginary parts of $ L$ and $ t$.
The segments joining intersection points, $\d _{12} ^A ,\ \d _{23} ^B
$, decompose into the longitudinal components, $ \e ^A_{12} L_A , \ \e
^B _{23} L_B $, and transverse components, $ d^A _{12} = \eta _{12}^A
t _A, \ d^B _{23}=\eta _{23}^ B t_B$.  The equations for the branes
$A,\ B$ are then represented as, $ X _{\a } = \xi _{\a } L _{\a } , \
[\xi _{\a } \in R,\ \a = A, \ B ]$ those of the wrapped cycles as, $ X
_{\a }= p_{\a } L_{\a } $, while the segments joining the brane
intersection points are decomposed in longitudinal and transverse
directions as, $\d _{\a }= \e _{\a } L _{\a } + d _{\a } \equiv \e
_{\a } L_{\a } + \eta _{\a } t _{\a } ,\ [\e_{\a } ,\ \eta _{\a } \in
R ] $ so that the grand tori cycles have squared length given by, $
(2\pi )^2 [(p_{\a } +\e _{\a } )^2 \vert L _{\a } \vert ^2 + \eta_{\a
} ^2 ]$.  In the present notations, the monodromy conditions in
Eq.(\ref{eqmon}) can be rewritten as, $ \sqrt 2 \D _{\calc } X = 2 \pi
v = 2 \pi (1-e^{2i \pi \t } ) [(p +\e ) L + \eta t ],\ [p\in Z ; \ \e
, \eta \in R] $
and the classical action may be expressed by the quadratic form \bea
&& S_{cl}=V_{11} \vert v_A \vert ^2 +V_{22} \vert v_B \vert ^2 + 2 \Re
(V_{12} v_A v_B ^\star ) \cr && = V '_{11} [ (p _A+\e _{12}^A )^2
\vert L_A \vert ^2 + d_{12} ^{A 2} ] + V' _{22} [(p _B+\e _{23}^B )^2
\vert L_B \vert ^2 + d_{23} ^{B2} ] \cr && + 2 \Re [ V ' _{12} ((p
_A+\e _{12} ^A ) L_A + d_{12}^A ) ((p _B+\e _{23}^B ) L_B + d_{23}^B )
^\star ] ,\label{sec2.clVV} \eea where $ V_{ij}, \ [V'_{ij} = V_{ij}
\prod _I (2\sin \pi \t ^I) ,\ (ij) =[11,22,12] ] $ are defined in
Eqs.~(\ref{app.V11}) in terms of the Hypergeometric functions of the
integration variable $x$ with parameters depending on the angles $ \t
, \ \t '$.  Useful limiting formulas for these functions are quoted in
Eqs.~(\ref{app.0G2}) and (\ref{app.1G2}).  The extra factors $ 2\sin
(\pi \t ^I ) $ in the  primed coefficients $V ' _{ij}$, arise as a result of
expressing the global monodromies in terms of closed rather than open
contours.  The low energy field theory limit of the classical
partition function factor is determined by the end point regions of
the $x$-integral.  These contribute an infinite series of pole terms
associated to the exchange of massive string excitations, with
infinite singular terms with respect to the kinematical variables $s,
t, u$, occurring whenever massless poles are exchanged in the relevant
channels.  Upon including the classical action contributions, the
classical partition function must be transformed by the Poisson
resummation formula prior to the analytic continuation of the
$x$-integral, in order to ensure the instanton series convergence.
Considering, for instance, the unequal angle case, $\t \ne \t '$,
where the massless gauge boson pole occurs only in the $s$-channel via
the $x^{-1}$ singularity, then the use of the limiting formulas in
Eqs.~(\ref{app.0G2}) and (\ref{app.1G2}) shows that $V_{22}$ is the
only function in the classical action, Eq.~(\ref{sec2.clVV}), which
vanishes in the limit $x\to 0$, hence indicating the need to perform
the Poisson formula resummation on the $p_B$ series.  The resulting
modified low energy representation of the classical partition function
factor in the case, $ d_A \ne 0,\ d_B = 0,$ reads as \bea && \sum
_{cl} e^{-S_{cl} } = \sum _{p_{A} , p_B \in Z } e^{- [ (p_A +\e _A) ^2
+ d_A ^2 ] \vert L_A \vert ^2 V ' _{11} - (p_B +\e _B ) ^2 \vert L_B
\vert ^2V ' _{22} - 2 (p_A +\e _A) (p_B +\e _B) \Re (L _A L ^\star _B
V'_{12} ) } \cr && = e^{- V '_{11} \vert d_A \vert ^2 } ({\pi \over V
'_{22} \vert L_B \vert ^2 })^\ud \sum _{p'_B, p_A} e^{-V '_{22} \vert
L_B \vert ^2 (\Im (\a _{BA} ) )^2 } e ^{-{\pi ^2 p_B^{'2} \over V
'_{22}\vert L_B \vert ^2} } e^{-2i\pi p'_B \Re (\a _{BA} ) } e^{ -(p_A
+\e _A)^2 \vert L_A \vert ^2 ( V '_{11} -{ \vert V '_{12} \vert ^2
\over V '_{22} } ) } ,\cr && [\a _{BA}= \e _B +(p_A+\e _A) { V
'_{12}L_A \over V '_{22} L_B } ,\ V '_{ij} \equiv V _{ij} \prod
_{I=1}^3 2\sin (\pi \t ^I) ]
.  \label{eq.partpoisson} \eea The analogous Poisson resummation at
the end point, $x\to 1$, performed on the sum over $p_A$, is obtained
from the above formula by substituting, $ A \leftrightarrow B, \
V_{11} \leftrightarrow V_{22} $.

We consider at this point the  brane stack parallel splitting process
which realizes the unified gauge symmetry breaking.  Since the
resulting massive gauge boson arises from the open strings stretched
between distant pairs of brane substacks, its mass $M_X $ is related
to the minimal interbrane transverse distance, $ d_A $, by the
familiar term in the string mass squared spectrum, $ M _X^2 \propto {
d _A ^{2} } $. For our configuration $ DABC$ of branes, the
relationship can be derived from the $s$-channel mass spectrum of the
open string sector $(B, D)$ by examining the contribution to the
classical action in Eq.~(\ref{sec2.clVV}) from the term $ V_{11}$.
Identifying the leading term in the limit $x\to 0$ of the classical
factor $x$-integrand as, $ e^{-S _{cl} } \sim x^{M_X^2} $, leads to
the result for the $(B,D)$ sector gauge boson squared mass \bea &&
\lim _{x\to 0} e^{-S _{cl} } = e^{- \vert d_A \vert ^2 V '_{11} }
= \prod _I e^{- \vert d _A^{I} \vert ^2 \sin ^2 (\pi \t ^I ) \ln (\hat
\d _I/x) } \ \Longrightarrow \ M_X ^2 = \sum _I \sin ^2 (\pi \t ^I )
\vert d_A ^{I} \vert ^2 , \label{sec3.wils}\eea where $\hat \d _I =
\hat \d (\t ^I, \t ^{'I} ) $ is the auxiliary angle dependent
parameter defined by Eq.(\ref{app.0G2}).  The consistent
implementation of the broken gauge symmetry by the brane displacement
requires including by hand in the string amplitude the extra
normalization factor, $ \prod _I (\hat \d _I ) ^{\sin ^2 (\pi \t ^I )
d_A ^{I2} }$, as is needed to cancel the prefactor of $ x^{M_X^2} $.
Recall that the overall constant normalization was previously
determined by matching the gauge coupling constant in the large radius
limit after factoring the classical partition function out of the
$x$-integral.  With the same normalization prescription based on the
identification of the gauge coupling constant, the partition function
must then include the extra factor, $ \prod _{I } (\hat \d ^I )^{
\vert d_A ^I \vert ^2 \sin ^2 (\pi \t ^I ) }. $

\subsubsection{Special  configuration with   same pairs of  
conjugate intersection angles}

We now specialize the results of the previous subsection to the
simpler case involving two fermion pairs of equal angles, $\t ^I= \t
^{'I}$, which is realized by the brane configuration with $ D=B,\
C=A$.  The derivation is straightforward provided that due care is
taken in dealing with the limit, $\t ^I \to \t ^{'I}$.  The combined
contributions to the four fermion string amplitude from quantum and
classical terms yields the formula \bea && \cala ' ( f_1 f_2 ^\dagger
f_3 f_4^\dagger )
= 2\pi g_s \int _0 ^1 dx \prod _I \bigg ({ \sin (\pi \t ^I)\over F(x)
F(1-x) } \bigg ) ^ {\ud } \cr && \times \bigg [x ^{- s -1 } (1- x )
^{- t -1}\cals _1 \calT _1 \sum _{p_A, p_B \in Z } e^{-S^{(1)} _{cl} }
- x ^{- t-1 } (1- x ) ^{- s -1} \cals _2 \calT _2 \sum _{p_D, p_C\in
Z} e^{-S^{(2)} _{cl} } \bigg ], \cr && \bigg [F(x)= F(\t , 1-\t; 1;
x),\ S ^{(1)} _{cl} = {\pi } \sin \pi \t ^I [ \vert p_A L_A + d^
A_{12} \vert ^ 2 {F(1-x ) \over F(x) } + \vert p_B L_B + d^ A_{23}
\vert ^ 2 {F(x) \over F(1-x) }] ,\cr && S ^{(2)} _{cl} = {\pi } \sin
\pi \t ^I[ \vert p_D L_D + d ^ D_{14} \vert ^ 2 {F(1-x ) \over F(x) }
+ \vert p_C L_C + d^ C_{43} \vert ^ 2 {F(x) \over F(1-x) } ] \bigg ]
\label{eqstreq}\eea
where we have used the result, $ I(x) \to 2 F(x) F(1-x)$, and the
abbreviated notation for the Hypergeometric function, $F(x) = F (\t ,
1-\t ; 1; x)$.  The equality of the direct and reverse orientation
terms enclosed inside the brackets is established by using the change
of integration variables, $ x\to (1-x)$.
The classical contributions consist of two multiplicative factors
associated to the winding and momentum states with respect to the
large 2-d tori generated by the lattice vectors, $ L_A ,\ L_B $.  The
squared mass spectrum of the $ (A,A) $ and $(B,B)$ open string sectors
are deduced by examining the end point regions $ x\to 0$ and $ x\to 1$
of the $x$-integral which select the $t$-channel and $s$-channel
poles.  Before showing this explicitly, we rewrite the lattice
summations over wrapped cycles in a compact form by introducing the
Jacobi theta function with moduli parameter $\tau $, defined by the
familiar series representation \bea && \vt [{ \t \atop \phi }] (\nu ,
\tau ) = \sum _{n\in Z} q ^{(n+\t ) ^2 /2 } e ^{2 i \pi (n+\t ) (\nu +
\phi ) }, \ [q = e ^{2 i \pi \tau } ,\ \vt [{ \t \atop \phi }] (\tau )
=\vt [{ \t \atop \phi }] (\nu =0 , \tau ) ] .  \eea
The resulting formula for the string amplitude reads \bea && \cala '
= 2\pi g_s \int _0 ^1 dx \bigg [ \cals _1 \calT _1 x ^{-s-1} (1-x)
^{-t-1} \vt[ { \e ^A _{12} \atop 0}] (\tau _A) \vt [{ \e ^B _{23}
\atop 0}] (\tau _B) \cr && - \cals _2 \calT _2 x ^{-t-1} (1-x) ^{-s-1}
\vt [{ \e ^B _{14} \atop 0}] (\tau _B) \vt [{ \e ^A _{43} \atop 0}]
(\tau _A ) \bigg ] I^{-\ud }(x) , \cr && \bigg [ I(x) = 2 F(x) F(1-x)
,\ \tau _A (x) = i \sin (\pi \t ^I ) \vert L_A \vert ^2 {F(1-x) \over
F(x)} , \cr && \tau _B (x) = i \sin (\pi \t ^I) \vert L_B \vert ^2
{F(x) \over F(1-x) } , \ \e ^ {A,B} _{ij } = {d^{A,B} _{ij } \over
\vert L_{A,B} \vert }
\bigg ] . \eea The duality transformation formula for the theta
function, $\vt[ { \e \atop 0 } ] (\tau ) = (-i \tau ) ^{-\ud } \vt[ {
0 \atop \e }] ( - {1\over \tau } ), $ accomplishes the same task as
the Poisson resummation formula.  At $x \to 0$, the theta function
factors with argument, $ \tau _A (x) \to i \infty $, are safe, while
those with argument, $ \tau _B (x) \to i 0 $, are unsafe, hence
requiring the use of a duality transformation to avoid the singular
behavior from the factor, $ F ^{-1/2} (1-x) $, as needed to interpret
the field theory limit in terms of an infinite series of $s$-channel
poles.  For $\e _B \ne 0$, the same argument with $ \tau _A (x) $ and
$ \tau _B(x)$ interchanged leads to a series of $t$-channel poles.
The following two representations of the string amplitude, obtained by
applying the duality transformations on $ \tau _B (x)$ and $ \tau _A
(x)$, achieve the $x$-integral convergence at small $x$ and small
$1-x$, respectively, \bea && \bullet \ \cala ' = \int _0^1 dx {2\pi
g_s \over \vert L_B \vert F(x) } \bigg [ \cals _1 \calT _1 x^{-s-1}
(1-x)^{-t-1} \vt [{ \e ^A _{12} \atop 0}] (\tau _A) \vt [{0 \atop \e
^B _{23} }] (-{1\over \tau _B}) \cr && + \cals _2 \calT _2 x^{-t-1}
(1-x)^{-s-1} \vt [{\e ^A _{43} \atop 0}] (\tau _A) \vt [{0 \atop \e ^B
_{14} }] (-{1\over \tau _B}) \bigg ] , \cr && \bullet \ \cala ' = \int
_0^1 dx {2\pi g_s \over \vert L_A \vert F(1-x) } \bigg [\cals _1
\calT_1 x^{-s-1} (1-x)^{-t-1} \vt [{0 \atop \e ^A }] (-{1\over \tau
_A}) \vt [{\e ^B _{23} \atop 0}] (\tau _B) \cr && - \cals _2\calT _2
x^{-t-1} (1-x)^{-s-1} \vt [{0 \atop \e ^A _{43} }] (-{1\over \tau _A})
\vt [{\e ^ B _{14}\atop 0}] (\tau _B) \bigg ] .\eea

Substituting now the $x$-integrand in Eq.~(\ref{eqstreq}) by its
leading term in the limit $x\to 0$ gives the low energy expansion of
the amplitude, \bea && \cala ' \simeq {2\pi g_s \over \vert L_B\vert }
\sum _{p_A, p_B} \bigg ( \cals _1 \calT _1 {\prod _I \d _I^ {-M
^{I2}_{A _{12} ,B _{23} } } e ^{2i \pi p_B \e ^{B} _{23}} \over -s +
\sum _{I} M ^{I2}_{A _{12},B} } - \cals _2 \calT _2 {\prod _I \d _I^
{-M ^{I2}_{A_{43},B_{14} } } e ^{2i\pi p_B \e ^{B}_ {14}} \over -t +
\sum _{I} M ^{I2} _{B _{14},A}}\bigg ) , \cr && [ M ^{I2} _{A _{ij}
,B}= (p_A +\e ^{AI} _{ij} )^2 \sin ^2 (\pi \t ^I) \vert L^I_A \vert ^2
+ {p_B^2 \over \vert L^I_B \vert ^2 } ,\cr && M ^{I2}_{B_{ij} ,A}=
(p_B +\e ^{BI} _{ij} ) ^2 \sin ^2 (\pi \t ^I) \vert L^I_B \vert ^2 +
{p_A^2 \over \vert L^I_A \vert ^2 } ] \eea where we have refrained
from writing the suffix $I$ on $ p_{A,B} ,\ L_{A,B} $ and $ \tau
_{A,B} (x) .$ It is interesting to note that if the reverse
orientation term were evaluated after performing the change of
integration variable, $ x\to (1-x)$, one would obtain the equivalent
representation involving only the $s$-channel poles, \bea && \cala '
\simeq 2\pi g_s \sum _{p_A, p_B} \bigg ( {\cals _1\calT _1 \over \vert
L_B\vert } { \prod _I \d _I^ {-M ^{I2}_{A _{12} ,B_{23} } } e ^{2i \pi
p_B \e ^{BI} _{23}} \over -s + \sum _{I} M ^{I2}_{A_{12},B} } - {
\cals _2 \calT _2 \over \vert L_A \vert } { \prod _I \d _I^ {-M
^{I2}_{B _{14} ,A_{43} } } e ^{2i \pi p_A \e ^{AI} _{43}} \over -s +
\sum _{I} M ^{I2}_{B _{14},A} } \bigg ) .\eea The existence of
distinct representations of string amplitudes as dual infinite series
of poles is a familiar consequence of the world sheet duality
symmetry.  This makes the comparison with the field theory limit
appear rather subtle, as will be discussed in the next section.  The
squared mass spectrum of states in the $s$-channel, $ M^2 _{A,B} $,
include momentum and winding modes from the $(B,B)$ and $(A,A)$
sectors. The pole positions reproduce the squared mass spectrum for
the momentum modes along the N directions longitudinal to brane $B$,
and winding modes along the D directions transverse to brane $A$.  The
spectrum in the $t$-channel $ M^2 _{B,A} $, is similar with $A,\ B$
interchanged.  The structure of the compactification mass spectrum is
formally equivalent to that for rotated branes parallel along some
$T^2_I$ torus, with the roles of the torus cycles $ e_1 ^I,\ e_2^I$
played here by the two brane sides, $ L_A,\ L_B$. The momentum modes
are associated with the cycle $L_A $ and the winding modes with the
transversally projected distance between the branes $A$ and $B$.  The
above string squared mass spectrum conforms with the familiar formula,
$ M^2 = \sum _{p, \ s \in Z } {p ^2+ s^2 (r_1 r_2 \sin \a ) ^2 \over
\vert L\vert ^2 } , $ for open strings stretched between parallel
$D1$-branes wrapped around the torus generated by cycles of length $L$
and shape angle $\a $.  We recommend Ref.~\cite{angel05} for further
discussion of the string mass spectrum in intersecting brane models.

\subsection{Processes  with  two pairs  of  fermion and  scalar
superpartner modes}
\label{sub22}

In the supersymmetric unified models, alongside with the $B,\ L$
number violating contributions of D term type to four fermion
subprocesses exchanging colored gauge bosons, F term type
contributions can occur from tree level subprocesses exchanging
colored higgsino modes between two pairs of massless matter fermions
and sfermions, $ \cala (\psi \psi \phi \phi )$.  The dominant chiral F
term operators of dimension $5$ are of form, $[QQQL]_F $ and $
[U^cD^cU^cE^c]_F$.  We study here the string theory predictions for
supersymmetric models by focusing on the configurations with two
conjugate pairs of localized open string modes with intersection
angles $\t $ and $\t '$.  The tree level contributions to the D and F
term operators can be identified by considering in turn the two
transition amplitudes on the disc surface, \bea && \cala _V (\psi _1
^\dagger \psi _2 { \phi _3 '} ^ {\dagger } \phi ' _4 )= \int {\prod _i
dx_i \over V_{CKG} } < V _{-\t } ^{(-\ud )} (x_1) V _{\t } ^{(-\ud )}
(x_2) V _{-\t '} ^{(-1)} (x_3) V _{\t' } ^{(0)} (x_4) > ,\cr && \cala
_S(\psi _1 ^\dagger \phi _2 \psi _3 ^{'\dagger } \phi '_4 )= \int {
\prod _i dx_i \over V_{CKG} } < V _{-\t } ^{(-\ud )} (x_1) V _{\t }
^{(-1)} (x_2) V _{-\t '} ^{(-\ud )} (x_3) V _{\t ' } ^{(0)} (x_4) > ,
\label{ampsv} \eea where we have signalled the vector and scalar
character of the two couplings by the suffix labels, $V, \ S $.  The
correlator involves same inputs for the vertex operators as those
introduced in the preceding subsection.

Let us start with the amplitude $\cala _V$.  The massless fermion
modes and the low-lying scalar modes are assigned the $SO(10) $ weight
vectors, $ r^A (1) = (+++,-+), \ r^A (2) = (---,--), \ r^A (3) =
(111,00), \ \tilde r^A (4) = (-1-1-1,00) + ,(000,10) ,$ implying the
squared masses, $ M_3^2 = M_4^2 = 1 -\ud (\vert \t '_1 \vert + \vert
\t '_2 \vert +\vert \t '_3 \vert ) $.  Note that $\tilde r^A (4) $
refers to the weight vector shifted by picture changing.  Since the
massless fermions have opposite space-time helicities, the only
allowed Dirac spinor matrix element is the Lorentz vectorial coupling,
$ \bar u_1 \g ^\mu k_{4\mu } u_2$.  This structure can also be
inferred by making use of the operator identity, $ u_1 ^{\dot \a } u_2
^\a S_{\dot \a } (x_1) S_{\a } (x_2)\psi _\mu (x_4) = {1\over \sqrt 2}
u_1 ^T C\G _\mu u_2 \to {1\over \sqrt 2}\bar u_1 \g _\mu u_2 $.  Since
the only relevant term in Eq.~(\ref{eqvertx2}) for the picture changed
vertex operator $ V _{\t' } ^{(0)} (x_4) $ is that involving the
spinor term $ (k\cdot \psi ) \psi ^\mu $, the same correlator factor,
$ Z (x)$, with the ground state twist operators only, appears as in
the four fermion amplitude.  The resulting chirality diagonal string
amplitude given by \bea && \cala ' _V = C_V (\calT _1+\calT _2) (\bar
u_1 (k_1) \g _\mu k_4 ^\mu u_2 (k_2) ) \cr && \times \int _0^1 dx x
^{-s-1} (1-x) ^{-t -1 + \ud \sum _I (\t ^I -\t ^{'I} ) } I ^{-\ud }
(x) e^{-S^{(1)}_{cl} } , \eea has indeed the Dirac spinor structure
expected from a gauge boson and gaugino exchange amplitude.  For
massless scalars, setting, $ \ud \sum _I \t ^I = \ud \sum _I \t ^{'I}
= 1 $, consistently with the assumed supersymmetry of the model,
reduces $\cala '_V$ to an amplitude of similar form to the four
fermion amplitude.

We discuss next the amplitude, $ \cala _S $, in the low energy limit,
$ k_i \to 0$.  One expects this to contribute to F term operators
since the incoming massless fermions have the same helicity.  In the
picture changed operator, $ V _{\t ' } ^{(0)} (4) $, we need only
retain the terms associated with the spinor fields along the internal
space directions, $ \psi ^J (z)$.  The following choice for the
$SO(10)$ spinor and vector weights of the states is then practically
forced on us, \bea && r ^A (1) =
r^A (3) = (+++, \mp \pm ),\ r^A (2) = (\underline{-1,0,0}; 0,0), \cr
&& \tilde r^A (4) = (-1,-1,-1; 0,0) +(\underline{1,0,0}; 0,0) , \eea
where the constant shift in the spinor weight, $ \tilde r^A (4) $,
arises from the picture changing, $ \psi ^J (z) \dh \bar X ^ J (z) \s
_ {\t '} (x_4) \simeq (z-x_4) ^{-\t ^{ 'J} } \tilde \tau _ {\t ^{'J} }
(x_4) e^{i H_J(z) } $, and the underlines refer to the possible
permutations of the entries in correspondence with the choice of the
index $J$.
We make henceforth the definite choice, $J=3$, which corresponds to
using the vector weights for scalar modes, $r ^A (2) = (00-1,00), \
\tilde r^A (4) = (-1-1-1,00) + (001,00) = (-1-10,00)$, with squared
masses determined by the condition of unit total conformal weights,
\bea && 0= h(V _\t (2) )-1 = \ud (\vert \t ^1 \vert +\vert \t ^2\vert
-\vert \t ^ 3 \vert ) - M^2_2 ,\cr && 0= h(V _{\t '} (4) )-1 = 1 -\ud
( \vert \t ^{'1} \vert +\vert \t ^{'2} \vert +\vert \t ^{'3} \vert )
-M^2_4 . \eea The coordinate twist field correlator includes two
factors, $Z _I (x_i) = <\s _{-\t ^I } (1) \s _{\t^I } (2) \s _{-\t ^
{'I} } (3) \s _{\t ^ {'I} } (4) > ,\ [I=1,2]$ identical to the
previously studied correlator factor, and one new factor involving a
single excited state twist field, $\tilde Z _{J=3} (x_i) = <\s _{-\t
^J} (1) \s _{\t ^J} (2) \s _{-\t ^{'J}} (3) \tilde \tau _{\t ^{'J}}
(4) > $.  For the present, we introduce the following shorthand
notation for the latter four point correlator with a single excited
twist field, \bea && \tilde Z_J (x) \equiv < \s _ { -\t ^J } (x_1) \s
_ {\t ^J } (x_2) \s _ {-\t ^{'J} } (x_3) \tilde \tau _ {\t ^{'J} }
(x_4)> \cr && = f _J (x_i ) < \s _ { -\t ^J} (x_1) \s _ {\t ^J} (x_2)
\s _ {-\t ^{'J} } (x_3) \s _ {\t ^{'J} } (x_4)> .\label{eqcorrextw}
\eea Upon performing the familiar gauge fixing choice of insertion
points, $x_1=0,\ x_2=x,\ x_3=1,\ x_4=X \to \infty $, extracting out
the appropriate factors of $X$ by writing, $ \tilde Z_J (x) / Z_J (x)
\to f_J (x)$, and using the known correlator factors \bea && < e ^{ -
{\varphi (0) \over 2} } e ^{ - {\varphi (x)} } e ^{ - {\varphi (1)
\over 2} } > = x^{-\ud } (1-x) ^{-\ud },\cr && \prod _{I=1,2} (<ssss>
\vert _{I} ) <ssst> \vert _{I=3} = x ^{- \sum_I \t^I (\t ^I -\ud ) +
(\t ^3 -\ud ) } (1-x) ^ {- \sum_I \t^I (\t ^ {'I} -\ud ) + (\t ^{'3}
-\ud ) } ,\eea one finds the string amplitude in the form \bea &&
\cala '_S = i C_S (\bar u_1 (k_1) u_2 (k_2)) (\calT _1 + \calT _2) \cr
&& \times \int _0^1 dx x ^{-s-1} (1-x) ^{-t -1 + \ud [( \t _1 +\t _2
-\t _3 ) - ( \t '_1 +\t '_2 -\t '_3 )] } f_3 (x) \prod _I \bigg ( I _I
^ {-\ud } (x) e ^ {- S _{cl} } \bigg ) .  \label{eq5pamp}\eea The
explicit dependence on the intersection angles in the exponent of
$(1-x)$ cancels out in the case of interest involving supersymmetric
cycles.

To reach a concrete final result, we modify here our initial choice of
the initial states so as to deal with the more tractable correlator
involving a conjugate pair of excited coordinate twist operators
rather than a single one as in the above case.  This choice is
motivated by the fact that the correlator of interest, $\tilde Z '_
{J} (x_i) \equiv <\s _{-\t ^J } (x_1) \tau _{\t ^J } ( x_2) \s _{-\t
^{'J } } (x_3) \tilde \tau _{\t ^{'J }} (x_4) > $, can be more readily
accessed within the formalism set up in Appendix~\ref{apptwist}.  In
fact, the latter correlator would arise if one assigned to the scalar
mode $\phi _2$ in $\cala _S$ of Eq.~(\ref{ampsv}) the choice of vertex
operator \bea && V ^{(-1)} _ \t (x_2) = \l ^a e ^{-\varphi } \tau _{\t
^J } s ^{r_v } _{\t ^J } \prod _{I \ne J } \s _{\t ^I} s ^{r_v } _{\t
^I} e ^{ik\cdot X }, \cr && [r _v ^A (2) = (00-1,00), \ M^2 _2 = \ud
(\vert \t ^1 \vert +\vert \t ^2 \vert +\vert \t ^ 3\vert )] \eea where
we have displayed the weight vector of the mode $\phi _2 $ and the
formula of its squared mass.
 For the remaining modes, $ \psi ^\dagger _1,\ \psi ^ {'\dagger } _3
 ,\ \phi ' _4$, in $\cala _S$, we continue using the same inputs as
 above with the choice of complex plane, $ J=3$.  The excited state
 twist field correlator, $ \tilde Z '_ {J} (x_i) $, can be evaluated
 in terms of the ground state correlator, $ Z_ {J} (x_i) $, by
 considering the representation for the ratio of these functions \bea
 && \tilde f '_J (x) = { \tilde Z ^{'} _J (x_i) \over Z _{J} (x_i) }
 \equiv { < \s _ { -\t ^J } (x_1) \tau _ {\t ^J } (x_2) \s _ {-\t
 ^{'J} } (x_3) \tilde \tau _ {\t ^{'J} } (x_4)> \over < \s _ { -\t ^J
 } (x_1) \s _ {\t ^J } (x_2) \s _ {-\t ^{'J} } (x_3) \s _ {\t ^{'J} }
 (x_4) > } \cr && = -\ud \lim _{z \to x_2, w \to x_4} (z-x_2) ^{1-\t
 ^J} (w-x_4) ^{\t ^{'J} } g _J(z,w), \eea where $ g _J(z,w) $ is the
 same correlator denoted as $ g(z,w) $ in Appendix~\ref{apptwist}.  We
 next perform the same choice of $SL(2,R)$ gauge group fixing of the
 $x_i$ variables and extract out the appropriate $X$ factors from the
 two partition functions, $ \tilde Z ^{'} _J \sim X^ {2\t ' -2 } ,\ Z
 _{J} \sim X^ {\t (1-\t ') } $.  A simple calculation yields the
 following result for the ratio of correlation functions \bea &&
 \tilde f'_3 (x) = \tilde C ' x^ {1-\t ^3 } (1-x) ^ {1-\t ^{'3 } } \dh
 _x \ln I_3(x) , \eea where the calculable normalization factor $
 \tilde C '$ will be left unspecified. Combining this with the
 familiar results for the correlator factors, \bea && < \prod _i
 e^{ik_i\cdot X } > = x ^{-s + M_2 ^2 } (1-x ) ^{-t + M_2 ^2 } ,\cr &&
 <ssst >\vert _J = x ^{-(\t ^J -\ud ) (\t ^J -1 ) } (1-x) ^{- (\t ^J
 -1 ) (\t ^{'J} -\ud ) } , \eea yields the final formula for the low
 energy string amplitude \bea && \cala ' _S = \tilde C_S (\calT _1 +
 \calT _2) (\bar u_1 (k_1) u_2(k_2) )\cr && \times \int _0 ^1 dx x^{
 -s } (1-x) ^{-t + \ud \sum _I (\t ^I -\t ^{'I } ) } {\dh \over \dh x
 } (\ln I_3(x) ) \prod _I ( I _I ^{-\ud } (x) e ^{-S ^I _{cl} (x) } )
 .  \eea This result exhibits the chiral structure expected from the
 exchange of a fermion mode. At this point, we remark that since $ M_2
 ^2 = O(m_s ^2)$ is finite, the process at hand is energetically
 forbidden so that the interest of the present calculation is academic
 at best.  One can suppress $ M_2 ^2 $ by choosing vanishing small
 angles, $\t ^I $, but at the price of dealing with delocalized modes
 $\psi _1 $ and $\phi _2$.
 
Before closing this discussion, we briefly indicate one possible route
to evaluate the partition function $\tilde Z _{J} $ in
Eq.~(\ref{eqcorrextw}).  For this one may use the limiting
representation, \bea && \tilde Z _J (x_i) = \lim _{x_5 \to x_4} (x_5
-x_4)^ {\t ^ {'J} } < \dh \bar X ^J (x_5) \s _{-\t ^J} (1) \s _{\t ^J}
(2) \s _{-\t ^{'J}} (3) \s _{\t ^{'J}} (4) > ,\eea and evaluate the
resulting five point correlator by a similar method to that used by
Fr\"olich et al.,~\cite{fro099} for the open string modes with mixed
ND boundary conditions.  Alternatively, one could consider the bilocal
correlators, $ \tilde g(z,w),\ \tilde h(\bar z,w)$, obtained by
inserting the quadratic products, $ \dh _z X ^J \dh _w \bar X^ {J} ,\
\dh _{\bar z } X ^J \dh _w \bar X^{J} $ in the correlator for $\tilde
Z _J (x_i) $, and apply the energy source approach reviewed in
Appendix~\ref{apptwist} by writing the general representation on the
functions consistent with the constraints.
We leave to a later work the feasible task of implementing these
calculations.

\subsection{Higher order processes  with baryon  and lepton number
  non conservation}
\label{sub23}

In spite of the stronger suppression of baryon number violating
processes initiated by the dangerous operators of dimension $\cddd
\geq 7$, the study of these contributions in grand unified theories is
motivated by the need to test variant gauge unification schemes
involving lower mass scales and different selection rules on $B, \ L$
non conservation~\cite{weinbergs,wilczee}.  For orientation, we
provide a brief overview of the baryon number violating processes from
higher dimension operators in Appendix~\ref{bnvnd}.  In the present
subsection, we present a qualitative discussion of the string
amplitudes in intersecting brane models associated with the $\cddd =7$
local operators coupling three quarks with single lepton and Higgs
boson and the $\cddd =9$ local operators coupling six quarks.  Our
presentation here will remain at a general level without commitment to
any specific model.

\subsubsection{Five point  amplitudes} 

With hindsight from the general structure of $\cddd =7$ operators, we
consider the tree level five point amplitude involving the localized
open string modes of four matter Dirac fermions and a single scalar
boson, $ \cala _5 = \cala ( \psi _1 \psi_2 ^ \dagger \psi _3 \psi _4 ^
\dagger \phi _5 )$, defined by
\bea && \cala _5 = \sum_{perms} \int \prod _i {dx _i \over V_{CKG} }
<V ^{(-\ud )} _{-\t } (x_1) V ^{(-\ud )} _{\t } (x_2) V ^{(-\ud )}
_{-\t _3} (x_3) V ^{(-\ud )} _{\t _4} (x_4) V ^{(0)} _{\t _5} (x_5)
>. \label{eqvo5} \eea For an acceptable embedding of the disk onto the
internal $T^6$ torus, one must require that the modes intersection
angles obey the condition, $ - \t _3 + \t _4 + \t _5 =0$.  In the low
energy limit of interest, $ k_i \to 0$, only the terms in the picture
changed operator, $V ^{(0)} _{\t_5} (x_5) $, depending on the internal
space directions, contribute. The following unique choice of weight
vectors for the massless modes, consistent with $H_A$-momentum
conservation and the GSO projection (odd number of $-\ud $ entries),
must be assigned to the vertex operators in Eqs.~(\ref{eqvo5}), \bea
&& - r ^A (1)= - r ^A (3)= (---,\pm \mp ) , \ r ^A (2)= r ^A (4)=
(---,\pm \pm ) , \cr && \tilde r ^A (5)= (-100, 00 ) + (100, 00 )=
(000,00) . \eea The string amplitude can be expressed in abbreviated
form as \bea && \cala ' _5 = C _5 (\calT _1 + \calT _2 ) (u _1^ {\a
_1} \cdots u _4^ {\a _4} ) \int {\prod _{i=1}^5 dx _i \over V_{CKG} }
< e ^{- {\varphi (x_1) \over 2 } } \cdots e ^{- {\varphi (x_4) \over 2
} } > \cr && \times < S _{\a _1 } (x_1) \cdots S _{\a _4 }(x_4) >
\bigg ( \Phi _{t} ^1 (x_i) \Phi _{\tilde \tau } ^1 (x_i) \bigg ) \prod
_{I=2,3} \bigg ( \Phi _s ^I (x_i) \Phi _\s ^I (x_i) \bigg ) <\prod
_{i=1}^5 e^{i k_i \cdot X_i (x_i) } > + \textit{perms} , \cr && \bigg
[\calT _1 = Tr(\l _1 \cdots \l_ 5),\ \calT _2 = Tr(\l _5 \cdots \l_
1),\ \Phi ^I _{ t \choose s} = <s _ {-\t ^I } (x_1) s _ {\t ^I } (x_2)
s _ {-\t _3^I } (x_3) s _ {\t _4 ^I } (x_4) { t _{\t _5 ^{I=1} } (x_5)
\choose s _ {\t _5 ^{I =2,3} } (x_5) } >,\cr && \Phi _{\tilde \tau
\choose \s } ^I = <\s _ {-\t ^I } (x_1)\s _ {\t ^I } (x_2) \s _ {-\t
_3 ^I } (x_3)\s _ {\t _4 ^I } (x_4) {\tilde \tau _{\t _5 ^{I=1} }
(x_5) \choose \s _ {\t _5 ^{I=2,3} } (x_5) } > ] . \label{eqamp4} \eea
With the M\"obius group gauge fixing choice, $ [x_1,\ \cdots , x_5] =
[0, x, y, 1, X\to \infty ]$, the correlator factors for the ghost,
spin and spinor twist field operators are evaluated by means of the
familiar rules \bea && < e ^{- {\varphi (x_1) \over 2 } } \cdots e ^{-
{\varphi (x_4) \over 2 } } > = [x y (y - x) (1-x) (1-y) ] ^{-1/4 } ,
\cr && (u _1^ {\a _1} \cdots u _4^ {\a _4} ) < S _{\a _1 } (x_1)
\cdots S _{\a _4 }(x_4) > = y ^{-\ud } (1-x) ^{-\ud } (\bar u_1 (k_1)
\g ^\mu u_2 (k_2)) (\bar u_3 (k_3) \g _\mu u_4 (k_4)) , \cr && \Phi ^I
_{ t } = \Phi ^I _{s}
= x^{- (\t ^I -\ud ) ^2 } y^ {(\t ^I -\ud ) (\t ^I _3-\ud ) } (y-x) ^
{ -(\t ^I -\ud ) (\t ^I _3-\ud ) } (1-x) ^ { (\t ^I -\ud ) (\t ^I _4
-\ud ) } (1-y) ^ {- (\t ^I_3 -\ud ) (\t ^I _4 -\ud ) } . \cr && \eea
We shall not attempt here an exact evaluation of the five point
coordinate twist field correlators, $ \Phi _{\tilde \tau \choose \s }
^I $, because of the significant labor involved in this task.
Nevertheless, the $n$ point correlators of the coordinate twist fields
are expected to have the general structure~\cite{bersh87,dixontasi}, $
<\prod _{i=1} ^n \s _{\t _i } (z_i) > = C _\s \prod _{i\ne j} (z_i
-z_j)^{-\ud (1-\t _i ) (1-\t _j )} \text{Det} ^{-\ud } (W ) $, where
$W$ denotes the period matrix, whose $ (n-2)\times (n-2)$ entries give
the period integrals of the $(n-2)$ independent holomorphic
differential forms over the $(n-2)$ independent cycles circling the
pairs of insertion points in the cut complex plane.  This result
motivates us in introducing the following definitions, obtained by
including only the explicit pair contraction factors with the
dependence on angles determined by the conformal symmetry \bea && \Phi
_{\tilde \tau \choose \s } ^I
= x^{-\t ^I (1-\t ^I ) } y ^{\ud (\t ^I (1-\t ^I _3) + \t ^I _3(1-\t
^I ) ) } (y-x)^{- \ud (\t ^I (1-\t ^I _3) + \t ^I _3(1-\t ^I ) ) }
(1-x)^{ \ud (\t ^I (1-\t ^I _4) + \t ^I _4(1-\t ^I ) ) } \cr && \times
(1-y)^{- \ud (\t ^I _3 (1-\t ^I _4) + \t ^I _4(1-\t ^I _3 ) ) }
{\tilde \calf _1 (x,y) \choose \calf _{2,3} (x,y) } , \eea where
$\tilde \calf _1 (x,y) , \ \calf _{2,3} (x,y) $ include the
determinants of the period matrices for the correlators, $\Phi
_{\tilde \tau } $ and $\Phi _{\s } $.
Combining now the various contributions to the five point string
correlator, we deduce the final form of the amplitude \bea && \cala '
_5 = C _5 (\calT _1 + \calT _2) (\bar u _1 (k_1) \g^\mu u_2(k_2))
(\bar u _3 (k_3) \g _\mu u_4 (k_4)) \cr && \times \int _0^1 dx \int
_x^1dy x ^{-1} (y - x) ^{-1} (1 -y) ^{-1} \tilde \calf _ 1 \calf _ 2
\calf _3 e ^{-S_{cl} } + \textit{perms} ,\eea where we have explicitly
displayed the reduced amplitude associated to the reference cyclic
ordering of the insertion points, $ x_1 \leq \cdots \leq x_5$.  Upon
associating the five modes in $\psi _1 \psi_2 ^ \dagger \psi _3 \psi
_4 ^ \dagger \phi _5 $ to the sequence of open string sectors $ (D,A)
(A,D) (D,B) (B,C) (C,D)$, one sees that only the reduced amplitude
with the trivial permutation $(12345)$ is non-vanishing if the branes
$ D, A, B, C$ are all distinct.  In the case $ B=A$, the reduced
amplitude with the permutation $(14532)$ would be added, and in the
case $ B=A,\ C= D$, the reduced amplitudes with the permutations
$(14532)$ and $(12534)$ would be added.

\subsubsection{Six point  amplitudes}

We turn next to the string amplitude for the six quark subprocesses
initiating the $ \D B =- 2 , \D L=0$ processes of $N-\bar N$
oscillation and two nucleon disintegration.  The discussion developed
by Kostelecky et al.,~\cite{kost86,kost87} is followed to some extent,
since we only focus here on the low energy limit of six fermion tree
level string amplitudes involving three conjugate pairs of localized
fermion modes.  The assignment of CP gauge factors and flavor and
color quantun numbers will remain implicit, without reference to any
specific model. The relevant six point amplitude, $ \cala _6 = \cala
(\psi _1^\dagger \psi _2 \psi _3^\dagger \psi _4 \psi _5 ^\dagger \psi
_6 ) $, admits the following representation in terms of the correlator
with vertex operators inserted on the disk boundary \bea && \cala _6 =
\sum _{perms} \int { \prod _{i=1}^ 6 dx_i \over V_{CKG} } < V ^{(-\ud
)} _{-\t } (x_1) V ^{(-\ud )} _{\t } (x_2) V ^{(-\ud )} _{-\t '} (x_3)
V ^{(-\ud )} _{\t '} (x_4) V ^{(-\ud )} _{-\t '' } (x_5) V ^{( +\ud )}
_{\t'' } (x '' ) > , \cr && \label{eqweights} \eea where the massless
fermion modes are assigned the unique choice of spinor weights, $
(---,\mp \mp ) $ or $ (+++,\mp \pm ) .$
Without proceeding any further, it is easy to convince oneself that
the $H_A$-momentum conservation forces the low energy limit of this
amplitude to vanish.  Nevertheless, whether the $ \D B= - 2 $
processes are forbidden or not at tree level cannot be concluded until
one has examined the possibility that VEV induced contributions might
arise from higher order processes.  Hindsight from the unified field
theory models with left-right gauge symmetry~\cite{mohasenja},
suggests considering the higher order baryon number violating
operator, $ u ^c ud^c dd^c d \D ^l $, coupling six quarks with a
scalar mode carrying the $SU(3)_c \times SU(2)_L \times SU(2)_R$ gauge
group quantum numbers of dileptons, $\D _l \sim (1,1, 3) $.  The
corresponding $\cddd =10 $ operator would yield a finite contribution
to the $\cddd =9$ operators once the electric charge neutral component
of $ \D ^l = \tau ^a_R \D ^l_a $ acquires a finite VEV.  The six
quarks in the local operators occur in three pairs, two pairs or a
single pair of electroweak group singlets while the scalar boson
multiplet $\D ^l $ must have at least one component which is a singlet
under the Standard Model group.  Guided by these observations, we
consider the seven point string amplitude, $\cala _7 = \cala (\psi _1
\psi ^\dagger _2 \psi _3 \psi ^\dagger _4 \psi _5 \psi ^\dagger _6 \D
^l _7 ) $, with localized open string modes inserted on the disk
boundary, \bea && \cala _7
= \sum _{perms} \int {\prod _{i=1} ^7 dx_i \over V_{CKG} } < V ^{(-\ud
)} _{-\t } (x_1) V ^{(-\ud )} _{\t } (x_2) V ^{(-\ud )} _{-\t '} (x_3)
V ^{(-\ud )} _{\t '} (x_4) V ^{(-\ud )} _{-\t _5} (x_5) V ^{( +\ud )}
_{-\t _6 } (x_6) V ^{(0)} _{ \t _7 } (x_7) > , \eea subject to the
condition on the intersection angles, $ \t _7 = \t _5 + \t _6$, while
adhering to our convention that all the angles $ \t _i $ are positive.
The assignment of spinor weights, consistent with the $H_A$-momentum
conservation and the expected structure of the Dirac spinor matrix
element, allows for the unique choice \bea && r ^A ( 1) = (+++, \mp
\pm ),\ r ^A (2) = (---, \pm \pm ),\ r ^A ( 3) = (+++, \pm \mp ),\ r
^A (4) = (---, \mp \mp ),\cr && r ^A ( 5) = (+++, \pm \mp ) ,\ \tilde
r^A (6) = (+++, \mp \pm ) + (00 - 1 , 00) = (++-, \mp \pm ) , \cr &&
\tilde r^A (7) = (-1-1-1, 00 ) + (00 1 , 00) = (-1-1 0, 00 ) , \eea
where the scalar mode has squared mass, $ M^2 _7= 1- \ud (\vert \t _7
^1\vert + \vert \t _7 ^2 \vert + \vert \t _7 ^3 \vert ) $.  The single
surviving reduced string amplitude can be expressed in abbreviated
form as \bea && \cala ' _7 = C _7 (\calT _1 + \calT _2) [a_V (\bar u
_1 \g^\mu u_2) (\bar u _3 \g _\mu u_4) (\bar u _5 u_6) + a_S (\bar u
_1 u_3) (\bar u _2 u_4) (\bar u _5 u_6) ] \cr && \times \int [d wd u d
yd x ] < e ^{- {\varphi (x_1) \over 2 } } ....e ^{{\varphi (x_6) \over
2 } } > <s _{- \t ^I } (x_1) \cdots {\tilde t _{-\t ^I _6} (x_6) t
_{\t ^I _7} (x_7) \choose s _{-\t ^I _6} (x_6) s_{\t ^I _7} (x_7) } >
\cr && \times < \s _{- \t ^I } (x_1) \cdots {\tau _{-\t ^I _6} (x_6)
\tilde \tau _{\t ^I _7} (x_7) \choose \s _{-\t ^I _6} (x_6) \s _{\t ^I
_7} (x_7) } > + \text{perms} , \cr && [\calT _1 = Trace(\l _1 \cdots
\l _7),\ \calT _2 = Trace(\l _7 \cdots \l _1) ] \eea where the upper
and lower entries for the twist field correlators refer to $ I=3$ and
$I=1,2$, and the Lorentz covariant structure is described by the
calculable functions $ a_S, \ a_V$ which we shall not attempt to
determine here.  The M\"obius group gauge fixing of the insertion
variables is set as, $ [x_1,x_2,x_3,x_4,x_5,x_6,x_7] = [0, w, u, y, x,
1, X\to \infty ] $.  In order to obtain finite contributions to $\cala
_7$, the brane models must comply with restrictive constraints.  Among
these are the conditions that the scalar boson be a localized mode
from a non-diagonal open string sector and that the $D6$-brane
embedding of the world sheet boundary forms a 7-polygon which closes
up to finite transverse separations.  The vector like gauge
unification schemes are out of the game because of the automatic $ B+
L$ or $ B-L$ conservation present in these models.  While the
left-right symmetric gauge models are natural candidates, as
illustrated by the brief review in Appendix~\ref{bnvnd}, we find that
none of the intersecting brane models discussed in the
literature~\cite{ibanyuk,kokosusy02} can be extended so as to include
finite contributions to the $\D B= - 2$ seven point amplitude at tree
level.  Indeed, for the $D$-brane models with Pati-Salam gauge
symmetry, $ SU(4)_c \times SU(2)_L \times SU(2)_R$, no perturbative
open string mode can occur with the required gauge group
representation, $ (10, 1, 3)$, which includes the boson $ \D ^l$.  The
higher dimensional multiplets which arise in the realization presented
in Ref.~\cite{kokosusy02} carry rather the representation, $ (10, 1,
1)$.  For the minimal left-right symmetric model with gauge group, $
SU(3)_c \times SU(2)_L \times SU(2)_R$, a localized dilepton scalar
mode $\D ^l $ carrying the triplet representation under $SU(2) _R =
SU(2)_a $ can possibly arise in orbifold models from the non-diagonal
open string sectors of type $ (a, \T ^g a ) $.

We close at this point the discussion of string amplitudes for the
exotic baryon number violating processes.  Our admittedly unfinished
presentation here is due to the non-trivial structure of the higher
order correlators for the coordinate twist operators.  We have
attempted here to clarify the conditions leading to finite string
amplitudes for five, six and seven localized matter modes of similar
structure to the four fermion amplitudes.  For the $\cddd =9$ baryon
number violating operators, only the VEV induced contributions can
possibly be present in the intersecting brane models. As in the
effective field theory model context (cf. Appendix~\ref{bnvnd}),
finite amplitudes may arise from dimension $\cddd = 10$ operators
involving massive scalar multiplets in higher dimensional
representations of the unified gauge group having Standard Model
singlet scalar components which acquire a VEV at some larger mass
scales.

\section{Field theory limit of  string amplitudes}
\label{sect3}

\subsection{Effective  low energy field theory} 
\label{sub31}

Before going into the applications, it is important to identify the
relevant fundamental parameters and determine their relation to the
parameters of the world brane field theory. The underlying string
theory is characterized by the Regge slope mass scale parameter, $ \a
' \equiv 1/m_s^2 $, and the string coupling constant which is linked
to the dilaton field VEV, $g_s= e^{ <\Phi >}$.  Alongside with the
complex structure moduli, $U^I = \hat b^I + i {r_1^I\over r_2 ^I}$,
and the wrapped three-cycles winding numbers, $ (n_a ^I, m_a ^I)$, we
introduce here the two real K\"ahler moduli parameters, $R$ and $r
\equiv 1/M_c$, representing in an average sense the length scale radii
for the 6-d torus and the three-cycles wrapped by the $D6$-branes.
The volumes of the internal manifold and three-cycles, $V_X $ and $
V_{Q_a} $, are described in terms of these parameters as \bea && V_X
\equiv (2\pi ) ^6 R ^6 ,\ V_{Q_a} \equiv \prod _{I=1}^3 (2\pi L_a ^I)
= r^3 \prod _I (2\pi \call _a ^I) =(2\pi r )^3 \call _a, \cr && [\call
_a = \prod _I \call ^I_a,\ \call _a ^I = {L_a ^I \over r} = {r_1 ^I
\over r} ( n^{I2} _a + {\tilde m^{I2} _a \over U_2^{I2} } ) ^\ud , \
\tilde m^I _a = m^I _a - n^I _a U^I _1 ] ,\eea
where $Q_a$ denotes the three-cycle wrapped by the $D6_a$-brane.  To
establish contact with the low energy physics, we must also introduce
an infrared cutoff mass parameter, $M_X$, whose meaning will be
discussed at length shortly.  It is useful to keep in sight the
familiar formulas expressing the 10-d gauge and gravitational coupling
constants of type $II$ supergravity theory, $ g_{10} $ and $ \kappa
_{10} $, and the tension, RR charge and gauge coupling constant
parameters of $Dp$-branes, $\tau _p , \ \mu _p $ and $ g_{Dp} $, as a
function of the string coupling constants, \bea && \kappa _{10} \equiv
{\kappa \over g_s} = 8\pi ^{7/2} \a ^{'2} ,\ g_{10} = { g_{YM} \over
g_s ^{\ud } } = {2 \kappa _{10} \over \sqrt {\a '} } , \cr && \tau
_p\equiv {\mu _p \over g_s} = {1\over g_s \sqrt {\a '} (2\pi \sqrt {\a
'} )^p } = {\sqrt \pi \over \kappa (2\pi \sqrt {\a '} )^{p-3} } , \cr
&& g^{-2} _{YM, p} \equiv \ud g^{-2} _{Dp} = \ud (2\pi \sqrt {\a '}
)^2 \tau _p ={ (2\pi \sqrt {\a '} )^{2-p} \over 2\sqrt {\a '} g_s }
. \label{eqstccs}\eea The 4-d gravitational Newton coupling constant,
$ G_N $, and the gauge interactions coupling constants on the
$Dp_a$-branes world volume theories, $ g_a $, are given by dimensional
reduction as \bea && G_N^{-1} \equiv M_P ^2 \equiv {16\pi \over 2
\kappa _4 ^2} = {16\pi V_X \over 2 \kappa ^2} = { 8 m_s ^8 V_X \over
(2\pi ) ^6 g_s ^2 } , \cr && g^{-2} _{a} = {V_{Q_a} \over 2 g^2
_{Dp,a} }= {V_ {Q_a} \sqrt {\a '} \over 2 g_s (2\pi \sqrt {\a
'})^{p-2} } = {V_{Q_a} m_s ^{p-3}\over 2(2\pi )^{p-2} g_s }, \eea
where the enhanced gauge symmetry case for $Dp$-branes overlapping the
$Op$-planes is obtained by replacing, $ g_a ^2 \to g_a ^2/2$.  For
reference, we also display below the inverse relations expressing the
fundamental string parameters as a function of the effective low
energy parameters for the case of $D6$-branes with $ p=6 $, \bea && \
m_s ={2\pi g_s ^{1/3} \a _a ^{-1/3} \over V_{Q_a} ^{1/3} }= { \a _a
M_P \over \sqrt {8 \l_a }} , \quad g_s = { m_s ^3 \a _{a} V_{Q_a}
\over (2\pi ) ^3 } = { \a _{a} ^{ 4 } M_P ^3 V_{Q_a} \over 8 ^{3/2}
(2\pi ) ^3 \l _a ^{3/2} }
, \cr && [\a _a \equiv {g ^2_{a} \over 4 \pi } = {g ^2_{X} \over 4 \pi
  k_a } ,\ \l _a = {\l \over k_a ^2} \equiv {V_X\over V_{Q_a} ^2 }
  \equiv {R^6\over r ^6 \vert \call _a \vert ^2} = \bigg ( {M_P g_a^2
  \over 8\sqrt 2 \pi m_s } \bigg ) ^2 ] . \eea
The dependence on the gauge group factors included in the parameters,
$ k_a = {V_{Q_a} \over V_{Q} } $, with $ V_Q$ denoting the volume of
some reference three-cycle, allows one to write the string gauge
unification relations as, $ k_a \a _a = \a _X$.  The gauge
interactions coupling constant, $g_X $, may be traded with either
$g_s$ or $m_s$, by using the relation, $( m_s r ) \prod _I \call _a ^I
= {g_s ^{1/3} \a _X ^{ - 1/3} } $.  For fixed $\a _X $, increasing $
m_s r $, causes $g_s$ to increase.  The ratio between the manifold to
three-cycle radii, $ \l = V_X / V_Q ^2 = (R /r ) ^6$, controls the
relative strength of the gauge and gravitational interactions.  We
note that the duality between compactifications of M-theory on $G_2$
holonomy manifolds and of type $II$ string theory on Calabi-Yau
manifolds with intersecting $D$-branes suggests~\cite{friedman02} that
the values of $\l $ are restricted by an upper bound of order unity,
while the larger values of $\l $ are favored~\cite{sverck06} in order
to reduce the mass scale of the string theory axions, $F_a$.  However,
we do not gain much in the present work in including the gravitational
interactions input, since this trades $G_N$ for the bulk radius
parameter $R$ which has no direct impact on the open string
observables.  One cannot simultaneously match $G_N $ and $g_a$ to the
observed values without imposing a wide disparity between the bulk and
brane radius parameters, $ r/R << 1$, hence requiring finely tuned
shape moduli parameters, $ r_2 ^I/r_1^I << 1$.  This circumstance
rules out the TeV scale models with $D6$-branes, which require large
extra dimensions in the bulk.  Of course, the restriction to toroidal
models with $ r \simeq R$ need not be in force in models with lower
dimensional $ D5$- or $D4$-brane backgrounds, where the existence of
subtori unwrapped by $D$-branes gives the ability to arbitrarily raise
the bulk volume $V_X$ relative to $V_Q$.  For the $D5$-brane and
$D4$-brane models, the elimination of $ m_s$ in favor of $M_P$ would
yield the formulas for the string coupling constant \bea && g_s\vert
_{D5} = {\a _a V_Q m_s ^2 \over (2\pi )^2 } ={\a _a ^2 M_PV_Q ^2 \over
4\sqrt 2 \pi V_X ^\ud }, \quad g_s\vert _{D4} = {\a _a V_Q m_s \over
2\pi } = {\a _a ^{4/3} M_P^{1/3} V_Q ^{4/3} \over \sqrt 2 (2\pi
)^{1/3} V_X ^{1/6} }, \label{eqd5d4} \eea which are seen to exhibit a
variant dependence of $g_s$ on the gauge coupling constant, $\a _a $.

Let us now discuss the mass cutoff which is inevitably needed to
regularize the infrared divergences in string amplitudes.  In unified
gauge theories, this is naturally associated with the gauge symmetry
breaking mass scale parameter, $M_X$. Since the wrapped cycle volume $
V_{Q_a} $ fulfills the same role as $M_X$, it is convenient to
introduce the dimensionless parameter~\cite{KW03}, $ L(Q_a) = V_{Q_a}
M_X ^3 $.  The compactification mass parameter, $ M_c \equiv {1\over r
} = {2\pi \call ^{1\over 3} \over V_Q } ,\ [\call = {1\over r^3 }
\prod _I (n_a ^{I2} r_1^{I2} + \tilde m_a ^{I2} r_2^{I2} )^\ud ] $ is
linked to the unified symmetry breaking mass scales $ M_X $ through
the string threshold corrections to the gauge coupling constants, $\D
_a$.  The moduli dependent functions, $\D _a $, are commonly defined
by writing the one-loop renormalization group scale evolution for the
running gauge coupling constants as \bea && \D _a \equiv { (4\pi )^2
\over \bar g_a ^2(Q) } - { (4\pi )^2 k_a \over \bar g_X ^2 } - b_a
\log ({Q^2 \over m_s ^{'2} }) ,\ [k_a = { V_{Q_a} \over V_Q } ] \eea
where $b_a$ denote the one-loop slope parameters of the gauge group
factors, $G_a$, determined by the massless charged modes, and $ m'_s$
denotes the string theory effective unification mass, determined by
matching onto the low energy field theory.
We choose here to treat $M_X$ as a free parameter without imposing any
condition on its relative ordering with respect to the
compactification mass, $ M_c $, except for the request that both
masses are bounded by the string mass scale, $ (M_X , \ M_c ) \leq
m_s$.  Loosely speaking, the condition $ M_X < M_c $ implies that the
unified symmetry occurs in 4-d, while the condition $ M_X > M_c $
implies that that this occurs in 7-d.  From the interval of variation
assigned to $ M_c$ by the combined consistency conditions, $ g_s <1,\
m_s /M_c > 1$, yielding, $ (\a _X \vert \call \vert ) ^{1\over 3} <
{M_c \over m_s } < 1 $, we see that the possibility $ M_X \geq M_c$ is
not excluded.

To describe the ratio of the string theory amplitude to that of the
equivalent field theory, we follow the procedure  of Klebanov and
Witten~\cite{KW03}.  Ignoring momentarily the regularization of
divergences, we write the string amplitude as, $ \cala _{st}= 2 \pi
g_s I (\t ) \calT \cals / (2m_s ^2) $, where the extra factor $\ud $
accounts for the orientifold projection, the $x$-integral $I (\t ) $
is a calculable function of the dimensionless free parameters, $ m_s r
= m_s /M_c ,\ s= m_s /M_X $, and $\calT ,\ \cals $ contain the
dependence on the gauge and Dirac spinor wave functions.  The
equivalent field theory amplitude is now assumed to include the same
factors, $ \cala _{ft}= 2\pi \a _X \calT \cals / M_X^2 $.  In
correspondence with the different prescriptions of identifying the
effective parameters, we deduce the following three predictions for
the ratio of string to field theory amplitudes, \bea && \calr
_{s/f}\equiv {\cala _{st} \over \cala _{ft} } = {\pi g_s /m_s^2 \over
2\pi \a _X /M_X^2} I (\t) \cr && = \bigg [ { g_s ^{1/3} \a _a ^{-1/3}
V_Q ^{2/3} M_X ^2 \over 8 \pi ^2 } \ , \ {M_X^2 m_s V_Q \over 2(2\pi
)^3 } \ , \ { \a _a V_{Q_a} M_P M_X ^2 \over 32 \sqrt 2 \pi ^3 \l _a
^{\ud } } \bigg ] I (\t ) , \label{eqratsf} \eea where the successive
entries inside brackets are obtained by eliminating $ g_s$ alone, $
m_s$ alone, or $g_s $ and $m_s$ together, in favor of the low energy
gauge and gravitational coupling constants.  It is instructive to
compare these results with the ratio of M-theory to field theory
amplitude~\cite{friedman02} \bea && \cala _{M} = C _M { g_7 ^2 M_{11}
\over 4 \pi } J_\mu J ^\mu \ \Longrightarrow \ \calr _{M/f} = {\cala
_{M} \over \cala _{ft} } \simeq {1 \over 2 } C_M L^{2/3}(Q) \a _X
^{-{1 \over 3 } } , \ [ M _{11} ^ {-9} = 2 \kappa _{11} ^2 (2 \pi ) ^8
] . \eea
Getting an estimate for the coefficient $C_M$ was the main motivation
of Ref.~\cite{KW03}.  The appearance in the string theory amplitude at
fixed $g_s$ of the same power dependence on the 4-d gauge coupling
constant, $ \cala \propto \a _X ^{2\over 3} $, joins with the fact
that in the large radius limit of the string and M-theory models the
gauge unification occurs in 7-d.  A useful discussion of this issue is
presented by Burikham~\cite{burik05}.  That this property is specific
to the $D6$-brane models appears clearly from the different dependence
on the 4-d coupling constants of $g_s$ in Eq.(\ref{eqd5d4}) for the
$D5$-brane and $D4$-brane models.  The $D6$-brane case is known to be
unique in admitting a purely geometrical lifting to the M-theory
models~\cite{blumshiu05}.

\subsection{Regularization of string  amplitudes} 
\label{sub32}

The enhancement of string amplitudes reflected by the non-analytic
dependence on the 4-d gauge coupling constant takes place naturally in
the large compactification radius limit, $r =1/M_c \to \infty $. The
effect is maximized by considering fermions localized at vanishingly
small distances relative to the compactification radius~\cite{KW03}.
It is important to realize that what justifies considering directly
the ratio of amplitudes in Eq.(\ref{eqratsf}) is the assumption that
the string amplitude is self-regularized. However, that the large
compactification radius limit of the string amplitude is only finite
in the case of $D6$-branes can be easily seen by rewriting the
$x$-integral truncated to the leading contribution at $x\to 0$ as \bea
&& \pi ^{L \over 2} \int _0 ^1 dx x^{-s-1} (\ln (1/x) )^{- {L\over 2
}} = \int d^L \vec q {1\over \vec q ^2 -s } ,\eea where the right hand
side exhibits the sum over propagators for the plane wave modes along
the $L=3$ effectively flat directions.  The convergence at $q\to 0$
would not hold for $L < 3$.  (Of course, the factor $ \ln 1/x$ slows
the convergence of the integral near the endpoint, $x=0$, but the
problem this would pose is only of technical order.)  In the $D5$- and
$D4$-brane models with the $x$-integral at $ L =2 $ or $ 1$, the
singularities at $x\to 0$ are seen to be not integrable ones.

Owing to the slowly convergent $x$-integrals, it is not clear whether
the enhancement survives finite compactification radii.  Indeed, as
just noted, the integrable singularity at $x=0$ is specific to the
case of $D6$-branes where the infrared singularity at $\vec q\to 0$ is
cancelled by the integration measure $ d^3 q$. At finite
compactification radius, the massless gauge boson modes in the $s-$
and $t-$channels are separated by a finite mass gap from the massive
string modes, and the presence of infrared divergences from massless
pole terms is then unavoidable.  The pole terms from exchange of
string modes can be separated out by means of the familiar analytic
continuation method.  While the non-trivial functional dependence of
the $x$-integrand from the coordinate twist field correlator precludes
using a fully analytical procedure, it is still possible to consider a
semi-numerical regularization of the $x$-integral where one removes by
hand the massless $s$- and $t$-channel gauge boson poles originating
from the contributions to the $x$-integral of form, $ \int _0 dx x^{-s
-1} \calf _0 (x) $ and $ \int ^1 dx (1-x)^{-t -1} \calf _1 (x) $,
where $ \calf _0 (x) ,\ \calf _1 (x) $ represent the limiting forms of
the remaining factors in the integral at $ x\to 0 $ and $ x\to 1 $.

We now discuss two distinct infrared regularizations that will be used
in the sequel to obtain numerical predictions in the finite
compactification radius case.  In the first prescription, the
regularized amplitude is constructed by subtracting out by hand the
massless gauge boson pole terms in the relevant channels and
substituting these by the corresponding massive pole contributions,
$1/s \to 1/(s-M_X^2 ) $.
We thus write the string amplitude as, $ \cala _{st} = \pi g_s \a '
[\cals _1 \calT_1 I_1 + \cals _2 \calT_2 I_2] $, and substitute for
the direct and reverse orientation integrals, $ I_1, \ I_2 $, the
subtraction regularized integrals with the massless poles removed,
\bea && I _1(\t ) \to I _{1} ^{reg}(\t ) = I_1 -I_{1,0} -I_{1,1}, \
[I_{1,0} = {\sqrt {\a '} \over \vert L_B \vert (-s) } ,\ I_{1,1} =
{\sqrt {\a '} \over \vert L_A \vert (-t) } ] \cr && I_2 (\t ) \to I
_{2} ^{reg}(\t ) = I_2 -I_{2,0} -I_{2,1},\ [I_{2,0} = {\sqrt {\a '}
\over \vert L_A \vert (-t) } ,\ I_{2,1} = {\sqrt {\a '} \over \vert
L_B \vert (-s) } ] \eea where the indices $ 0$ and $ 1$ in $ I_{i, 0}
,\ I_{i, 1}, \ [i=1,2] $ refer to the limits $ x\to 0$ and $x\to 1$.
In the case, $ \e _A = \e _B=0 $, where massless poles are present in
both $s$- and $t $-channels, the regularized string amplitude is
defined schematically as, \bea && \cala = [\cala - \cala \vert _{ x
\to 0} - \cala \vert _{ x \to 1} ] + \cala _s ^0 + \cala _t ^0 , \quad
\bigg [\cala _s ^0 = { C_B \over \vert L_B \vert ( s- M_X ^2) } , \
\cala _t ^0 = {C _A \over \vert L_A \vert ( t- M_X ^2) }, \cr && \cala
\vert _{ x \to 0} = {C _B \over \vert L_B \vert } \int _0 ^1 dx x
^{-s-1} ,\ \cala \vert _ {x \to 1} = {C _A \over \vert L_A \vert }
\int _0 ^1 dx (1-x) ^{-t-1} \bigg ] \eea where the constant
coefficients $ C_A,\ C_B$ are determined from the limits at the end
points of the $x$-integral.  We now assume that the field theory
amplitude has the same dependence on the Dirac spinor and gauge wave
functions. Using, for simplicity, the approximate equalities, $\vert
L_A \vert =\vert L_B\vert = \vert L \vert $, \ $ I_{1} ^{reg} = I_{2}
^{reg} = I ^{reg}$, and $ \a _X = \a _a , \ m_X = M_X$, we can express
the ratio of string to field amplitudes in the fixed $g_s$
prescription as \bea && \calr _{s/f} = {\pi g_s \a '\over \vert L
\vert } \bigg [ \cals _1 \calT_1 \bigg ( I _{1} ^{reg} \vert L \vert +
{\sqrt {\a '} \over -s +M_X^2 } +{\sqrt {\a '} \over -t +M_X^2 }\bigg
) \cr && + \cals _2 \calT_2 \bigg (I _{2} ^{reg} \vert L \vert +
{\sqrt {\a '} \over -t +M_X^2 } +{\sqrt {\a '} \over -s +M_X^2 } \bigg
) \bigg ] \bigg [ 2\pi \a _a \bigg ( {\cals _1 \calT _1 \over -s
+M_X^2 } + {\cals _2 \calT _2 \over -t +M_X^2 } \bigg ) \bigg ]^{-1}
\cr && \simeq 1 + \ud g_s ^{1\over 3} \a _X ^ {- {1\over 3 } } \vert L
\vert ^{2\over 3} M_X^2 I ^{reg} (\t ) . \eea In the fixed $m_s$
prescription, the ratio would instead read as, $ \calr _{s/f} \simeq 1
+ \ud M_X ^2 m_s \vert L \vert I ^{reg} (\t )$.  For the case with non
coincident intersection points, $ \e _A = 0, \ \e _B \ne 0 $, the $t
$-channel pole is absent, so that using the above prescription without
the terms $ \cala _ {\vert x \to 1} $ and $\cala _t ^0 $, yields the
ratio, $\calr _{s/f}= 1 + M_X ^2 m_s \vert L \vert I ^{reg} .$ An
analogous result holds in the case, $ \e _A \ne 0, \ \e _B =0 $.

The second, perhaps more natural, regularization procedure is based on
the description of the unified gauge symmetry breaking by the deformed
configuration obtained by splitting a finite distance apart the
unified gauge theory $D6$-brane stack into two stacks.  The transverse
displacement parameter is related to the symmetry breaking mass scale
$M_X $ by the leading contribution at $x\to 0$ in the classical action
factor, as displayed by Eq.~(\ref{sec3.wils}).  This amounts to moving
along the Coulomb branch of the gauge group moduli space of vacua and
is T-dual to a Wilson flux line around the cycles normal to the brane
stack.  Denoting the resulting multiplicatively regularized
$x$-integrals as $ \hat I _{1,2} ^{reg} $, and using the same
simplifying assumptions as above, leads to the following regularized
ratio of string to field theory amplitudes in the fixed $g_s$
prescription, \bea && \calr _{s/f}= { \cala _{st} \over \cala _{ft}} =
{ \ud m_s \vert L \vert ( \cals_1 \calT_1 \hat I _{1} ^{reg} + \cals_2
\calT_2 \hat I _{2} ^{reg} ) \over [ {\cals_1 \calT _1 \over -s +M_X^2
} + {\cals_2 \calT _2 \over -t +M_X^2 } ] } \simeq \ud g_s ^{1\over 3}
\a _X ^{-{1\over 3} } \vert L \vert ^{2\over 3} M_X^2 \hat I ^{reg} .
\label{eq.dispresc} \eea The ratio in the fixed $m_s $ prescription would instead read, $
\calr _{s/f} \simeq \ud M_X^2 m_s \vert L \vert \hat I ^{reg} .$
Recall that the regularized integral, $ \hat I ^{reg}$, now includes
the brane induced form factor, $ \prod _I ( \hat \d _I ) ^{\sin ^2
(\pi \t ^I ) d _A ^{I2} }$, as needed to restore the correctly
normalized gauge coupling constant.  In the case with equal angles,
the correction factor simplifies to $ (\hat \d ) ^ {M^2_X}$.  The
numerical comparison of the above discussed regularization
prescriptions will be presented in the next section.

To conclude, we comment on the calculation of the ratio of amplitudes
in orbifold models.  The vertex operators of the $(a,b)$ modes must
now be summed over the orbifold group images, $ V_{(a,b)} \to \sum _g
V_{(a,b _g)}$.  We choose not to include the normalization factor
$1/\sqrt N $ at this stage, since the amplitude normalization will be
determined in this case by comparison with the gauge coupling
constant.  Other conventions would lead to the same end result.  Upon
combining the summations over the orbits in the various vertex
operator factors, the compatibility conditions from the target space
embedding on the brane boundaries leaves a single orbifold group
summation of equal reduced amplitudes, which then introduces an
overall factor $N$.  One must now recall that promoting the torus to
an orbifold entails a reduction of the torus volume by the group order
factor $N$.
The gauge and gravitational interaction parameters are expressed via
dimensional reduction by same formulas as for tori, with the
three-cycle volume identified as, $ V_Q ^{orb}= {V_Q ^ {tor} \over N
}$.  This change can be implemented by using either $( g_a
^{-2})_{orb} = {( g_a ^{-2})_{tor} \over N} $ and $ \ ( M_P ^2 )_{orb}
= {( M_P ^2 )_{tor} \over N} $, with fixed string parameter, $g_s$, or
$(g_s)_{orb} = { (g_s )_{tor} \over N} ={ g_a ^2 V_Q ^{tor} m_s^3\over
2 (2\pi )^ 4 N } $, with fixed gauge and gravitational coupling
constants.  Repeating the calculation of the string amplitude
factorization on the gauge boson pole term, one arrives at the
formula, $g_s = {m_s ^3 V_{Q_a} \a _a \over (2 \pi )^3 N } $, with the
understanding that the three-cycle volume is that evaluated for the
torus, $ V_{Q_a} = (2\pi r )^3 \prod _I ( L ^I_a ).$ One may thus
summarize the conversion from torus to orbifold descriptions by the
following schematic correspondence \bea && (\cala _{st})_{tor} = {2\pi
g_s \over 2 (V_Q ) _{tor} } \calT \cals I (\t ) \ \Longrightarrow \
(\cala _{st})_{orb} = {2\pi g_s N \over 2 (V_Q)_{tor} } \calT \cals I
(\t )
. \eea This shows that upon passing from tori to orbifolds, one may
retain the same normalization constant, $ C= 2\pi g_s$, while using
the modified formula for the string coupling constant, $ g_s = m_s ^3
\a _X \vert L \vert /N $.

\subsection{Large compactification radius  limit} 
\label{sub33}

To prepare the ground for the calculation of four fermion amplitudes
in semi-realistic models, we present the results obtained in the large
compactification radius limit using the self-regularized $x$-integral.
The ratio of string to field theory depends then on the branes
intersection angles only. Lacking a simple analytic approximation for
the $x$-integral, we perform the quadrature numerically.  The relevant
integral $ I (\t ) $ in the equal angles case, $\t ^I =\t ^{ 'I}$,
with the classical action factors omitted, is evaluated numerically
for the toy orientifold model~\cite{KW03} realizing the $SU(5)$ group
unification with the mirror pair of $D6/D6'$-brane stacks, allowing
for a single massless $10$ matter multiplet.  One advantage of this
simple model is in studying the anticipated enhancement effect in the
amplitude, $ 10 \cdot {10}^{\dagger } \cdot 10 \cdot {10}^{\dagger }
$, as a function of the brane-orientifold angle only.  The amplitude
in Eq.~(\ref{eqstreq}) can be expressed in terms of the $x$-integral
\bea && {\cala ' \over 2 C \cals _1 (\calT _1 + \calT _2) } \equiv I
(\t ) = \int _0 ^1 {dx \over x} \prod _I \bigg ( {\sin \pi \t ^I \over
F(x) F(1-x) } \bigg )^{1/2} , \label{eqxintnum} \eea where we used the
identity, $ [x(1-x)]^{-1} = x^{-1} +(1-x)]^{-1}$, and the symmetry of
the integrand under $x\to (1-x)$.  We pursue here the study in
Ref~\cite{KW03} by evaluating $ I(\t ) $ for generic three-cycles, not
obeying the supersymmetric restriction on the angles, $\sum _I \t ^I =
0 $.  As already noted, the infrared finiteness of the integral does
not dispense us from taking care of the sensitive integrations at the
end points of the integration interval.
Reasonably accurate numerical values for this integral can be obtained
by the simple procedure which consists in subtracting the leading term
at $x\to 0$ of the integrand and adding back its contribution to the
integral, as illustrated in the following formula \bea && I(\t )
\equiv \int _0 ^1 dx \calf (x) = \int _0 ^1 dx [\calf (x)- \calf _0
(x)] + \int _0 ^1 dx\calf _0 (x) \equiv [I(\t ) -I_0 (\t ) ] + I_0 (\t
), \cr && [\calf (x) = x^{-1} \prod _I ( {\sin \pi \t ^I \over F(x)
F(1-x) } )^{1/2} ,\ \calf _0 (x) = \lim _{x\to 0} \calf (x) ,\cr && I
_0 (\t ) \equiv \int _0 ^1 dx \calf _0 (x) \equiv \int _0 ^1{ dx \over
x} \prod _I \bigg [{\pi \over \ln \d _I - \ln x } \bigg ] ^{\ud }
].\eea The same subtraction procedure can be used for the $x=1$ end
point, with the limiting function $ \calf _0 (x)$ substituted by $
\calf _1 (x) = \lim _{x\to 1} \calf (x) = \calf _0 (1-x) $.  The
subtracted integral displayed in the second entry of the above
equation has a simple representation in the two special cases
involving three equal angles, $\t ^1=\t ^2 =\t ^3 =\t $, or two
unequal angles, $\t = \t ^1=\t ^2 \ne \t ^3$, respectively.  The
resulting analytic formulas for these two cases are given by \bea &&
\bullet \ I _0 (\t , \t , \t ) = 2 \pi ^{3/2} (\ln \d )^{-3/2} , \cr
&& \bullet \ I _0 (\t , \t , \t _3) = {\pi ^{3/2} \over \vert \ln (\d
_3 /\d ) \vert ^\ud } \bigg [- \T _H (\D ) \ln { 1- (1- { \ln \d \over
\ln \d _3 } ) ^\ud \over 1+ (1- { \ln \d \over \ln \d _3 } ) ^\ud }
\cr && + \T _H (- \D ) [ \pi - 2 \arctan \bigg ({1\over -1+ { \ln \d
\over \ln \d _3 } } \bigg ) ^\ud ] \bigg ] , \eea where $\T _H $
denotes the Heaviside function ($\T _D (y ) =1 $ for $ y>0$ and $0$
for $ y< 0$) with $ \D \equiv \ln {\d _3 \over \d } $ positive and
negative in correspondence with the cases, $ \d < \d _3$ and $ \d > \d
_3$.

\begin{center}  \begin{figure} 
\epsfxsize= 6 in
\epsffile{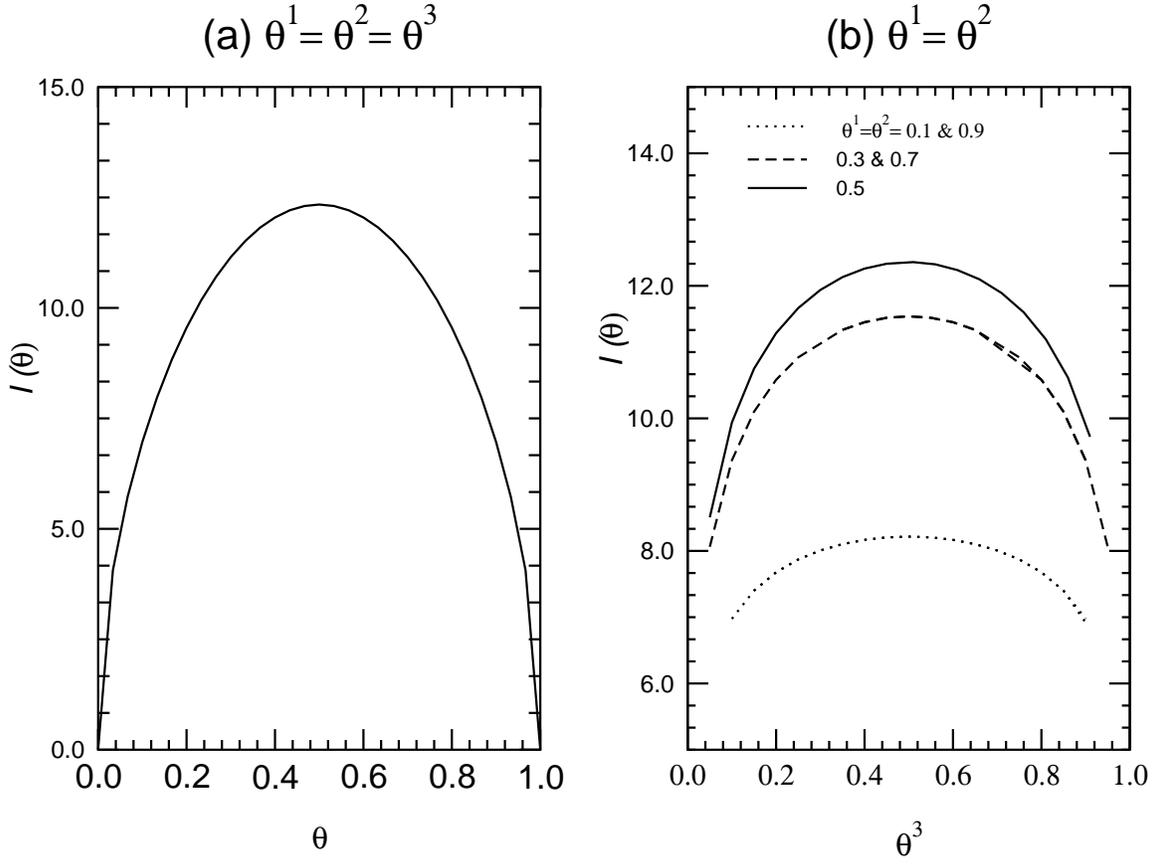}
\caption{\it Plot of the $x$-integral $I(\t ) $ defined in
Eq.~(\ref{eqxintnum}) for the four fermion string amplitude in the
large radius limit as a function of the brane intersection angles in
the equal angle case $\t = \t '$. The left hand panel $(a)$   shows, for the
case with three equal angles,   a plot of $I(\t ) $ versus the
common angle, $\t = \t ^1= \t ^2= \t ^3 $.  The right hand panel $(b)$
 shows, for the case with two equal angles,  plots of $I(\t ,\t , \t ^3)
$ versus $\t ^3$,  at three values of the two equal angles, $\t ^1= \t
^2 = (0.1, 0.3, 0.5)$.  Because of the symmetry of the integral under
$ \t ^I \to 1 - \t ^I $, same results apply in the cases involving the
angles, $\t ^1= \t ^2 = (0.9, 0.7, 0.5)$.}
\label{figasam01}
\end{figure} \end{center}

The numerical results for the integral $I (\t ) $ as a function of the
single variable intersection angle are displayed in
Figure~\ref{figasam01}.  We note that the integral rapidly vanishes at
the end points, corresponding to parallel or anti-parallel brane
stacks, and is maximized at the midpoints, $ \t ^I = \ud $, as
expected from the symmetry under $\t ^I \to (1-\t ^I)$, at fixed $I$.
Away from the end points, the integral vary slowly and monotonically
inside the interval, $ I(\t ) \simeq (6\ - \ 12 ) $ for $ \t \in [0,
\ud ] $.  Recall that for supersymmetric cycles satisfying $\sum _I \t
^I = 2 $, the integral was found to vary inside the range, $ I(\t )
\simeq (7\ - \ 11 ) $.  No significant differences thus arise in going
from the case of supersymmetric cycles~\cite{KW03} to that of
non-supersymmetric cycles.


\section{Baryon number violation in  gauged unified orbifold models}
\label{sect4}

\subsection{$Z_3$ orbifold    $SU(5)$  unified  model}
\label{sub41}

We here employ the formalism developed in Section~\ref{sect2} to
examine the enhancement of string amplitudes for nucleon decay
processes at finite compactification radius.  We consider two type
$II$ string theory realizations of $SU(5)$ gauge unification using
$D6$-branes.  The first uses a non-supersymmetric $Z_3$ orbifold
model~\cite{Blumenhagen:2001te} and the second a supersymmetric $ Z_2
\times Z_2$ orbifold model ~\cite{Cvetic:2001nr}.  To complement our
discussion of the solutions for the non-chiral spectrum of the first
model, we summarize in Appendix~\ref{appz3} the properties of models
using the $Z_3$ orbifold with $D6$-branes, based on the work of
Blumenhagen et al.,\cite{Blumenhagen:2001te}.

\subsubsection{Low lying mass spectrum} 

The $SU(5)$ gauge unified $Z_3$ orbifold model of Blumenhagen et
al.,\cite{Blumenhagen:2001te} uses the minimal setup of two $D6$-brane
stacks $ N_a=5,\ N_c=1$ with effective wrapping numbers, $ (Y_a , Z_a
) = (3, \ud ), \ (Y_c , Z_c ) = (3, - \ud )$, solving the RR tadpole
cancellation condition, $ \sum _\mu N_\mu Z _\mu =2$.  This realizes
the gauge group $SU(5) \times U(1)_a \times U(1)_c $ with three chiral
matter fermion generations, $ F _i , \ \bar f _i , \ \nu ^c _i $,
localized at the brane intersections and several copies of non-chiral
fermions, $ H _i , \bar H _i $ and $ K_{a,l} ,\ K_{b,l} $. The gauge
group representation content and multiplicity of these modes is
displayed in the table below, in correspondence with the open string
sectors to which they belong.
\vskip 0.5 cm

\begin{center}
\begin{tabular}{c|ccc|cc} 
Mode & $F$ & $\bar f $ & $ \nu ^c $ & $ H+\bar H $ & $K_a$ \\ &&&&& \\
\hline &&&&& \\ Brane & $ (a',a) $ & $ (a,c) $ & $ ( c,c') $& $(a',c)
$ & $ (a,a)$ \\ &&&&& \\ Irrep & $3 (10,1)_{2,0} $ & $ 3 (\bar
5,1)_{-1,1} $ & $ 3(1,1)_{0,-2} $ & $ n(5,1)_{1,1} + n(\bar
5,1)_{-1,-1}$ & $I^{Adj} _a (24,1)_{0,0} $ \\
\end{tabular}
\end{center}
\vskip 0.5 cm

A vertical bar has been inserted to separate the chiral from the
vector representations.  The complete spectrum for matter fermions is
of the form, $ (3 + \D _F) F +\D _F \bar F,\ (3 + \D _{\bar f}) \bar f
+ \D _{\bar f} \bar f ,\ (3 + \D _{\nu ^c }) \nu ^c + \D _{\nu ^c }
\bar \nu ^c $, where the model dependent integers $\D _F ,\ \D _{\bar
f},\ \D _{\nu ^c } $ count the multiplicities of mirror vector
generations.
Of the two Abelian gauge group factors, the linear combination with
charge, $ Q_{an}= 5 Q_a+Q_c$, is anomalous while the unbroken
orthogonal combination, $ Q_{free} = {Q_a\over 5} -Q_c$, is anomaly
free.  When the unified gauge group breaks down to the Standard Model
group, the leftover unbroken Cartan subalgebra generator of $SU(5)$
combines with the anomaly free gauge charge so as to yield the
unbroken gauge symmetry, $U(1)_{B-L}$.  The anomaly free gauge
symmetry $U(1)_{free}$, which assigns charges $Q _{free} = {2 \over 5}
(1, -3, -2 ) $ to the representations $ (10, \bar 5, 5)$, precisely
identifies with the accidental symmetry of the minimal $SU(5)$
unification related to the $Y$ and $B-L$ charges as, $ Q _{free} + {4
\over 5} Y - Q _{B-L} =0 .$

\begin{center}
\begin{figure}  
\epsfxsize=  5 in
\epsffile{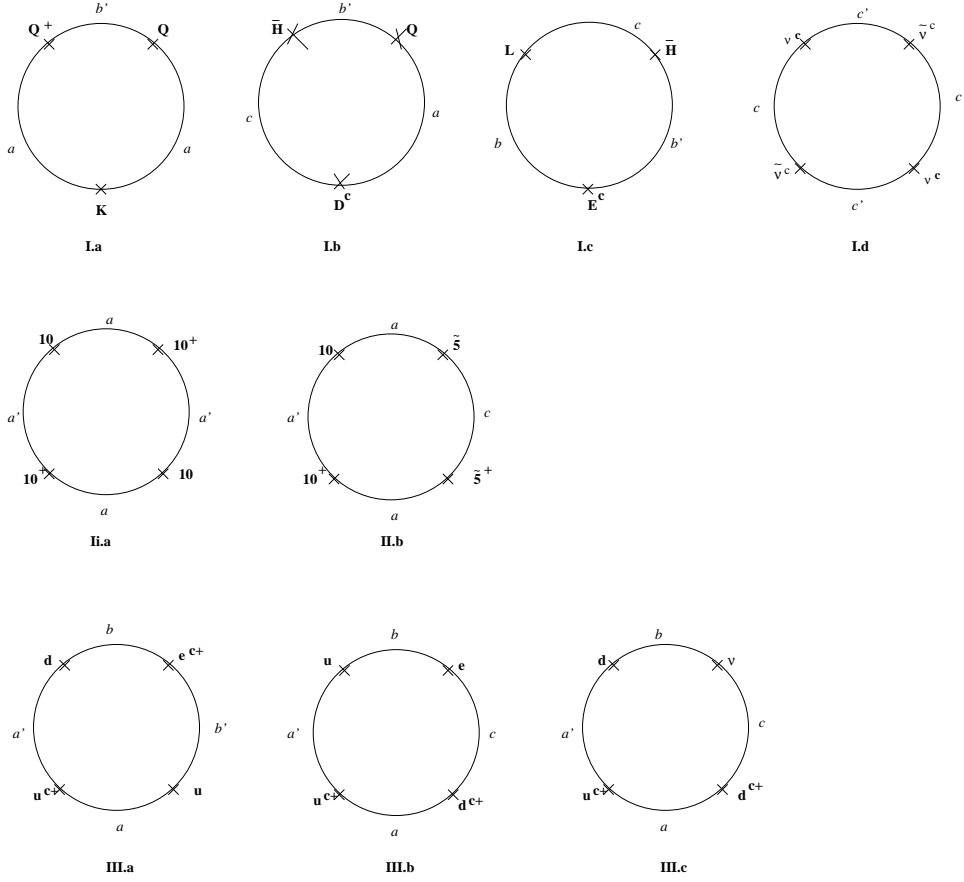}
\vskip 1 cm
\caption{\it Representative disk couplings for three and four point
string amplitudes in the $SU(5)\times U(1)$ gauge unified model of the
$Z_3$ orbifold discussed in Subsection~\ref{sub41}.  The arc segments
on the disk boundary are labelled by the branes, $a,\ b, \cdots $ and their mirror
images,  $ a',\ b',\cdots $, and the crosses by
the open string modes.  In the first array ({\bf I}), the coupling
$I.a$ refers to the Dirac mass generation for localized fermion and
scalar modes in bifundamental and adjoint representations, the
couplings $I.b$ and $I.c$ to the trilinear Yukawa coupling of down
quarks and leptons with the electroweak Higgs boson, and the coupling
$I.d$ to the quartic coupling, $\tilde \nu ^ {c \dagger } \tilde \nu ^
c \nu ^{c\dagger } \nu ^c$.  In the second array ({\bf II}), the
couplings $II.a,\ II.b$ represent the  baryon number
violating subprocesses in the $SU(5)$ symmetry limit of the $Z_3$
orbifold model for the configurations $10 \cdot {10} ^\dagger \cdot 10
\cdot {10} ^\dagger $ and $10 \cdot \bar 5 \cdot 10 ^\dagger \cdot
\bar 5 ^\dagger $.  In the third array ({\bf III}), the couplings $
III.a , \ III.b,\ III.c$ refer to the contributions to the baryon
number violating local operators in the quark and lepton fields, $
O_{e_L ^c} ,\ O_{e_R ^c} ,\ O_{\nu _R ^c} $, listed in
Eq.(\ref{eqndop}) of Appendix~\ref{bnvnd}.  The particle modes are
assigned to brane intersection points in the convention where one
navigates the disk boundary clockwise.}
\label{figz3disc}
\end{figure}
\end{center}

The scalar modes content of the model includes the ingredients needed
to accomplish the Higgs mechanism breaking of the various gauge
symmetries.  The scalar field VEVs for the $ K_{a,l} $ singlet
components of the adjoint representation modes of $SU(5)$, are needed
to break the unified $SU(5) _a$ gauge symmetry at the unification mass
scale $M_X$; those for the modes $ \nu ^c _i$ are needed to break the
$ U(1)_{B-L} $ gauge symmetry at some low intermediate mass scale; and
those for the modes $ (H_i , \ \bar H_j) $ are needed to break the
electroweak $SU(2)_L$ gauge symmetry at the Fermi mass scale.  The
presence of tachyon or low-lying scalar modes $K_{a,l}$ with mass
considerably lower than the string mass parameter, is facilitated by
the fact that these modes arise in massless chiral supermultiplets.
The $U(5)$ singlet components of the $ K_{a,l} $ scalars have the
ability to produce large Dirac masses for the various vector pairs of
spin-half fermions.  The tree level string couplings of the matter
fermions, represented by the disk configuration in the amplitude $I.a
$ of Figure~\ref{figz3disc} and the similar couplings for the higgsino
and gaugino like fermions, $ \tilde H_i + \tilde {\bar H}_i,\ \tilde
K_{a, l} $, produce the Yukawa trilinear interactions, \bea &&
L_{EFF}= h ^F _{ij,l} \tilde F_i \tilde F _j ^\dagger K_{a, l} + h
^{\bar f} _{ij,l} \tilde {\bar f }_i \tilde {\bar f }_j ^\dagger K_{a,
l} + h ^H _{ij,l} \tilde H_i \tilde {\bar H }_j K_{a, l} + h ^K
_{mn,l} \tilde K_{a, m} \tilde K_{a, n} K_{a, l} , \eea which are
necessary to decouple the extra fermion modes.  For this to occur, it
is necessary that the fermion mass matrices, $ h ^{F } _{ij,l} < K_{a,
l} > ,\ h ^{\bar f } _{ij,l} < K_{a, l} > ,\ h ^{ H } _{ij,l} < K_{a,
l} > $ and $ h ^{ K } _{mn,l} < K_{a, l} > $, have rank not smaller
than the vector representation multiplicities, $ \D _F, \ \D _{\bar
f},\ n $ and $ I^{ \bf Adj } _a $.

The intermediate mass scale breaking of $U(1)_{B-L}$ can be
accomplished by the localized scalars $\tilde \nu ^c$ of the $ (c,c')$
sector, which enter with multiplicity $3$.  For a tachyon mode to be
present, one must cancel the negative contribution to the string
squared mass depending on the angles by the positive contribution
depending on the branes $ c,\ c' $ distance.  For this, one must
require that the brane $c$ does not pass through the origin of the
coordinate system.  While the Dirac and Majorana mass matrices of the
left chirality neutrinos vanish, finite Majorana masses can occur for
the right chirality neutrinos via the quartic coupling, $ \tilde \nu ^
{c} \tilde \nu ^ {c} \nu ^c \nu ^c $, once the $\nu ^c$ tachyon scalar
raises a VEV.  The corresponding correlator, $ < V_{-\t } ^{(-\ud )}
(\nu ^c ) V_{\t }^{(0)} ( \tilde \nu ^ {c \dagger } ) V_{-\t }^{(-\ud
)} (\nu ^{c \dagger } ) V_{\t }^{(-1)} ( \tilde \nu ^ {c} ) > $,
represented by $I.d$ in Figure~\ref{figz3disc}, is expected from the
structure of the string amplitude in Eq.~(\ref{eq5pamp})
to give a contribution to the Majorana neutrino mass of form, $ m_{\nu
^c} \sim {g_s < \tilde \nu ^ {c} > ^2 \over m_s } $.

Although the $Z_3$ orbifold brane setup of interest does not preserve
any supercharges, one may still use the freedom in the
moduli parameters to dynamically select a vacuum having low-lying or
tachyon scalars with the appropriate gauge quantum numbers.  For the
localized modes in non-diagonal sectors, $(a,b)$, this possibility can
be explored by considering the stability tetrahedron with edges traced
in the space of the brane pair intersection angles, $ \t ^I _{ab}$, by
the four equations expressing the vanishing of the scalar modes
minimal squared mass in Eq.(\ref{sec2.tetra}).  The preserved
supercharges on the tetrahedron faces, edges and vertices generate the
supersymmetries, $ \caln = 1, 2,4$. The faces separate the inside
domain where all four squared masses are positive from the outside
region where at least one squared mass is negative.  Clearly, one
should avoid scalar tachyons with the gauge quantum numbers of matter
fermions, $F,\ f $, but would welcome those with Standard Model
singlet components whose condensation can accomplish the Higgs
mechanism symmetry breaking~\cite{iban0}.

The unified $SU(5)_a$ gauge symmetry breaking down to the Standard
Model corresponds to the deformation described by the brane splitting,
$a \to a + b$.  Following Blumenhagen et
al.,\cite{Blumenhagen:2001te}, we consider the setup consisting of the
three stacks, $ N_a=3,\ N_b=2,\ N_c=1$, with the effective wrapping
numbers, $ (Y_a , Z_a ) = (Y_b , Z_b ) = (3,\ud ) $, solving the RR
tadpole cancellation condition.  While the Coulomb branch deformation
is restricted to mutually parallel $D$-branes in the orbifold
equivalence classes, $[a]$ and $ [b]$, one may also consider the
deformation leading to non-parallel $a$ and $ b$ brane stacks.
Removing from the gauge symmetry group, $ U(3)_a \times U(2)_b \times
U(1)_c $, the anomalous Abelian factor, $ Q_{an}= 3Q_a +2Q_b+Q_c$,
leaves the extended Standard Model model gauge group, $SU(3) \times
SU(2) \times U(1)_Y \times U(1) _{B-L} , \ [ Y= -{1\over 3} Q_a + \ud
Q_b ,\ Q_{B-L}= -{1\over 6} (Q_a -3Q_b +3Q_c) ]. $ The localized
fermions consist of three chiral generations of quarks, leptons and
antineutrinos, $(q_i , \ u^c_i, \ d^c _i , \ l_i, \ e^c_i, \ \nu ^c_i)
$, and several copies of adjoint representation modes, $ K_{a,l} ,\
K_{b,l} $, with model dependent multiplicities, as displayed in Table
\ref{tablesmz3}.
We have not included in the table the vector pairs of fermion modes in
the sectors $(b',c) + (a',c)$ with gauge quantum numbers, $\tilde H_i
\sim [(1,\bar 2) _{0,1,1} + (\bar 3,1 ) _{0,1,1} ],\ \bar {\tilde H }
_i \sim [(1, 2) _{0,-1,-1} + (3,1 ) _{0,-1,-1} ] , \ [i=1, \ \cdots ,
n]$. In spite of the vanishing chiral multiplicity, $ I ^{\chi }
_{b'c} =0 $, one expects that the massless spin half modes and the
low-lying complex scalar modes in these sectors enter with a finite
non-chiral multiplicity of the form, $ I _{b'c_{g_1}} =- I
_{b'c_{g_2}} = n ,\ [g_1\ne g_2]$ characterized by the non-vanishing
integers, $n$.  The presence of tachyons among the low-lying
electroweak doublet complex scalars in the modes, $ H_i + \bar H_i$,
is needed to accomplish the electroweak gauge symmetry breaking, as
indicated already.  To relate the electroweak Higgs mechanism to the
brane recombination process, $b '+ c \to e$, one would need a fine
compensation in the tachyon mode mass squared between the negative
contribution from the angle dependent terms, $\a ' M^2 = \min [ \ud
\sum _I (\pm \vert \t ^I _{b'c} \vert ),\ 1 -\ud \sum _I \vert \t ^I
_{b'c} \vert ] $, and the positive contribution from the interbrane
distance.  In spite of the analogy with the models incorporating an
$\caln =2$ supersymmetry subsector~\cite{iban0} or with the
supersymmetry-quiver models (first article in Ref.~\cite{iban1}),
involving $D6$-branes wrapping parallel one-cycles, the present model
features specific differences, unlike the claim made by Axenides et
al.,~\cite{axeni03}.  In particular, the conjugate modes, carrying the
representation $ n (\bar N_b ,\bar N_c) + n ( N_b ,N_c) $, are neither
$\caln =2$ hypermultiplets, since the cycles in the three planes are
not parallel, nor $\caln =1$ chiral supermultiplets, since none of the
wrapped cycles are supersymmetric.

The candidate electroweak Higgs boson modes, $ H_i \ \bar H_i $,
account only in part for the quark and lepton mass generation.  The
graphs $I.b$ and $I.c$ in Figure~\ref{figz3disc} represent the
trilinear Yukawa couplings of down quarks and leptons with the
electroweak Higgs boson, $ Q _{b'a } D^ {c \dagger } _{a c_{g }} (\bar
H ) _{c _{g } b' } $ and $L _{b c } (\bar H )_{c b'_{g} } E^c _{b'_{g}
c} $.  One sees that the trilinear Yukawa couplings exist only for the
down-quarks and charged leptons, $ q d^c \bar H $ and $ l e^c \bar H$,
while the corresponding ones for the up-quarks, $q u^cH$, are
forbidden. Indeed, the scalar mode with the needed quantum numbers, $
H \sim (1, \bar 2) _{-3,-1,0} $, cannot be realized as a perturbative
open string mode.  The obstruction to generate masses for both the up
and down quarks is a general feature of the minimal type models which
realize the right handed triplet quarks as the antisymmetric
representation of $SU(3)_c$.  The same problem also occurs in models
using the $Z_2\times Z_2 $ orbifold~\cite{Cvetic:2001nr} and the
$Z_3\times Z_3$ orbifold~\cite{kokor04}.  In the flipped $SU(5)$
model, it is rather the down-quarks and leptons Yukawa couplings that
vanish.  One way to account for the top quark mass here is by
appealing to the dynamical gauge symmetry breaking or to composite
Higgs boson mechanisms.  For completeness, we note that to remedy the
shortcoming of intersecting brane models in favoring a separable
structure of the quark and lepton mass matrices with rank one, recent
proposals consider using a hypercolor type scheme~\cite{kita04}, loop
corrections~\cite{abelgoodsell06}, or higher dimensional
operators~\cite{dutta05}.


\subsubsection{Unified   gauge model   and Standard Model realizations}

To each set of effective wrapping numbers $(Y_\mu , Z_\mu )$ of a
given lattice type, there arise a large number of realizations
described by the wrapping numbers $(n _\mu ^I, m _\mu ^I)$ solving
Eqs.~(\ref{appzy}).  Each solution comes with a specific non-chiral
spectrum and $D$-brane intersection angles, constituting input data
indispensable for evaluating string amplitudes.  As found by
Blumenhagen et al.,~\cite{Blumenhagen:2001te} for the model with three
brane stacks, $ (n_\mu ^I, m_\mu ^I),\ [\mu = a, b, c] $ each pair
$(Y_\mu , Z_\mu )$ admits some $36$ solutions, hence yielding a total
of $ 4\times 36 ^3$ realizations for the four inequivalent lattices.
However, only a few solutions gave acceptable mass spectra for the
scalar modes.  Some of the natural constraints to impose on the
physical solutions include the absence of tachyon scalar modes with
the quark and lepton quantum numbers, and the presence of at least a
single scalar tachyon among the electroweak singlet bosons $\tilde \nu
^c \sim (c,c') $, charged under $U(1)_{B-L} $, and the electroweak
doublet bosons $\bar H \sim (b',c)$.

The scan over solutions for the wrapping numbers of the extended
Standard Model with three brane stacks, is conveniently performed by
means of a numerical computer program.  We study the two cases
involving parallel and non-parallel $D6_a/D6_b$-brane stacks.  We
begin with the simpler case of parallel branes $ a $ and $ c$, whose
discussion overlaps that of the gauge unified models.  After
elimimating the solutions with tachyon scalars carrying quark and
lepton quantum numbers, we could not find any solution satisfying the
requirement that tachyon scalars be present for both $\tilde \nu ^c $
and $H, \ \bar H $ modes.  However, since several solutions exist with
tachyon scalars in either the $\tilde \nu ^c _i $ or $H _i, \ \bar H
_i $ modes, we consider the minimal retreat from our initial goal
allowing for the solutions with at least one tachyon among the modes
$\tilde \nu ^c _i$.  Restricting the numerical search to the interval
of integer wrapping numbers, $ [-6, +6]$, we find $16$ and $10$
inequivalent solutions for the $T^6/Z_3$ lattices ${\bf A A A }$ and
${\bf B B B}$, but no solutions for the lattices ${\bf A A B }$ and
${\bf A B B}$.  The wrapping numbers for the solutions of the
$D6_a/D6_c$-brane setup with parallel $a, \ b $ stacks are quoted in
Table~\ref{tablez3}.  The displayed information on the intersection
numbers for members of orbifold orbits, $ I_{a'a_g} , \ I_{ac_g} , \
I_{cc'_g} , \ I_{a'c_g} ,\ [g=0,1,2] $ fully determines the non-chiral
spectrum.  There is no need here in quoting the intersection angles
since these are easily determined from the displayed information.
(Note that $\t _{\mu '} = -\t _{\mu } .$) In the various realizations
of the unified model, the number of mirror matter generations take
values in the range, $\D _F \simeq [ 0 \ -\ 5 ],\ \D _{\bar f} \simeq
[ 0 \ -\ 3] ,\ \D _{\nu ^c } \simeq [ 0 \ -\ 5 ] $, the vector
higgsino modes, $ H_i, \ \bar H _i ,\ [i=1, \cdots n]$ appear with the
finite multiplicities, $ n=2, 6, 7$, and the gaugino like modes,
$K_{x, l} ,\ [l =1, \cdots , I^{\bf Adj} _x ,\ x =a, b ] $, enter with
the uniform value of the multiplicity, $ I^{\bf Adj} _a = I^{\bf Adj}
_b = 7$.  Note that none of the solutions in Table~\ref{tablez3}
satisfies the property that the brane pairs, $b',\ c $, are parallel
along a single complex plane.  In the cases where tachyon modes for
$H_i +\bar H_i$ appear, the squared mass of these modes sets at the
typically large value, $ \a' M^2 (H ) \simeq - 0.11$, thus justifying
our previous claim that no $\caln =2$ subsectors arise in the present
model.


We next examine the Standard Model realizations with non-parallel
stacks, $(n_b^I , m_b ^I) \ne (n_a^I , m_a ^I) $.  Requiring the
absence of scalar tachyons with quark and lepton quantum numbers,
yields some $50$ solutions for lattice ${\bf A A A }$, $40$ solutions
for lattice ${\bf B B B}$, and no solutions for the lattices ${\bf AAB
}, \ {\bf ABB}$.
The maximally constrained selection of solutions (Case I), requiring a
nearly massless or tachyon scalar for the modes $\tilde \nu ^c$ and $H
+\bar H$, retains 4 solutions for lattice ${\bf BBB}$.  The less
constrained selection of solutions (Case II), requiring a nearly
massless or tachyon scalar for the modes $\tilde \nu ^c$, retains 5
solutions for lattice ${\bf AAA }$ and 10 solutions for lattice ${\bf
BBB}$.  We display in Table~\ref{tabz3scan} the solutions found for
the wrapping numbers of the three non-parallel stacks realizing the
extended Standard Model in the partially and fully constrained
searches of Cases II and I.


We now discuss the conditions set on the effective string mass scale,
$m'_s$, by the low energy data for the $SU(3)\times SU(2)\times U(1)_Y
$ gauge coupling constants, $g_a $.  The hypercharge embedding, $ Y=
-{1\over 3} Q_a +\ud Q_b$, implies the relation, $ g_1 ^{-2}\equiv g_Y
^{-2}= {2 \over 3} g_ 3 ^{-2} + g_2 ^{-2} $, which can be compared
with experiment by substituting the one-loop running coupling
constants, \bea && 4\pi \bigg [ {2 \over 3 \a _3 (Q) } + { 2\sin ^2 \t
_W (Q) -1\over \a (Q) }\bigg ] = B \log {Q^2 \over m_s ^{'2} } + \D ,
\cr && [B = {2 \over 3} b_ 3 +b_2 -b_1, \ \D = {2 \over 3} \D _ 3 +\D
_2 -\D _1] \eea subject to the assumption that we deal with a
non-supersymmetric unified model with no new physics thresholds
between $Q= m_Z$ and $ m'_s $.
We can estimate $m'_s$ by substituting the experimental values, $ \a
(m_Z) =127.95 ,\ \sin ^2 \t _W(m_Z)= 0.23113 ,\ \a _s(m_Z) = 0.1172, \
[m_Z= 91.187 \ GeV]$ while using suitable inputs for the slope
parameters $b_a$ of the massless modes below $ m'_s $.  (The slopes
are calculated from the general formula, $ b_a = {1\over 6} (-c _a(
R_{S} ) - 2 c _a( R_{F} ) + 11 c _a( R_{V} ) ) $, with $c _a$ denoting
Dynkin indices for complex scalar, Weyl fermions and vector modes.)
Since all the above listed solutions feature equal volumes for the $a,
\ b,\ c$ cycles, we expect to find, as in the conventional field
theory type unification, poorly convergent running coupling constants
indicative of a low unification mass scale.  It is known, however,
that the presence of massive charged vector modes can significantly
improve the compatibility with data.  Thus, assuming that a new
threshold is present at some mass, $m_A < m' _s$, modifies the above
quoted formula to \bea && 4\pi \bigg [ {2 \over 3 \a _3 (Q) } + {
2\sin ^2 \t _W (Q) -1\over \a (Q) } \bigg ] - (B+\d B) \log {Q^2\over
m_A^2}= B \log {m_A ^2\over m_s ^{'2}} + \D , \eea where the slope
parameter correction, $\d B = {2 \over 3} \d b_ 3 + \d b_2 - \d b_1, $
includes the contributions from the vector modes that decouple at
$m_A$.
We now test this relation by assuming the Standard Model evolution in
the interval, $Q \in [m_Z , m_A]$ with slope parameters, \bea && b_3 =
11 -{4F\over 3},\ b_2 = {22\over 3} -{4F\over 3} - {N_h \over 6},\ b_1
= {20\over 9} F - {N_h \over 6} \ \Longrightarrow \ B= {44\over 3},
\eea and including in the interval $ Q\in [m_A, m'_s]$ the extra
massless vector multiplets, $ (3 + \D ) (F + \bar f + \nu ^c ) + \D
(\bar F + f + \bar \nu ^c ) + I^{Adj} _ {a } K_a + I^{Adj} _ {b } K_b
+ n H + n\bar H $, with slope parameters, \bea && \d b_3 = -3 I_a
^{Adj} -n ,\ \d b_2 = -2 I_a ^{Adj} -n ,\ \d b_1 = - {5n\over 6} \
\Longrightarrow \ \d B= -2 ( I_a ^{Adj} + I_b ^{Adj}) + {5n\over 6}
. \eea Assigning the tentative values, $ I_a ^{Adj}= I_b ^{Adj} =n =
[0,2,4,7] $ and $ {m_A \over m'_s} = 10^{-2} $, we find that the
predicted string compactification scale sets at the values: $ m'_s
\simeq [5. \times 10^{13},\ 3.7 \times 10^{14},\ 2.7 \times 10^{15} ,\
5.3 \times 10^{16}] \ $ GeV.  These results explicitly demonstrate how
a mild hierarchy for the intermediate mass threshold can significantly
pull the string mass scale down.

\subsubsection{Non-supersymmetric flipped $SU(5) \times U(1)_{fl} $ model}

We here consider the flipped gauge unified model of the $ Z_3$
orbifold constructed by Ellis et al.,~\cite{ellis03} (Model I in Table
II of their paper) constructed with three $D6$-brane stacks $ N_a=5, \
N_b =1,\ N_c =1$ with effective wrapping numbers, $ (Y_a,Z_a )= (3,\ud
) ,\ (Y_b,Z_b )= (0, -{1\over 3} ) ,\ (Y_c,Z_c)= (3,-{1\over 6} ) $,
which realize the gauge group, $SU(5)\times U(1)_{a} \times
U(1)_{b}\times U(1)_{c} $.
The chiral spectrum of localized fermions is displayed in the table
below in correspondence with the group representations and the
assigned open string sectors.

\vskip 0.5 cm
\begin{center}  
\begin{tabular}{c|cccccc}   
Mode & $ F $ & $ \bar f $ & $ e^c $ & $ \bar H $ & $H $& $ S $ \\
\hline &&&&&& \\ Sector & $ (a',a) $ & $ (a,c) +(a,b) $ & $ (c',c) +
(b',c) $ & $ (a',b) $ & $(a',c ) $ & $ (b,c) $ \\ &&&&&& \\ Irrep & $
3(10)_{2,0,0} $ & $ 2(\bar 5)_{-1,0,1} $ & $ 2(1)_{0,0,-2} $ & $
1(\bar 5)_{-1,-1,0}$ & $ 1(5) _{1,0,1}$& $ 1(1) _{0,1,-1}$ \\ & &$
+1(\bar 5)_{-1,1,0}$ & $ +1(1)_{0,-1,-1}$ &&&
\end{tabular}
\end{center} 
 \vskip 0.5 cm

On side of the anomalous symmetry group, $ Q_{an}= 5Q_a +Q_c$, one
finds the unbroken gauge symmetry group, $SU(5)\times U(1)_{fl} \times
U(1)_{free} ,\ [Q_{fl} = \ud (Q_a -5 Q_b-5 Q_c) ,\ Q_{free} = Q_b ] $
with three chiral matter generations, $ (F+\bar f +e^c)\sim (10 + \bar
5 + 1) $, along with massless higgsino like modes, $ (\bar H + H+
S)\sim (\bar 5 + 5 + 1)$.  The light or tachyon scalars with same
quantum numbers as the higgsino modes are needed for the Higgs
mechanism breaking of $U(1) _{free}$ and the electroweak $SU(2)_L$
gauge symmetries.  The Yukawa coupling, $S H\bar H$, decouple the
higgsino modes once the scalar $S$ mode raises a VEV.  The absence of
low-lying scalars carrying the representations $ 10, \ \overline {10}
$ means that one must resort to a higher dimensional Higgs mechanism
to accomplish the unified gauge symmetry breaking.

We have developed a computer aided scan of the solutions for the three
branes wrapping numbers selecting the realizations with scalar
tachyons absent for quark and lepton modes but present for the $H,\
\bar H$ and $S$ modes.  These conditions select only the $T^6/Z_3$
lattices, $ {\bf ABB ,\ BBB } $.  The maximally constrained selection
(Case I) with tachyon or nearly massless $ H, \bar H, S$ scalar modes
gives 6 solutions for $ {\bf ABB} $ and no solutions for $ {\bf BBB}$.
The least constrained selection (Case II), with tachyon or nearly
massless scalar modes for $ S$ only, retains 35 solutions for $ {\bf
ABB} $ and 2 solutions for $ {\bf BBB}$.  We display in
Table~\ref{scanz3flip} a representative sample of the solutions found
for the lattice $ {\bf ABB} $ in Case I and the lattice $ {\bf BBB} $
in Case II.


\subsubsection{Enhancement effect in four fermion   baryon number
violating processes}


We now discuss the two-body nucleon decay amplitudes in the $Z_3$
orbifold model realization of $SU(5)$.  The four fermion amplitudes
displayed in Figure~\ref{figz3disc} by the graphs $ II.a $ and $ II.b$
refer to the configurations $ 10 \cdot {10}^\dagger \cdot 10 \cdot
{10}^\dagger $ and $ 10 \cdot {\bar 5}^\dagger \cdot \bar 5 \cdot {10}
^\dagger $, which contribute to the proton decay channels, $ p\to \pi
^0 + e^c _L$ and $ p\to \pi ^0 + e^c _R$. The corresponding amplitudes
in the broken unified gauge symmetry version are displayed by the
graphs $ III.a,\ III.b$ of Figure~\ref{figz3disc}, with graph $ III.c
$ referring to the neutrino emission decay channel.

The numerical calculations are performed by setting the unified
coupling constant at the value, $ \a _X = {1\over 25 } $, assuming
that the wrapped cycles have the same radii, $r^I =r $, and regarding
the gauge unification and compactification mass scales, $ s = {m_s
\over M_X } $ and $ m_s r = {m_s \over M_c} $, as free parameters.
For simplicity, we further set $ M_{GUT} = M_X$.  The winding numbers
and angles of the wrapped three-cycles are determined for each
solution.  The same numerical value for the three-cycles volume,
$\vert \call _a \vert \equiv \prod _I (L_ a ^I /r ^I) = \sqrt 7$, is
found in all the solutions. The low energy constraints allow then to
express the string coupling constant by the numerical relation, $ g_s
\simeq 0.11 \ (m_s r)^3 /N $.  The weak coupling condition on the
string coupling constant entails an upper bound on the wrapped
three-cycle radius \bea && g_s = {\a _X m_s^3 \vert L \vert \over N} =
{\a _X (m_s r)^3 \vert \call \vert \over N } \leq 1 \ \Longrightarrow
\ m_s r \leq ( {N\over \a _X \vert \call \vert } )^ {1\over 3} \simeq
2.\ N ^ {1\over 3} ,\eea where $N$ denotes the orbifold group order.
Using these inputs in Eq.~(\ref{eqratsf}), yields the following
approximate numerical formulas for the ratio of string to field theory
amplitudes in the fixed $g_s$ and $m_s$ cases \bea && \calr _{s/f}
\equiv {\cala _{st} \over \cala _{ft} } \simeq 6.88 \ g_s ^{1/3} ({\a
_X \over 0.04} )^{-1/3} ({2 M_X \over m_s})^2 (m_s r)^2 ( { \hat I
^{reg} (\t ) \over 10}) \cr && \simeq 3.30 \ ({2 M_X \over m_s })^2
(m_s r)^3 ({I ^{reg} (\t ) \over 10}) , \label{eq.ratiosf} \eea where
we have set the reference value of the regularized $x$-integral at the
approximate value obtained in the large radius, $ I^{reg} = 10$, as
follows from Figure~\ref{figasam01}, while ignoring momentarily its
dependence on the parameters, $ M_X = m_s /s $ and $ m_s r $.  We note
that the ratio has a fast power growth, proportional to $(m_s r)^2$ in
the fixed $g_s$ and to $(m_s r)^3$ in the fixed $m_s$ case, if one
discounts the suppression effect from the classical action factor in $
\hat I^{reg} $.  In comparison, the analysis of Klebanov and
Witten~\cite{KW03} with the reference value of the Wilson line
parameter set at, $L(Q)= 8$, predicts the ratio, $\calr _{s/f}\simeq
1.5\ ({L(Q) \over 8 } )^{2 \over 3} g_s^{1 \over 3} ({\a _X \over
0.04} )^{-1/3} ({I (\t ) \over 10}) $. The fact that this is a factor
$4$ smaller than the result in Eq.~(\ref{eq.ratiosf}) reflects on the
uncertainties in the input parameters.  It is also interesting to
compare with the ratio predicted by making use of the gravitational
interactions input, $\calr _{s/f}\simeq 2.3 \ 10^{-3} ({M_P\over M_{X}
\sqrt \l } ) ({L(Q) \over 8 }) ({\a _X \over 0.04})({I (\t ) \over
10}) $, since this suggests that the compactifications with small
wrapped cycles, hence close to the upper limit on $\l ^{1\over 6} =
R/r $ of order unity, would not support an enhancement effect.

\begin{figure} 
\begin{center} 
\epsfxsize= 5 in
\epsffile[40 230 550 570]{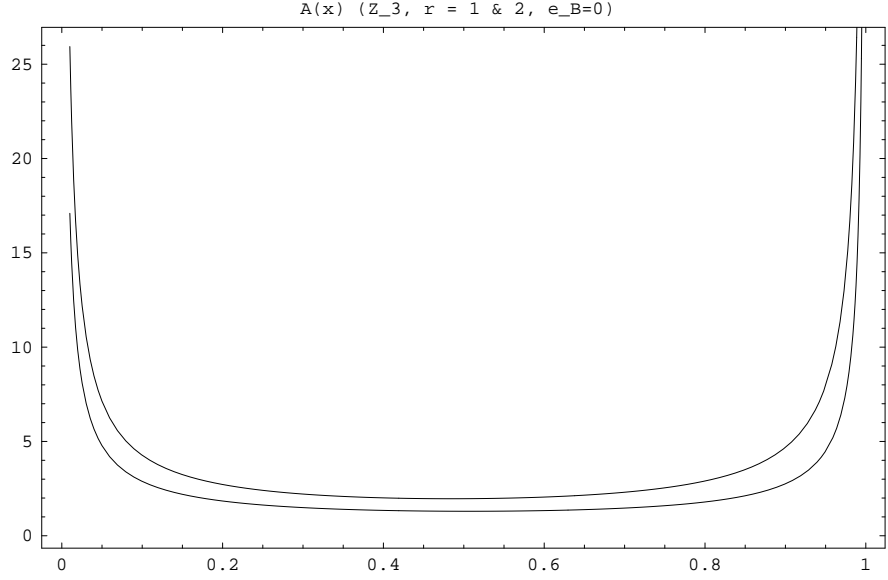}
\vskip 0.2 cm
\caption{\it Plot as a function of $x$ of the string amplitude
$x$-integrand $ 2 \tilde I(\t ; x) $ in Eq.~(\ref{eqxintplot}).  The
regularized integrand is obtained by subtracting the leading end point
singular terms, $ [\tilde I(\t ; x) - \tilde I(\t ; x=0) -\tilde I(\t
; x=1) ] $, associated with the exchange of massless gauge boson
poles.  The present case refers to solution $V$ of the $Z_3$ orbifold
model in Table~\ref{tablez3} for the equal angle amplitude, $10 \cdot
10 ^\dagger \cdot 10 \cdot 10 ^\dagger $, at coincident intersection
points, $\e_A = \e _B =(0,0,0)$, for  the two values of the wrapped
three-cycle radius, $m_s r=1$ (upper curve) and $m_s r=2$ (lower
curve).}
\end{center}
\label{figx2}
\end{figure}

A brief aside on numerical issues is in order.  For a good convergence
of the series sums in the classical partition function, the summation
labels should cover the range, $\vert p_A \vert ,\ \vert p_B \vert
\leq 6$, throughout the interval of radii, $ m_s r\in [1,4]$, The
larger is $m_s r$, the slower is the convergence of the zero mode
summations.
The sigma model perturbation theory sets the lower bound, $ m_s r > 1
$, while the string perturbation theory sets the upper bound, $g_s
\approx 0.11 \ (m_s r)^3 /N < 1 $.  Nevertheless, we shall consider
small excursions at small and large radii for the sake of illustrating
the rapid growth of string amplitudes in these forbidden regions.  To
simplify the implementation of the displaced brane regularization, we
shall also assume that the transverse displacement $d_A^I$ takes place
in a single complex plane with fixed $I$, with the leading term in the
classical action factor of Eq.(\ref{sec3.wils}) reading as, $ ( \hat
\d _I ) ^{ M_X^2} e ^{ V ^I_{11} d_A ^{I2} }$.  To illustrate the slow
numerical convergence of the $x$-integral, we show in
Figure~\ref{figx2} a plot of the integrand in the subtraction
regularization prescription using the definition \bea && 2 \tilde I(\t
; x) \equiv {\dh \over \dh x} [ \cala ' _{st} / ( C \cals _1 ( \calT
_1 +\calT _2) )] = [x(1-x)]^{-1} ({ \sin \pi \t \over F(x) F(1-x) })
^{\ud } e ^{-S_{cl}} ,
\label{eqxintplot} \eea  
which generalizes the definition in Eq.~(\ref{eqxintnum}) for $ I(\t
)= \int _0^1 dx \tilde I(\t ; x).$
in the finite radius case.  We see that after removing the end point
singularities due to the massless poles, the integrand still picks up
its leading contribution from narrow intervals close to the end
points.  This feature holds true at all the relevant values of $m_sr$.
The slow convergence of the $x$-integral is a critical slowing factor
in the cases involving distinct pairs of intersections angles, $\t \ne
\t '$.
   
Turning now to the predictions for the ratio of string to field theory
amplitudes, $\calr _{s/f}$, we examine first how the results vary for
different realizations of the unified $Z_3$ model. Since the
intersection angles in the various solutions follow repetitive
patterns, we need to perform independent calculations only for a few
cases.  The ratio $\calr _{s/f}$ calculated with the parameter values,
$ m_sr=1, \ s = m_s/M_X=1,\ \e _A = \e _B = (0,0,0) $, is displayed in
the following table for the representative sample of four solutions
covering the complete set in Table~\ref{tablez3}.
\vskip 0.5 cm
\begin{center} 
\begin{tabular}{|c|c|c|c|}  
\hline Solution & $\t ^I _{a'a}$ & $\calr _{s/f}$ (Pole Subtraction) &
$\calr _{s/f}$ (Brane Displacement) \\ \hline $ {\bf AAA }$ &&& \\ $ I
$ & $ (0.879, 0.333, 0.667) $ & $ 3.15 $ & $3.16$ \\ $ II $ & $
(0.545, 0.667, 0.667) $ & $ 5.93 $ & $ 2.48 $ \\ $ III $ & $ (0.667,
0.667, 0.545) $ & $ 5.66 $ & $ 2.64$ \\ $ V $ & $ (0.333, 0.667,
0.879) $ & $5.93$ & $2.67 $ \\ \hline $ {\bf BBB } $ &&& \\ $ I $ & $
(0.454, 0.333, 0.333) $ & $ 7.54 $ & $3.06$ \\ $ II $ & $ (0.333,
0.667, 0.121) $ & $ 7.02 $ & $3.14$ \\ \hline \end{tabular}
\end{center}  \vskip 0.5  cm

We see that small variations in the angles can have a significant
impact on the predictions.  The sensitivity is stronger in the
subtraction than in the displacement regularization scheme.  The near
equality of the results in the two prescriptions for solution $I$ must
be viewed as a coincidence.  Although the angle and volume parameters
are nearly the same for all solutions, there are important variations
in the wrapping numbers and hence in the fixed complex plane volume
factors, $ L_A ^I$, from one solution to the other. Indeed, we find
different predictions for solutions with same angles but different
wrapping numbers.  From the formal relation between the $x$-integrals
in the subtraction and displacement regularization prescriptions,
$\hat I^{reg} = I^{reg} /(1+ \ud M_X^2 m_s \vert L \vert I^{reg} )$,
one infers that $ \hat I^{reg} < I^{reg}.$ In fact, we notice that the
predictions for $\calr _{s/f}$ is a factor $ 2\ - \ 3 $ larger in the
subtraction regularization scheme than in the displacement
regularization scheme.  Since the displaced brane prescription is more
realistic, we conclude that the enhancement effect is of typical order
of magnitude, $ \calr _{s/f} \simeq 3. \ (M_X /M_c )^2$.

\begin{center} \begin{figure}
\epsfxsize= 6in
\epsffile[30 10 570 520]{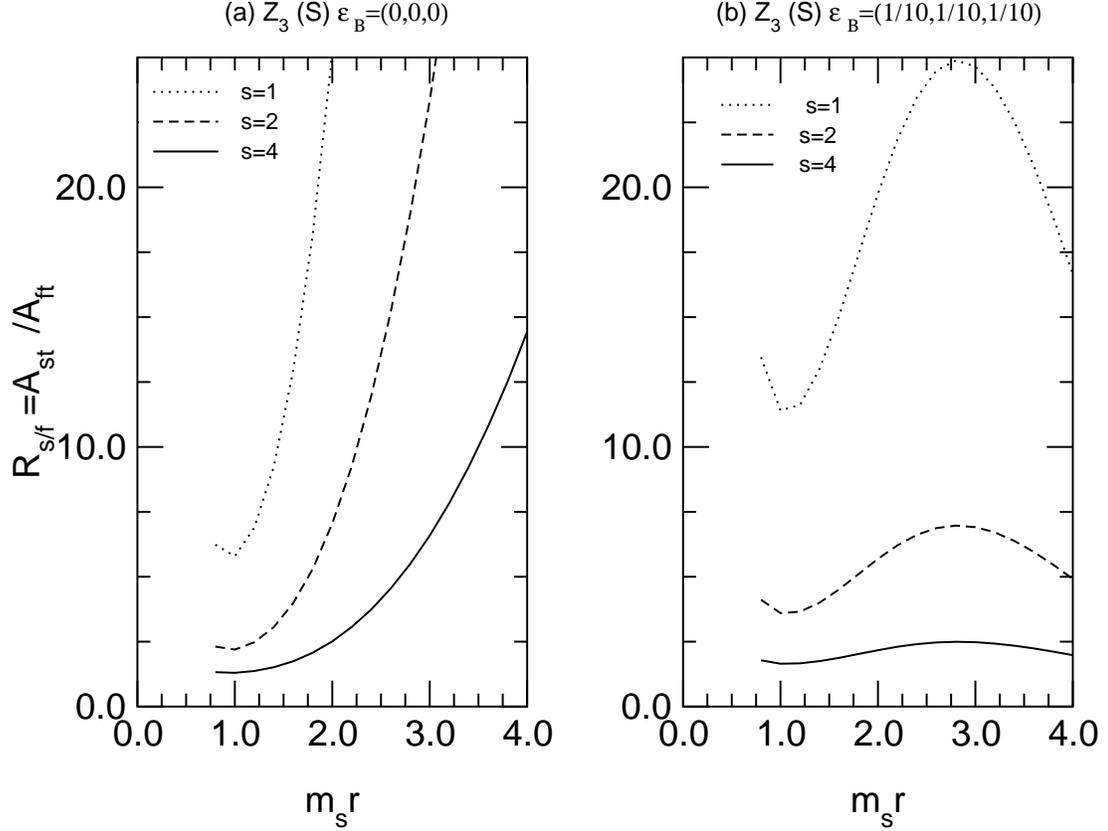}
\vskip 0.5 cm
\caption{\it The ratio of string to field theory amplitudes, $ \calr
_{s/f}$, in the subtraction regularization prescription for the $Z_3$
orbifold model is plotted as a function of the inverse
compactification mass parameter $ m_sr $ for three values of the
inverse unification mass parameter, $ s \equiv m_s/ M_X = 1, 2, 4 $.
The results refer to solution $V$ of lattice $\bfA \bfA \bfA $ in
Table~\ref{tablez3}, described by the wrapping numbers, $n_a
^I=n_b^I=(0,1,2),\ n_c^I=(-1,-1,0),\ m_a^I= m_b^I=(1,0,1),\
m_c^I=(-2,0,1)$, and the intersection angles $\t ^I_{a'a}= (0.333,
0.667, 0.879) $.  The nucleon decay amplitude for the equal angle
configuration $\t = \t '$ with matter mode representations, $10 \cdot
{10} ^\dagger \cdot 10 \cdot {10} ^\dagger $, is evaluated in the
broken unified gauge symmetry model.  The left hand panel $(a)$ shows
the case with coincident intersection points, $\e _A= \e _B = (0,0,0)
$ and the right hand panel $(b)$ the case with distant intersection
points, $\e _A= (0,0,0) , \e _B= (1/10 ,1/10,1/10) $.}
\label{figsrz3}
\end{figure}
\end{center}

\begin{center}
\begin{figure} 
\epsfxsize= 6in
\epsffile[30 1 570 500]{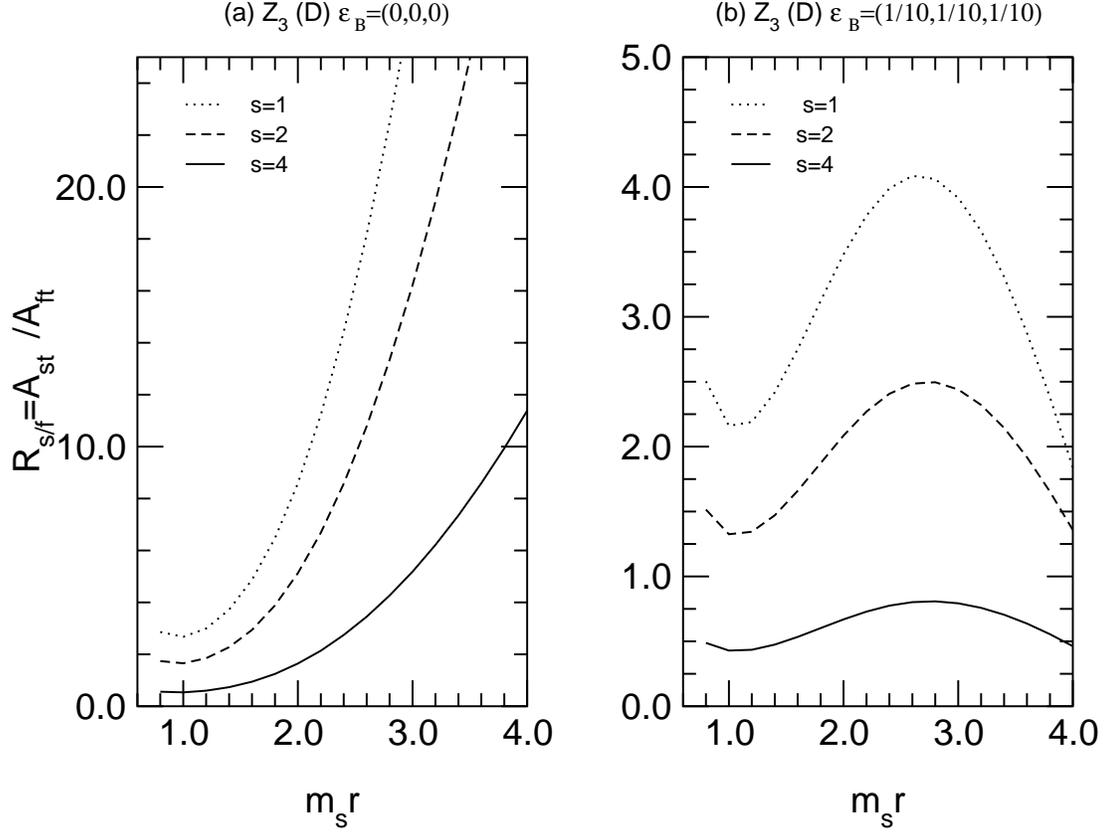}
\caption{\it The ratio of string to field theory amplitudes, $ \calr
_{s/f}$, in the displaced regularization prescription for the $Z_3$
orbifold model is plotted as a function of the inverse
compactification mass parameter, $ m_sr = m_s/M_c$, for three values
of the gauge unification mass, $ s \equiv m_s/ M_X = 1, 2, 4 $.  The
results for the string amplitude refer to solution $V$ of the lattice
$\bfA \bfA \bfA $, with the wrapping numbers $n_a ^I=n_b^I=(0,1,2),\
n_c^I=(-1,-1,0),\ m_a^I= m_b^I=(1,0,1),\ m_c^I=(-2,0,1 )$ and
intersection angles $\t ^I_{a'a}= (0.333, 0.667, 0.879) $.  The
nucleon decay string amplitude in the equal angle configuration $\t =
\t '$ in the matter mode representations, $10 \cdot 10 ^\dagger \cdot
10 \cdot 10 ^\dagger $, is evaluated in the unified gauge symmetry
broken phase with the displaced brane separation adjusted to the
infrared cutoff mass, $M_X \equiv m_s /s $, by means of the formula, $
M_X ^2 =\sum _I \sin ^2 (\pi \t ^I) \vert d ^I \vert ^2 $.
The left hand panel $(a)$ shows the case with coincident intersection
points, $\e _A= \e _B = (0,0,0) $, and the right hand panel $(b)$ the
case with distant intersection points $\e _A= (0,0,0) , \e _B= (1/10
,1/10,1/10) $.  }
\label{figmrz3}
\end{figure}
\end{center}

We now discuss the dependence on $m_s r $ based on predictions
displayed in Figures~\ref{figsrz3} and~\ref{figmrz3} for the ratio of
string to field theory amplitudes in equal brane angle configurations.
The plots show a rapid rise of the ratio with increasing $m_s r$.  In
both the subtraction and displacement regularizations, starting from
$O(1)$ values near $m_sr=1$, the ratio $\calr _{s/f}$ rapidly
increases to $O(10)$ for $m_sr \simeq 2 $.  Note that further
increasing $ m_s r $ would clash with the perturbative constraint on
$g_s$.  The amplitudes ratio rapidly increases with increasing $ M_X$
at fixed $r$. It is larger for $ M_X > M_c$, corresponding to the
region in the plots, $ r > s$, lying on the right hand side of the
vertical line, $ r =s$.  The comparison between Figures~\ref{figsrz3}
and~\ref{figmrz3} shows that the ratio is smaller over the full
interval of $ m_s r$ by a factor of order $2\ -\ 4$ in the brane
displacement regularization relative to the pole subtraction
regularization.

The ratio of string to field theory amplitudes is strongly reduced for
intersection points separated by a finite longitudinal distance, $\e
_B\ne 0$ or $ \e_A \ne 0 $.   
In the coupling, $ 10_{aa'} \overline{10}^\dagger _{a'a} \bar 5 _{ac}
\bar 5 _{ca} $, the distance $\e _B $ between the $ \bar 5 $ and $10$
modes is generically non-vanishing, while $\e _A =0$.
This is expected from the exponential
dependence of the classical factor.  The results illustrate the level
of suppression that may be expected for the flavor non-diagonal
nucleon decay processes involving second generation leptons and/or
quarks.  We see on the predicted ratios displayed in the right hand
panels $(b)$ of Figures~\ref{figsrz3} and~\ref{figmrz3}, that a
significant suppression occurs even for the moderate value, $\e _B =
({1\over 10}, {1\over 10}, {1\over 10} ) $.  Comparing the results in
the left and right hand panels $ (a)$ and $(b)$ of
Figures~\ref{figsrz3} and Figure~\ref{figmrz3}, we note that the two
regularization schemes give a similar dependence on the parameters,
with a stronger reduction for increasing $ m_s r$ taking place in the
displaced brane regularization.  Following an initial rise, the ratio
undergoes a drastic reduction for radii, $ m_s r > 2$.  The fact that
near $ m_s r \simeq 1 $, the ratio starts out larger for $\e _B\ne 0 $
than for $\e _B =0$, comes about because of cancellation effects in
the series sum over zero modes due to the nontrivial complex phases
introduced by using the Poisson resummation formula.  The latter
factor is a counterpart of the overlap integral between the localized
modes wave functions encountered in the context of the field theory
orbifold models.  Since the flavor changing nucleon decay processes
are drastically reduced except in narrow intervals close to $ m_s r
=1$ or for $ M_X /m_s > 1/3 $, one may expect a strong sensitivity to
the quark and lepton flavors in these models.

\begin{center} \begin{figure} 
\epsfxsize= 6in
\epsffile[30 1 570 500]{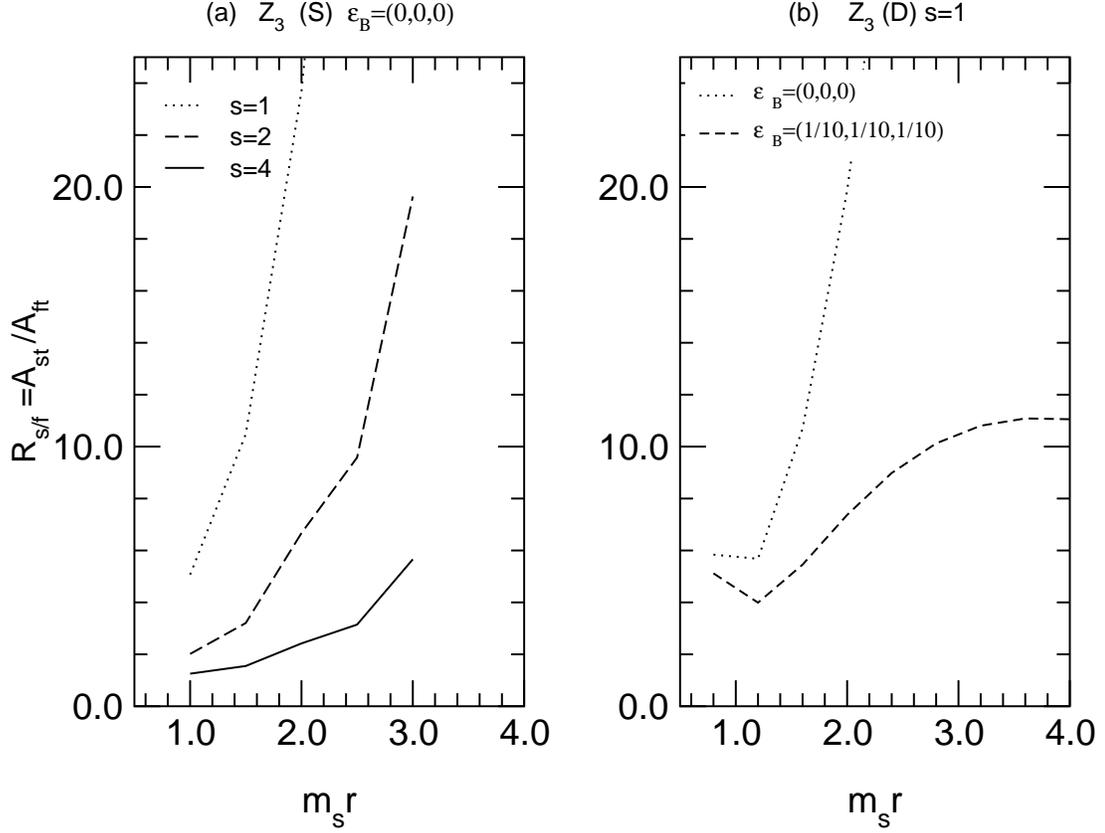}
\caption{\it Plots of the ratios of string to field theory amplitudes,
$ \calr _{s/f}$, in the displaced regularization prescription for the
$Z_3$ orbifold model as a function of the inverse compactification
mass parameter $ m_s r = m_s/M_c$ for solution $V$ of the $\bfA \bfA
\bfA $ lattice with wrapping numbers $n_a=n_b=(0,1,2),\ m_a=
m_b=(1,0,1),  \  n_c=(-1,-1,0),\ m_c=(-2,0,1) $ and distinct intersection
angles for the $10 , \ \bar 5 $ modes, $\t _{a'a} ^I = (0.333, 0.667,
0.879),\ \t _{bc}^I = (0.227, 0.333, 0.560) $.  The nucleon decay
string amplitudes for the matter representations, $10 \cdot 10
^\dagger \cdot \bar 5 \cdot \bar 5 ^\dagger $, refer to the unequal
angle case, $\t \ne \t '$. The left hand panel $(a)$ shows results
using the subtraction regularization prescription with the values of
the unification mass scale, $ s \equiv m_s/ M_X = 1, 2, 4 $, at
coincident points $\e _A = \e _B = (0,0,0)$. The right hand panel
$(b)$ shows results using the displacement regularization prescription
at $s = m_s/ M_X= 1$ with the intersection point distance parameters
set at, $ \e _A= \e _B = (0,0,0) $ and $\e _A= (0,0,0) , \e _B= (1/10
,1/10,1/10) $, respectively.
}
\label{figsarz3}
\end{figure}
\end{center}

The ratio of amplitudes in the unequal angle case is plotted as a
function of $m_s r$ in Figure~\ref{figsarz3}.  We see that although
somewhat weaker, the enhancement effect is still of same order as in
the equal angle case.  The comparison between the left and right hand
panels $(a)$ and $(b)$ again shows that a drastic reduction takes
place in the non-diagonal configuration even with moderately distant
intersection points. We note that  for distant modes at large  values of the 
$ s $  parameter, the ratio takes values below unity.  
That the string theory mechanism studied here
might introduce a reduction rather than an enhancement joins with the
conclusion reached by Acharya and Valandro~\cite{acharyadro05} upon
evaluating the contribution from the tower of Kaluza-Klein modes in a
field theory version of the M-theory amplitude compactified on the
solvable 7-d lens space manifolds, $ Q = S^3/Z_p $.

\subsection{Supersymmetric $SU(5)$ model of  $Z_2 \times Z_2 $ orbifold}
\label{sub42}

\subsubsection{Unified model  and threshold corrections}


We consider here the supersymmetric gauge unified models with
intersecting $D6$-branes constructed by Cvetic et
al.,~\cite{Cvetic:2001nr} for the $ Z_2 \times Z_2 $ orbifold.  To
avoid repetition, we briefly recall that the point symmetry group
consists of the two generators, $\t ,\ \o $, and allows the Hodge
numbers are, $ h^{(1,1)} = 3 ,\ h^{(2,1)} = 3 + 48 = 51$. The minimal
$SU(5)$ model is obtained with three $D6$-brane stacks, $ N_{a_1}
=10,\ N_{a_2} =6 ,\ N_{b}=16$, with wrapping numbers, $ [(n_a ^I ,
\tilde m_a ^I ) ] = [(1,1) \ (1,-1) \ ( 1, \ud ) ] ,\ [(n_b ^I ,\tilde
m_b ^I ) ]= [(0,1) \ (1,0) \ (0 ,- 1) ] $, satisfying the RR tadpole
cancellation conditions.  Each brane stack is made to preserve some
$\caln =1$ supersymmetry by by tuning the ratio of radii, $\chi _I $,
so as to satisfy the condition on the brane-orientifold intersection
angles, $\sum _I \t ^I _\mu = {1\over \pi } \sum _I \arctan {\tilde
m^I_\mu \chi _I \over n ^I_\mu } = 0 \ mod \ 2 ,\ [\chi _I = {r_2^I
\over r_1^I} ]$.  With the $D6$-branes assumed to pass by the orbifold
fixed points, one must impose on the CP factors the projection
conditions involving the gauge twist embedding matrices, $\g _{\t ,\mu
} ,\ \g _{\o ,\mu } $ alongside with the orientifold projection
condition involving $\g _{\O \calr , \mu }$.  We shall not further
elaborate on the dependence with respect to the gauge factors since
this should cancel out in the simple-minded definition we have adopted
for the ratio of string to field theory amplitudes.  The issue of the
string amplitude normalization has been examined
Refs.~\cite{Cvetic:2005lll,cvetter06}.

After the orbifold and orientifold projections, the resulting gauge
group is given by $ SU(5) _{a_1} \times SU(3)_{a_2}\times USp(16)
_b\times U(1)_{a_1} \times U(1)_{a_2}$, where the symplectic gauge
group factor arises from the $D6_b$-brane being parallel to the
$O6$-plane fixed under $ \O \calr \t \o $.  The massless modes from
the diagonal sectors include the adjoint representations, $ 3(24,1,1)
_{0,0} + 3(1,8,1) _{0,0} + 3(1,1,136) _{0,0}$, and three massless
adjoint representation $\caln =1$ chiral supermultiplets, $\Phi _1,\
\Phi _2,\ \Phi _3 $, which descend from the initial $\caln =4$ gauge
multiplet.  The non-diagonal sector includes four chiral generations
of matter chiral supermultiplets, $F_i,\ (f_{\a , i} +\bar f _a ) ,\
[i=1,\cdots 4;\ \a =1,2,3; \ a= 1,\cdots , 16] $ where the indices $
\a $ and $ a $ label the fundamental representations of $ U(3)$ and
$USp(16)$, along with Higgs boson chiral supermultiplets, $ C ^{'a} ,\
C ^\a _a $, as listed in the table below.

\begin{center}
\begin{tabular}{c|ccccc} 
Mode & $F$ & $ f _\a $ & $ \bar f _a $ & $ C ^{' \a } $ & $ C ^\a _{a
} $ \\ &&&&& \\ \hline &&&&& \\ Sector & $ (a_1,a'_1) $ & $ (a_1
,a'_2) $ & $ (a_1 ,b) $& $(a_2 ,a'_2) $ & $ ( a_2,b) $ \\ &&&&& \\
Irrep & $4 (10,1,1)_{2,0} $ & $ 4 (5,3,1)_{1,1} $ & $ 1 (\bar 5
,1,16)_{-1,0} $ & $ 4 (1,\bar 3 , 1)_{0,2}$ & $ 1 (1,\bar 3
,16)_{0,-1} $ \\
\end{tabular}
\end{center}
\vskip 0.5 cm

The intersection angles for the $10$ and $\bar 5$ modes are
parameterized by, $ F \sim 10: \ \t ^I _{a_1 a'_1} = \t ^I_{a'_1}- \t
^I_{a_1} = (-2\a ^1, 2\a ^2 ,-2 \a ^3 ),\ f\sim \bar 5 :\ \t ^I
_{a_1b}= \t ^I _{b} - \t ^I _{a_1} = (\ud -\a ^1, \a ^2 ,-\ud -\a ^3),
$ in terms of the angle parameters $\a ^I$ obeying the supersymmetry
condition, $ \a ^1-\a ^2 +\a ^3 =0 \ \text{mod} \ 2 ,\ [\pi \a ^{
I=(1,2)} \equiv \arctan (\chi _{ I=(1,2)} ) ,\ \pi \a ^{3} \equiv
\arctan {\chi _{3} \over 2 } ].  $ We only consider here the solution
defined by setting $\a _1, \ \a _2$, at fixed values.

The unified $SU(5)$ symmetry breaking to the Standard Model and the
breaking of $ U(3)$ and $ USp(16)$ are accomplished through the finite
VEVs of the adjoint representation scalar modes.  The $5,\ \bar 5$
mirror fermion generations decouple through the mass terms induced by
the trilinear superpotential, $ W = \l _ i f _{\a i } \bar f _a C ^\a
_{a}$, provided one assumes that the tachyon scalars, present among
the $C ^\a _{a } $, give the $ 12 \times 16$ mass matrix, $ \l _ i < C
^\a _{a } > $, rank 12.  The electroweak Higgs doublets arise from the
massless linear combinations of $f _{\a i } , \ \bar f _a $.  Since
the trilinear superpotential $ F _i \bar f _j \bar H$ only is allowed
by the gauge symmetries, while $ F _i F_j H $ is forbidden, one finds
again that the Higgs mechanism generates a mass matrix only for
down-quarks and leptons.

We now discuss an indicative prediction for the unified mass scale, $
M_X$, based on a rough calculation of the massive string threshold
corrections to the gauge coupling constants.  These have the additive
structure, $\D _a = \D _a ^{\caln =2} + \D _a ^{\caln =1} $, in
correspondence with the contributions from the string state subsectors
of $\caln =2, \ 1 $ supersymmetry. From the discussion in
Ref.~\cite{lustieb03} for type $II$ supersymmetric intersecting brane
models, we borrow the following schematic formulas for these two
components of the threshold corrections \bea && \D _a ^{\caln =2} =
\sum _{b, I} b ^I _{ab} [ \log [ {T_2^I \vert \eta (T^I) \vert ^4 } ]
+ \log ( 4\pi e^{-\g _E} V_a ^I ) ] ,\cr && \D _a ^{\caln =1} = \sum
_{b, I} b _{ab} \log \bigg ( {1 -\t _{ab} ^I \over 1 +\t _{ab} ^I }
\bigg ) , \ [ V_a ^I = { \vert L_a ^I \vert ^2 \over T_2^I } ] \eea
where $ b ^I_{ab}$ denote the slope parameters due to the $(a,b)$
modes with branes parallel in a single $T^2_I$,
$\g _E = 0.577...$ denotes the Euler constant and the correct use of
the formula for $ \D _a ^{\caln =1} $ requires that the interbrane
angles lie inside the range, $\vert \t ^I \vert < 1$.  A natural
definition for the effective string unification scale, $ m'_s $, can
be considered by absorbing inside the logarithmic term, $\ln Q^2 /m_s
^2$, the contributions from $\D _a ^{\caln =2} $ which are finite in
the large radius limit, thus yielding, $ m'_s = m_s \prod _I (4\pi
e^{-\g _E} V_a ^I ) ^ {- { \sum _b b ^I _{ab} / 2 b_a } } .$ One may
similarly transfer the explicit $T_2^I$ moduli dependent contributions
to $\D _a$ to the massless mode logarithmic term by introducing the
effective mass scale, $ \tilde m'_s = m_s \prod _I (4\pi e^{-\g _E}
L_a ^{I2} ) ^ {- { \sum _b b ^I_{ab} / (2 b_a ) } } $, which then
naturally identifies with the compactification scale, $\tilde m'_s
\simeq M_c$.  We now consider the parameterization of the string
threshold corrections, $ \D _a = -b_a \D + k_a \caly $, assuming that
the moduli independent universal constant terms in $\D _a $ are
absorbed into the string unified coupling constant, $ m'_s$.  The
redefined mass and coupling constant unification parameters can then
be expressed as \bea && {M_X \over m'_s} =e ^ {\D \over 2 } = e ^ { -
{\D _a \over 2 b_a }} ,\ {1\over g_X ^{2} } \to {1\over g_X ^{ 2} } +
{\caly \over (4\pi )^2 }. \eea

To obtain a rough estimate of the effect of $\caln =1$ sectors on the
ratio $ M_X / M_c$ for the unified brane stack $a$, we consider the
contribution to $\D _a ^{\caln =1}$ from a single brane stack $b$ for
the choice of interbrane angles, $\t ^I _{ab}= \pm {p \over 7} (
-{2\over 5} , {2\over 3} , -{4\over 15} ) ,\ [p=1, \cdots , 10]$.
With increasing $p \geq 1 $, the numerical results for $\D _a ^{\caln
=1} / b_{ab} $ start from the low values $ \pm 5.\ \times 10^{-4}
$  and increase rapidly up to the value, $ \pm 1.8 $, at $ p=10$.  It
is important that both positive and negative signs of $\D_a /b_{ab} $,
leading to $ M_X /M_c$ ratios smaller and larger than unity, can
occur.  The favored values lie in the range, $ M_X /M_c \sim \ud \ - \
2 $, although one may not exclude a pile up effect from several branes
which would widen this interval.

It is interesting to compare these results with those found in the
M-theory compactification on a $G_2$ holonomy manifold, $X_7 \sim Q
\times K3 $, using the lens manifold, $ Q= S^3/Z_q$, with $SU(5)$
gauge group broken by Wilson lines around $ Q $.  The predicted
threshold corrections~\cite{friedman02}, $\D _a = 2 \sum _i \calT _i
Tr _{R_i}(Q_a ^2 ) \simeq 10 k_a \calT _\o +{2 \over 3} b_a ( \calT _0
- \calT _\o ) $, involve the analytic torsion index, $\calT _{\o _i}
$, in the representations $\o _i$ of the subgroup $U(1)_Y$ commuting
with $SU(5)$.  Defining $ V_Q = (2\pi r)^3 = (2\pi /M_c )^3 $ and
using the numerical estimate, $ \ L(Q)= e ^{ \calT _\o - \calT _0 }= 4
q \sin ^2 (5 \pi w /q) \simeq 10 $, yields the prediction for the mass
ratio, $ {M_X \over M_c } \equiv {L^{1\over 3} (Q) \over 2 \pi } =
{e^{(\cali _\o -\cali _o)/ 3} \over 2 \pi }\simeq 0.3 $.

\subsubsection{Nucleon decay amplitudes}

\begin{center} \begin{figure}
\epsfxsize= 6in
\epsffile[30 1 580 500]{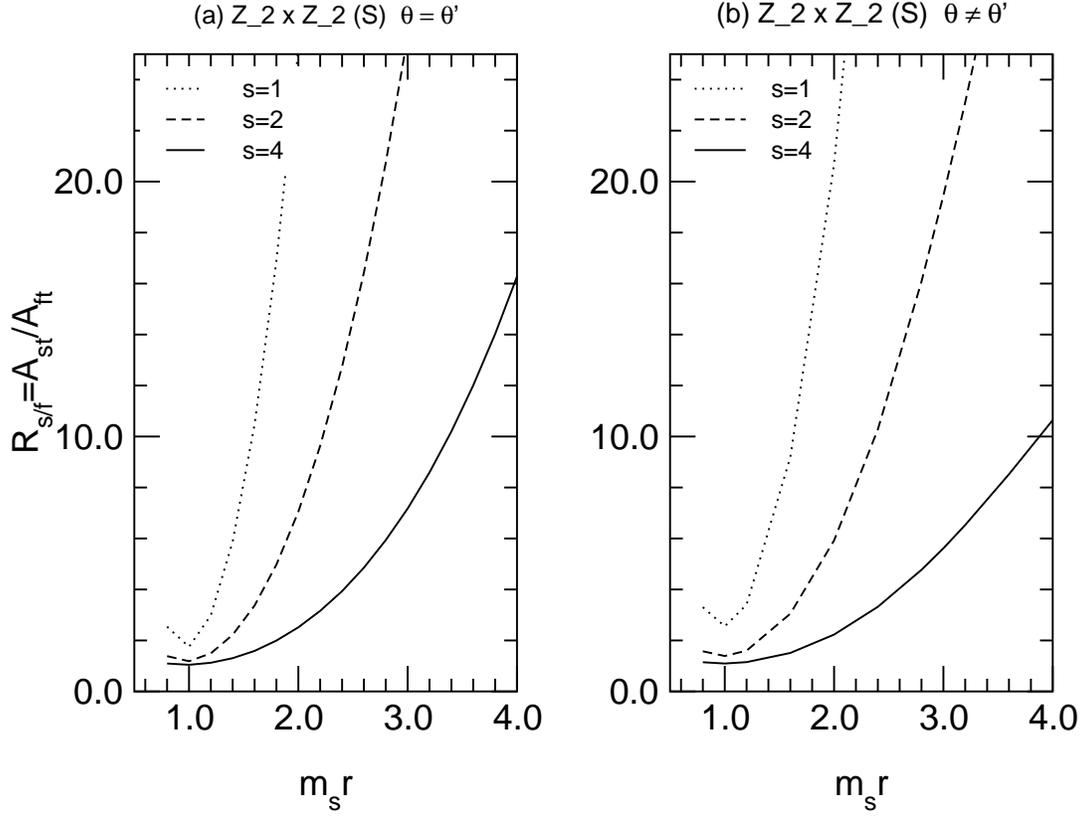}
\caption{\it Plots as a function of the radius, $m_s r$, of the ratio
of string to field theory amplitudes, $\calr _{s/f }$, for the
$Z_2\times Z_2$ orbifold supersymmetric $SU(5)$ gauge unified model
using the subtraction regularization prescription. We show results in
panel $(a)$ for the case involving equal brane intersection
angles, $\t ^I =\t ^{'I} $, and in panel $(b)$ for that involving  
unequal angles, $\t ^I \ne \t ^{'I}$.  The solution is described by wrapping numbers,
$n_a=(1,1,1),\ n_b=(0,1,0),\ m_a=(1,-1,\ud ),\ m_b=(1,0,-1)$, and
angle parameters, $\pi \a ^I \equiv \arctan (\tilde m _a ^I U_2 ^I /
n_a ^I ) $, set at the values, $ \a _2={1\over 3},\ \a _1 ={1\over
5},\ \a_3 =(\a _2 -\a _1)= {2\over 15} $, yielding the intersection
angles, $\t ^I = \t ^I _{aa' _1} = (3/5,2/3, 11/15)$ and $\t ^{'I} =
\t ^I _{a_1 b} = (3/10,1/3,11/30)$, for the modes, $10$ and $
\overline {5} $, after transformation to the interval $[0,1]$.}
\label{figsrz2z2}
\end{figure}
\end{center}

The numerical results for the ratio of string to field theory
amplitudes in the subtraction regularization procedure are displayed
in Figure~\ref{figsrz2z2}.  The panels $(a)$ and $(b)$ refer to the
equal and unequal intersection brane angle cases, associated to the
nucleon decay amplitudes in the configurations $ 10 \cdot 10^{\dagger
} \cdot 10 \cdot 10^{\dagger } $ and $ 10 \cdot \bar 5 \cdot
10^{\dagger } \cdot \bar 5 ^{\dagger } $.  The dependence on the
compactification radius is shown for a discrete set of values of $M_X$
at a fixed value of the free complex structure moduli, $\chi_ I $,
determined by the choice of $\a _1,\ \a _2 .$ The results are very
similar to those found previously for the $Z_3$ orbifold. We thus
conclude, in conformity with our previous conclusion from the study of
$ x $-integrals, that no essential difference arise for the ratio in
the supersymmetric models.  However, to be more conclusive in the case
of the dimension $ \cddd =5$ amplitudes, one should improve the
quantitative understanding of the correlators involving excited
coordinate twist fields.

We have used so far the information bearing only on the unified model
by assuming a schematic representation of the gauge symmetry breaking.
A fully consistent calculation of the four point baryon number
violating processes requires specifying the brane setup at low
energies.  Explicit realizations of the deformed vacua with displaced
brane stacks have been discussed in Ref.~\cite{Cvetic:2001nr}.  The
brane displacement regularization is explicitly realized by splitting
some $D6$-brane stack into separated quark and lepton substacks. One
can then adequately suppress the baryon number violating couplings of
$\cddd =6 ,\ 5 $ by increasing the distance between the substacks.  We
consider here the three family extended Standard Model defined in
Tables $IV$ and $V$ of Ref.~\cite{Cvetic:2001nr} with the setup of six
brane stacks, $ 8 D6_{a_1} ,\ 8 D6_{c_1} ,\ 4D6_{b_1 } ,\ 2 D6_{b_2},\
2 D6_{a_2}, \ 4 D6_{c_2} $, yielding the gauge group, $ USp(8) _{a_1}
\times USp(2) _{a_2} \times U(2)_{b_1} \times USp(2)_{b_2} \times U(4)
_{c_1} \times USp(4) _{c_2} $.  The brane stack splitting, $ 8
D6_{c_1} = 6 D6_{c_{1q} } + 2 D6_{c_{1l}} $, resulting in $ U(4) _{c_
{1q} } \to U(3) _{c_ {1q} } \times U(1) _{c_ {1l} } $, is used to
suppress the baryon number violating processes.  With the first gauge
factor broken as, $ USp(8) _{a_1} \to U(1)_{8} \times U(1)_{8'} $, the
hypercharge is embedded in the linear combination, $ Y= { Q_{c_{1q} }
\over 6 } - {Q_{c _{1l} } \over 2 } + { Q_8 + Q _{8'}\over 2 } . $ The
open string sector assignment for the quarks and leptons, $ q\sim
(b_1,c_{1q} ) +(b_1,c'_{1q} ) ,\ u^c \sim (a_1,c _{1q}) ,\ d^c \sim
(c_{1q}, a_1) , \ l \sim (c' _{1l} , b _1) + (c _{1l} , b _1) ,\ e^c
\sim (c_{1l}, a _1) $, allows to identify the configurations of open
string modes which contribute to the relevant $\cddd =6$ and $5$
dangerous operators, \bea && \bullet \ O_{e_L^c}
\sim u e^{c\dagger } u^{c \dagger } d \sim (b' _1, c_{1q}')(c'_{1l}, a
' _1)(a' _1, c' _{1q} ) (c'_{1q} , b'_1) , \cr && O_{e_R^c} \sim u e
u^{c\dagger } d ^{c \dagger } \sim (c_{1q} , b' _1)(b'_1,
c_{1l})(c_{1q} ,a_1) (a_1, c_{1q}) ,\cr && \bullet \ qqql \sim
ud\tilde u \tilde e \sim (b_1, c'_{1q} ) (b_1, c_{1q} ) (b_1, c'_{1q}
) (c'_{1l} , b_1 ),\cr && (u^c e^c \tilde u ^c \tilde d^c )\sim
(a_1,c_{1q} )(c_{1l}, a_1) (a_1,c_{1q}) (c_{1q} , a_1) . \eea The
above brane embeddings of the disk boundary show that while the
non-supersymmetric operators, $ O_{e_L^c} $ and $ O_{e_R^c} $, and the
supersymmetric operator, $(u^c e^c \tilde u ^c \tilde d^c )$, are
allowed, the supersymmetric operator, $qqql$, is disallowed.  Since
the one-loop dressing of the operator $u^c e^c \tilde u ^c \tilde d^c
$ involves the exchange of electroweak higgsinos, one concludes that
the nucleon decay from $\cddd =5$ operators should be adequately
suppressed in the present model.

\section{Discussion and conclusions}
\label{secconcl} 

We have studied in this work the nucleon decay processes in
semi-realistic string models of gauge unification. Our aim was to
assess the stringy enhancement effect caused in the M-theory and type
$II$ string theory models by the localization of matter modes in the
internal space manifold.  The effect has maximal strength in the large
radius limit.  In accounting for the effect, the reduced power
dependence of the string amplitude on the unified gauge coupling
constant, $\cala \propto \a _X ^{2/3} $, appears as a secondary
manifestation in comparison to the strong peaking of the $x$-integral
at the end point regions.  The leading contributions from these end
point regions arise from the towers of momentum modes propagating
along the internal space directions wrapped by the branes.

The localization of open string modes at brane intersection points is
formally analogous with that of closed string twisted modes at the
orbifold fixed points.  On practical grounds, an important difference
is that while the distance between intersection points may be very
small relative to the compactification distance scales, the fixed
points of orbifold groups are generically a finite distance apart.
Thus, the inevitable suppression from the classical partition function
factor in closed string amplitudes may be minimized in brane models
provided that the distances between intersection point are small
compared to the wrapped three-cycle radius.

The consistent discussion of string unified models requires
introducing an infrared matching mass scale in addition to the string
parameters $g_s ,\ m_s$ and $ r$.  This is identified here with the
unified gauge symmetry breaking mass scale, $M_X$, which may stand for
the open string moduli representing either the Wilson flux line or the
unified brane splitting.  Since we lack a fundamental understanding of
how the compactification and $D$-brane processes stabilize the closed
and open string moduli, we must regard $M_c $ and $M_X$ as free
parameters bounded by $m_s$ and differing by at most a few orders of
magnitude from each other and from $m_s$.  The relative ordering of
the parameters $ M_X $ and $ M_c$ has a crucial impact on the
enhancement effect.  While a definite relation between them is implied
by the heavy threshold corrections to the gauge coupling constants, we
lack a quantitative understanding of these effects to make a learned
choice on the relative order of $ M_X$ and $M_c$.  Although our study
of the size of $ M_X /M_c$ implied by the threshold corrections is not
fully conclusive, this seems to indicate that the ordering of these
scales can go both ways.

Our numerical predictions confirm that some enhancement of the string
amplitude may be present at finite radius.  The consitency
requirements select the intervals, $ M_X / m_s \in [ 1/4 \ - \ 1 ]$
and $ m_s r \in [1 \ - \ 4 ]$.  We find a significant growth of the
ratio with $ m_s r$, which is maximal for coincident intersection
points.  For distant intersection points the suppression effect from
the classical action factor limits the ratio to values of $ O(1)$.  If
$ M_X <M_c$, the smallest contribution found with lowest $ M_X$ leads
to ratios of order $ 2$ at coincident intersection points and order $
1$ for distant intersection points.  If $ M_X > M_c$, the ratio
attains $ O(10)$.  The same conclusion applies to the configurations
of brane pairs with same and distinct intersection angles.  Thus, the
effect is not restricted to the brane setup with same intersection
angles, as concluded from the study of Ref.~\cite{KW03}, based on a
brane setup which did not include the unequal angle configuration.

The application to the supersymmetric model using the $ Z_2 \times Z_2
$ orbifold indicates that a similar enhancement effect may be present
in the nucleon decay processes from $ \cddd =5$ operators.  No
essential difference appears in supersymmetric models, as also
indicated by the similar size of the $x$-integrals for the
supersymmetric and non-supersymmetric configurations of intersecting
angles.  Our study has an indirect bearing on the nucleon decay
processes described by the dimension $ \cddd = 7, \ 9 ,\ 10 $
operators, although we have not carried out explicit calculation of
the string ampliudes in these cases.  We found that restrictive
selection rules are set at the tree level by the $D$-brane embedding.
No finite contributions to these operators are found in the $D$-brane
models discussed in the present work.

In conclusion, our calculations confirm the presence of a small
enhancement effect which is maximal for diagonal configurations with
same intersection points, but not necessarily carrying the same group
representations.  No essential changes are observed upon passing to
supersymmetric amplitudes associated with the dimension $ 5$
operators.  The string amplitudes feature a rapid growth in the
regimes of large and small compactification radii relative to the
string length parameter in which the string and world sheet
perturbation theories are invalidated.  The suppression effect in the
case of distant intersection points is found to be substantial.  The
enhancement effect is also potentially present for the six fermion or
higher order processes contributing to the $\D B=-2$ amplitudes at
tree level.  However, the examination of proposed intersecting brane
models~\cite{kokogut02} realizing Pati-Salam or related gauge unified
models indicates that this source of baryon number violating is
generically suppressed because of the absence of the higher
dimensional matter modes needed to contribute to VEV induced $\D B= -
2$ operators.







\begin{acknowledgments} 
The author  would like to thank the referee for drawing
his attention to Ref.~\cite{higaki05}  and Dr. S. Wiesenfeldt  for  
a helpful correspondence on  recent results  in grand unification.
\end{acknowledgments} 

\appendix
\section{Vacuum  correlator  of coordinate twist fields}
\label{apptwist}

We review in the present appendix the calculation of four point
correlators for the open string coordinate fields obeying the twisted
boundary conditions appropriate to intersecting branes wrapped around
three-cycles.  The restriction to factorisable $T^6$ tori allows us to
specialize the discussion to a single $T^2 _I $ torus, parameterized
by the pair of complexified coordinate fields, $ X^{I} (z),\ \bar
X^{I} (z) $, with the relevant part of the stress energy generator
given by, $ T ^I(z) = - {2\over \a '} \dh _z X ^I (z)\dh _ z \bar X
^{I} (z) $.  For notational convenience, we restrict in the sequel to
a single complex plane and omit the space component indices $I $ by
identifying the string coordinates as, $ X= X^I,\ \bar X= \bar X^{I}
$.  We are interested in the correlators involving two distinct pairs
of conjugate twist fields, $\s _{\pm \t ^I} $, with opposite sign
angles, $ Z ^I(z_i)= <\s _{-\t ^I } (z_1) \s _{\t ^I } (z_2) \s _{-\t
^{'I } } (z_3) \s _{\t ^ {'I } } (z_4) > $. The calculations will be
pursued along same lines as in the conformal field theory approach
developed for the twisted sectors of heterotic string orbifold models
by Dixon et al.,~\cite{dixon87} and Bershadsky and
Radul~\cite{bersh87}.  Our presentation closely follows the work of
B\"urwick et al.,~\cite{burwick}.

\subsubsection{Quantum  partition function}

The energy source approach~\cite{dixon87} exploits the observation
that the world sheet stress energy tensor acquires a non vanishing VEV
in the twisted sector vacuum created by the primary twist fields.  One
is thus motivated to consider the five point correlator, $Z_T (z; z_i
)= {< T(z) \s _{-\t }(z_1) \s_{\t }(z_2) \s _{-\t '} (z_3) \s_{\t
'}(z_4) > \over < \s _{-\t } (z_1) \s_{\t } (z_2) \s _{-\t '}(z_3)
\s_{\t '} (z_4) > } $, obtained by inserting $T(z)$, along with the
two six point correlators involving the insertion of bilocal quadratic
products of the coordinate fields, \bea && g(z,w; z_i) \equiv g(z,w) =
{ < - {2\over \a '} \dh _zX (z) \dh _w \bar X (w) \s _{-\t } (z_1) \s
_{\t } (z_2)\s _{-\t '} (z_3) \s _{\t '} (z_4)> \over < \s _{-\t }
(z_1) \s _{\t }(z_2) \s _{-\t '}(z_3) \s _{\t '} (z_4)> } , \cr &&
h(\bar z,w; z_i ) \equiv h(\bar z,w) = {< - {2\over \a '} \dh _{\bar
z} X ( \bar z) \dh _w \bar X (w) \s _{-\t } (z_1) \s _{\t }(z_2) \s
_{-\t '} (z_3) \s _{\t '}(z_4) \over < \s _{-\t } (z_1) \s _{\t }(z_2)
\s _{-\t '} (z_3) \s _{\t '} (z_4) > }.  \eea

The correlators $g(z,w) ,\ h(\bar z,w)$ are meromorphic functions of
$z, \ \bar z $ and $w$, whose singularities are fully determined by
the operator product expansion of the primary twist fields, as given
in Eqs.~(\ref{app.XOPE}).
The fact that the short distance expansion in the limit $z\to w$ of
$g(z,w)$ exhibits a double pole term, allows one to relate this
function to the correlator $Z_T(z; z_i) $ as \bea && - {2\over \a '}
\dh X (z) \dh \bar X (w) = {1\over (z-w)^2 } + T(z) +\cdots \
\Longrightarrow \ g(z,w) = {1\over (z-w)^2 } + Z_T (z; z_i ) +\cdots ,
\eea where the dots represent the higher order terms in the expansion
in powers of $ (z-w)$.  Since no singularities are present in the
limit $\bar z\to w$, one concludes that the bilocal correlator $
h(\bar z , w) $ has no pole terms and so must tend to a constant in
this limit.  Writing the short distance expansion of $Z_T(z; z_i)$ in
the limit $ z \to z_I , \ $ for fixed index $I$, with the help of the
familiar operator product expansion \bea && T(z) \s _\t (z_I) = {h
(\s_\t ) \s_\t (z) \over (z-z_I) ^2 } + {\dh _ {z_I} \s _\t (z_I)
\over (z-z_I) } , \ [h(\s_\t ) = \ud \t (1-\t ) ] \eea yields the
differential equation relating the correlators $ Z (z_i) $ and $
Z_T(z; z_i) $ \bea && {\dh \over \dh z_I } \ln Z (z_i) = (z -z_I) Z_T
(z; z_i) - {h (\s_\t ) \over z-z_I} .\eea The holomorphy properties of
the functions $g(z, w)$ and $ h(\bar z,w) $, as the variables $z,\
\bar z $ and $w$ approach the insertion points $z_i$, allows one to
write the following general representations for these functions, \bea
&& g(z,w) = \o _ {\t , \t ' } (z) \o _ {1- \t , 1- \t ' } (w) \bigg [
{P(z, w) \over (z-w)^2 } + A (z_i) \bigg ] , \ h(\bar z,w) = \bar \o _
{1-\t , 1 - \t ' } (\bar z) \o _ {1- \t , 1- \t ' } (w) B (z_i) , \cr
&& \cr && \bigg [\o _ {\t , \t ' } (z) = (z-z_1 ) ^{-\t } (z-z_2 )
^{\t -1} (z-z_3 ) ^{-\t ' } (z-z_4) ^{\t ' -1} , \cr && \o _ {1-\t ,
1-\t ' } (w) = (w-z_1 ) ^{\t -1 } (w-z_2 ) ^{-\t } (w-z_3 ) ^{\t ' -1
} (w-z_4) ^{-\t '} ,\cr && \bar \o _ {1-\t , 1-\t ' } (\bar z ) =
(\bar z -z_1 ) ^{\t -1 } (\bar z -z_2 ) ^{-\t } (\bar z -z_3 ) ^{\t '
-1 } (\bar z -z_4) ^{-\t '},\cr && P(z, w) = \sum _{i, j =0} ^2
\a_{ij} w^i z^j = \a _{00} + \a _{01} z + \a _{10} w + \a _{20} w ^2
\cr && + (\a _{11} w + \a _{21} w ^2)z + (\a _{02} + \a _{12} w + \a
_{22} w ^2) z^2 \bigg ]
\label{app.gh} \eea
where the coefficients $A(z_i) ,\ B(z_i) $ (to be determined in the
sequel via the global monodromy conditions) and the coefficients $
\a_{jk} $ of the second order polynomial $P(z, w)$ are functions of
the insertion points $z_i$.  The polynomial $P(z, w)$ is determined by
the requirement that one reproduces the pole structure, $ g(z,w) \to
(z -w )^{-2}$, in the limit $ z\to w$.  Expanding the various factors
of $ g(z,w)$ in powers of $(w-z)$ leads to the two sets of relations:
\bea && \bullet \ P (z,z)=[ \o _ {\t , \t ' } (z) \o _ {1- \t , 1- \t
' } (z) ]^{-1} = \prod _i (z-z_i) \cr && \Longrightarrow \ \a _{00} =
z_1z_2 z_3z_4, \ \a _{10} + \a _{01} = - (z_2 z_3z_4 + z_3z_4 z_1+z_4
z_1 z_2+ z_1 z_2 z_3) ,\cr && \a _{11} + \a _{02} + \a _{20} = z_3 z_4
+ z_4z_1 + z_1z_2+z_2z_3 , \ \a _{21} + \a _{12} = -( z_1 +z_2+
z_3+z_4 ) ,\ \a _{22} =1, \cr && \bullet \ {P'(z,z) \over P(z,z) } = -
{\o '_ {1- \t , 1- \t ' } (z) \over \o _ {1- \t , 1- \t ' } (z) } =
\bigg [ {\t -1 \over z- z_1} + {\t ' -1 \over z-z_3} - {\t \over
z-z_2} - {\t ' \over z- z_4} \bigg ] \cr && \Longrightarrow \ \a _{10}
= \t z_3 z_4 (z_2-z_1) + \t ' z_1 z_2 (z_4-z_3) - z_2 z_4
(z_3+z_1),\cr && 2 \a _{21} + \a _{12} = -[\t (z_{1} - z_2) + \t '(
z_{3} -z_4) +2(z_2+z_4) + z_1+z_3 ],\cr && \a _{11} + 2\a _{20} = -[
(\t -1) (z_3z_4 +z_2z_4 +z_2z_3) + (\t '-1) (z_2z_4 +z_1z_4 +z_1z_2)
\cr && - \t (z_3z_4 +z_1z_4 +z_1z_3) - \t ' (z_2z_3 +z_1z_3 +z_1z_2) ]
, \eea allowing to express the solution for the $T(z)$-inserted
correlator as \bea && Z_T(z; z_i) = \o _{\t ,\t '} (z) \o '_{1-\t
,1-\t '} (z) P' (z,z) + \ud \o _{\t ,\t '} (z) \o ''_{1-\t ,1-\t '}
(z) P(z,z) \cr && + \o _{\t ,\t '} (z) \o _{1-\t ,1-\t '} (z) \bigg (
A(z_i) + \ud P''(z,z) \bigg ) \cr && = - \o _{\t ,\t '} (z) \o '_{1-\t
,1-\t '} (z) P ^2 (z,z) +\ud \o _{\t ,\t '} (z) \o ''_{1-\t ,1-\t '}
(z) P(z,z) \cr && + (A(z_i) +\a _{20} + \a _{21} z + z^2 ) P (z,z) ,
\eea with the primes standing for the derivative $\dh _w $, namely,
$P' (z,w) = \dh _w P (z,w) ,\ P'' (z,w) = \dh ^2_w P (z,w) $.
The equations for the coefficients, $ \a_{12} ,\ \a _{21} $, are
independently solved, with the solutions given by, $ \a _{12}= (\t -1)
z_1 -(\t +1) z_2 + (\t '-1) z_3 -(\t ' +1) z_4,\ \a _{21}= -[\t z_1
+\t ' z_3 +(1-\t ) z_2 +(1-\t ' ) z_4] .  $ The condition on the
double pole term gives eight linear equations for the nine
coefficients, $\a _{ij}$.  However, since the coefficient $\a _{20}$
can always be absorbed inside the function $A(z_i)$, this coefficient
remains arbitrary as long as $A(z_i)$ is unspecified.  We make the
convenient choice~\cite{burwick}, $\a _{20}= \ud [(\t +\t ') (z_1 z_3
- z_2 z_4) +(\t -\t ') (z_1 z_4 - z_2 z_3) + 2 z_2 z_4] $, which then
fully determines $P(z,w)$.  Evaluating the function $ g(z,w)$ by
taking first the limit, $ w \to \infty $, and next the limit, $ z \to
z_2 $, leads to the differential equation for the correlator of
interest \bea && {\dh \over \dh z_2} \ln Z (z_i) = {\t -\t ' \over 2}
{z_{34} \over z_{23} z_{24} } + { \t (1-\t ) \over z_{12} } + \t (1-\t
' ) \bigg ({1 \over z_{32} }+ {1 \over z_{24} } \bigg ) +{A(z_i) \over
z_{21} z_{23} z_{24} } , \eea where we use the notation, $ z_{ij}
\equiv z_i -z_j$, and have set $ z_J = z_2$.  Similar equations apply
by letting $z$ approach the other insertion points.  Thanks to the
invariance under the $SL(2,R)$ subgroup of the conformal group, one
can arbitrarily fix three insertion points.  With the choice of
insertion points along the real axis boundary, $ z_1= x_1= 0, z_3=
x_3= 1, z_4 =x_4= X \to \infty $, leaving the single real variable, $
z_2=x_2= x \in [-\infty ,\infty ]$, one can reconstruct the full
dependence by noting that the M\"obius group invariant functions of
four variables in $C$ can only depend on the harmonic ratio variable,
$ x= {z_{21} z_{34} \over z_{24} z_{31} } $. (Note that $ (1-x) =
{z_{23} z_{41} \over z_{24} z_{31} } $.)  The so far unspecified
coefficients of the polynomial $P(z,w)$ are now given by the simpler
formulas, $ \a _{20} = X [\ud (\t -\t ') + x(1 -{\t +\t ' \over 2} ) ]
, \ \a _{11} = X (-\t + \t ' +1 + x\t ') , \ \a _{02}= {X\over 2} [-x
+ (\t -\t ') (1-x) ] .  $ It is convenient to absorb the known
dependence on the variable $X $ by considering the reduced correlator
and coefficients identified by hats, $ Z(z_i) = \hat Z(x) (x-X)^{ \t
(1-\t ') } ,\ A(z_i) = (x-X) \hat A(x) ,\ B(z_i) = (x-X) \hat B(x)
$. The differential equation for the reduced correlator reads then as,
\bea && {\dh \over \dh x } \ln \hat Z (x)= -{\t -\t ' \over 2(1-x)}
+{\t (1-\t ') \over (1-x)} -{\t (1 -\t ) \over x} - {\hat A(x) \over
x(1-x) }.
\label{eqdiffz}\eea To proceed further at this point, one must
separate out the zero mode part of the coordinate field, associated
with the classical zero modes, from its quantum or oscillator
part. The additive splitting, $ X(z)= X_{cl} (z) + X_{qu} (z) $,
produces the multiplicative splitting of the correlator into classical
and quantum factors, $\hat Z (x)= Z_{qu} (x) Z_{cl} (x) ,\ [Z_{cl} (x)
=\sum _{cl} e ^{-S_{cl}(x)}] $ where the classical factor represents
the contribution from the world sheet instantons and anti-instantons,
defined as the Euclidean space coordinate field solutions of the world
sheet equations of motion, $ \dh_z X _{cl} (z) =0 ,\ \dh_{\bar z} X
_{cl} (\bar z) =0 $.
The general solutions with the appropriate singularities at the
insertion points, $ z_i$, are described in terms of the meromorphic
functions introduced earlier as, $ \dh_z X _{cl} (z) =b \o _{\t ,\t '}
(z) ,\ \dh_{\bar z} X _{cl} (\bar z) = c \bar \o _{1-\t , 1-\t '}
(\bar z) $, where the complex constants $b,\ c$ represent free
continuous parameters.

We postpone the discussion of the classical partition function until
the next subsection, and continue with the study of the quantum
factor.  The coefficients $ A (z_i), \ B (z_i)$ are determined by
imposing the global boundary conditions requiring that the quantum
components of coordinate fields, unlike the classical ones, are single
valued functions on the cut complex plane $C$.  For definiteness, we
consider the four point amplitude with the configuration of open
string sectors, $ < V _ {-\t , {(D,A)} } (x_1) V _ { \t , {(A,B)} }
(x_2) V _ {-\t ', {(B,C)} } (x_3) V _ { \t ' , {(C,D)} } (x_4) > $,
setting the boundary along the real axis of $C_+ $, with $ z_i = x_i.$
The embedding in $ T^2_I$ is described by the four-polygon displayed
in Figure~\ref{mapbr}, with vertices and edges given by the images of
insertion points, $ x_1,\cdots , x_4$, and arc segments, $ S_\a = (
x_1, x_2), \ ( x_2, x_3), \ ( x_3, x_4), \ ( x_4, x_1)$, along the
disk boundary.
On the four segments covering the periodic real axis, $[A,B,C, D]= [(
x_1, x_2), \ ( x_2, x_3), \ ( x_3, x_4), \ ( x_4, x_1)]$, the complex
coordinate fields obey the relations, \bea && e^{-i\phi ^I _S } \dh X
^I (z) - e^{i\phi ^I _S} \bar \dh \bar X ^I (\bar z) =0, \ e^{-i\phi
^I _S}\bar \dh X ^I (\bar z) - e^{i\phi ^I } \dh \bar X ^I (z) =0 , \
[S= A,B,C, D ] \label{eqappbcs} \eea with the interbrane angles
assigned in the various intervals as, $ \phi _{D} = 0,\ \phi _A = \phi
= \pi \t ,\ \phi _{B} = 0,\ \phi _{C} = \phi ' = \pi \t '. $ Since the
vector space of contours is of dimension $2$, it suffices to consider
the integrated form of the boundary conditions for two independent
contours.  We choose the contours $ C_1 , \ C_2$ along the branes $ A$
and $ B$, with angles $\phi _A =\pi \t $ and $\phi _B = 0 $.  To
establish contact with the formalism used for closed strings, it is
convenient to consider instead the associated closed contours (or
cycles) surrounding the intervals for the open contours, $C_i \in C _+
$.  The cycles $ \calc _i \in C$ are defined by adding to the contours
$C_i$ their reflected image $ C'_i$ with respect to the real axis
boundary, $ \calc _i = C_i - C'_i , \ [i=1,4]$.  This definition of
the fields dispenses one from the need to define the analytic
continuation of fields on the Riemann sheets of the cut plane $C$.  We
now consider the contour integrals on $ C_i$ for the sum of the two
boundary conditions in Eqs.(\ref{eqappbcs}), and reshuffle the two
pairs of terms so as to express these as cycle integrals on the basis
of independent cycles, $\calc _i $ by using the transformation under
the reflection, $ \dh X \to e ^{2i\phi _S} \dh \bar X $ and $ \bar \dh
X \to e ^{2i\phi _S} \bar \dh \bar X $.  The resulting integrated
boundary conditions on the coordinate fields \bea && 0= \int _{C_i} [
(\dh X - e^{2i\phi _S} \bar \dh \bar X ) + (\bar \dh X - e^{2i\phi _S}
\dh \bar X ) ] = \int _{C_i - C'_i} (\dh X + \bar \dh X ) = \int
_{\calc _i} (\dh X + \bar \dh X ) , \cr && \eea are then used to
express the trivial monodromy conditions on the quantum component of
the six-point correlators, $0= \int _{\calc _i} dz g(z, w) + \int
_{\calc _i} d\bar z h( \bar z, w) .$ This is our first encounter with
the mixed type correlator, $h(\bar z, w) $.  The two integrals along
the contours $C _1,\ C _2$ give the relations needed to determine the
two unknown coefficients, $A (z_i),\ B (z_i)$.  We note that $ \int
_{\calc _1} \sim \int _0^x ,\ \int _{\calc _2} \sim \int _x^1 .$
Substituting Eqs.~(\ref{app.gh}) for $ g (z, w) ,\ h (\bar z, w) $
into the integrals over the cycles $\calc _1,\ \calc _2 $, gives the
pair of equations for $i= (1,2)$, \bea && \hat A(x) \int _{\calc_i} dz
\o _{\t, \t '} (z) + \hat B(x) \int _{\calc_i} d\bar z \bar \o _{1-\t,
1-\t '} (\bar z) \cr && = - x {(\t -\t ') \over 2} \int _{\calc_i} dz
\o _{\t, \t '} (z) -(1 -\t ') \int _{\calc_i} dz (z-x) \o _{\t, \t '}
(z) = x(1-x) \dh _x \int _{\calc_i} dz \o _{\t, \t '} (z) , \cr &&
\eea where the last step is deduced by integration by parts.  The
resulting two linear equations for the coefficients $\hat A (x) ,\
\hat B (x)$, obtained with $ \calc _1,\ \calc _ 2$, can be expressed
by the matrix equation, \bea && \pmatrix {\int _{\calc_1} dz \o _{\t,
\t '} (z) & \int _{\calc_1} d\bar z \bar \o _{1-\t, 1-\t '} (\bar
z)\cr \int _{\calc_2} dz \o _{\t, \t '} (z) & \int _{\calc_2} d\bar z
\bar \o _{1-\t, 1-\t '} (\bar z) } \pmatrix {\hat A (x) \cr \hat B (x)
} = x(1-x) \dh _x \pmatrix { \int _{\calc_1} dz \o _{\t, \t '} (z) \cr
\int _{\calc_2} dz \o _{\t, \t '} (z) }. \eea These equations are
readily solved for $\hat A (x), \ \hat B (x) $, with the explicit
formula for $\hat A (x)$ reading as \bea && \hat A (x) = x(1-x){ (\dh
_x \int _{\calc_2} \o _{\t, \t '} ) (\int _{\calc _1} \bar \o _{1 - \t
,1 - \t '}) - (\dh _x \int _{\calc_1} \o _{\t, \t '} ) (\int _{\calc
_2} \bar \o _{1 - \t ,1 - \t '} ) \over (\int _{\calc _1} \bar \o _{1
- \t ,1 - \t '} ) ( \int _{\calc_2} \o _{\t, \t '} ) - ( \int _{\calc
_2} \bar \o _{1 - \t ,1 - \t '}) ( \int _{\calc_1}\o _{\t, \t '} ) } =
\ud x(1-x) \dh _x \ln I(x),\cr && [I(x) = { 1 \over (2i\pi )^2 } e ^{
i\pi (4-p) \t - 2i\pi \t ' } [ (\int _{\calc_1} \o _{\t, \t '})( \int
_{\calc _2} \bar \o _{1 - \t ,1 - \t '}) - ( \int _{\calc_2} \o _{\t,
\t '}) ( \int _{\calc _1} \bar \o _{1 - \t ,1 - \t '} ) ]
\label{eqcoeffa} \eea where to simplify the notations we have omitted
writing the integration measure factors $dz$ or $ d \bar z$ in the
cycle integrals. For the purpose of simplifying the final results, we
have set the so far undetermined constant normalization of $ I(x)$ to
a conveniently chosen value depending on the parameter $p$.  Explicit
analytic formulas for the cycle integrals will be presented
shortly. The total derivative structure of the function, $ \hat A (x)
/[x(1-x)] $, in Eq.~(\ref{eqcoeffa}) transforms the differential
equation for $Z _{qu} (x)$ in Eq.~(\ref{eqdiffz}) to the simple form,
$ \dh _x \ln (Z_{qu} (x) I^\ud (x) ) = -\t (1-\t ) /x + [\t (1-\t ')
+\t '(1-\t ) ] /[2 (1-x)] $, which is readily solved to give the final
form of the quantum partition function, \bea && Z_{qu} ^I(x) = C _\s
x^{-\t ^I(1-\t ^I) } (1-x) ^{ -\ud \t ^{'I}(1-\t ^I ) -\ud \t ^I (1-\t
^{'I} ) } I ^{-\ud }(x) , \label{eqsolz}\eea where we have reinstated
the complex plane label, $I$.  The constant $C _\s $ can be determined
by making use of the pole factorization of the correlator.  Note that
we have corrected a confusing misprint in Eqs.(3.21) and (3.37) of the
work by B\"urwick et al.,~\cite{burwick}.  These would have given a
wrong dependence of the exponent of $(1-x)$, lacking the symmetry
under $\t \leftrightarrow \t '$, which our above result exhibits
explicitly.  The integrals over the two independent cycles have the
following analytic forms \bea && \int _{\calc _1} dz \o _{\t , \t '}
(z) = 2\pi i e ^{2\pi i ( {\t ' \over 2} -\t ) } G_2 (x) , \cr && \int
_{\calc _2} dz \o _{\t , \t '} (z)= 2i \sin (\pi p \t ) (1-x) ^{\t -\t
'} e ^{i\pi [ (p-2) \t + \t ' ] } B_1 H_1 (1-x) , \cr && \int _{\calc
_1} d \bar z \bar \o _{1 - \t , 1- \t '} (\bar z) = 2\pi i e ^{2\pi i
( {\t ' \over 2} -\t ) } \bar G_1 (x) , \cr && \int _{\calc _2} d \bar
z \bar \o _{1 - \t ,1 - \t '} (\bar z) =- 2i \sin (\pi p \t ) (1-x)
^{\t ' -\t } e ^{i\pi [ (p-2) \t + \t ' ] } B_2 \bar H_2 (1-x) , \cr
&& \ \Longrightarrow \ I (x) = {\sin (\pi p \t ) \over \pi } (B_2 \bar
G_1 H_2 + B_1 \bar G_2 H_1 ) , \cr && \bigg [ B_1 = B(\t , 1 -\t ')
\equiv {\G (\t ) \G (1-\t ') \over \G (1 +\t -\t ')} ,\ B_2 =B(\t ', 1
-\t ) \equiv {\G (1-\t ) \G (\t ') \over \G (1 -\t +\t ')} ,\cr &&
G_1(x) = F(\t , 1 -\t ' ; 1 ; x) , \ G_2(x) = F(1-\t , \t ' ; 1 ; x) ,
\cr && H_1(x) = F(\t , 1 -\t ' ; 1 + \t -\t ' ; x) , \ H_2(x) = F(1-\t
, \t ' ; 1 -\t +\t '; x) , \cr && F(a,b;c; x)= {\G (c)\over \G (b)\G
(c-b) } \int _0^1 dt t^{b-1} (1-t)^{c-b-1} (1-xt)^{-a} \bigg ]
\label{app.ampqu} \eea where $B(a,b)$  and  $\G (z)$   designate the
Euler Beta and Gamma functions and $F(a,b;c;z)$ the Riemann
Hypergeometric function.  Although we have kept track in the above
formulas of the complex conjugate functions, signalled by the bar
symbol, this distinction is not relevant here to the extent that the
restriction of the various functions to the real axis are real.

\subsubsection{Classical  partition function}

We now return to the calculation of the classical action factor, $
Z_{cl} =\sum_{cl} e^{-S_{cl} } $, where the summation extends over the
non-trivial monodromies around the cycles $\calc _i$ of the coordinate
field classical solutions, $X_{cl} (z,\bar z)$. For orientation, we
recall that in the simpler case of $S^1$ circle compactification, the
classical summation extends over the 1-d lattice of solutions with
winding numbers, $ \sqrt 2 \D _ \calc X =\sqrt 2 \int _{S^1} d X = 2
\pi v ,\ [v \in Z]$.  In the $T^2_I$ torus case, the monodromies of
the coordinate $X ^I$ around the cycles, $ \calc_1,\ \calc_2 $, belong
to the 2-d grand lattice generated by the cycles along the
$D6_A/D6_B-$branes, images of $ \calc_1,\ \calc_2 $, possibly shifted
by the non-vanishing distance between distant intersection points.
Denoting the intersection points along the branes $D6_A$ and $D6_B$ by
$ f_1 ^A, \ f_2 ^A$ and $ f_2 ^B, \ f_3 ^B$, in correspondence with
the images of the insertion points $ x_1, \ x_2$ and $ x_3$, one can
express the non-trivial monodromy conditions on the classical
components of coordinate fields, $\D _{\calc _i} X ^I \equiv \int
_{\calc _1} ( dz \dh X ^I + d\bar z \bar \dh X ^I )$, for the cycle
basis $\calc _{1,2}$ mapped to the $D$-branes $A,\ B$, in terms of the
2-d lattice generated by the $D$-brane pair, $ L_A,\ L_B$, as \bea &&
\sqrt 2 \D _{\calc _1} X ^I \equiv 2\pi v_A = 2\pi (1 - e^{2 i\pi \t }
) (f^A _1 -f^A _2 + p_A L_A ) ,\cr && \sqrt 2 \D _{\calc _2} X ^I
\equiv 2\pi v_B = 2\pi (1 - e^{2 i\pi \t } ) (f^B _2 -f^B _3 + p_B L_B
),
\label{eqglomo} \eea 
where we have assumed, for simplicity, the brane lattice to be
generated by the integers $ p_A,\ p_B \in Z.$ The relation $ \D _{
\calc _i }X = \D _{ C_i - C'_i } X = (1-e^{2 i\pi \t } ) X ,\ [i=1,2]$
follows from the $2\pi \t $ rotation of coordinates along the
reflected contours.  If we had used the open contours in the upper
complex plane, $ C_ + $, as in the presentation by Cvetic and
Papadimitriou~\cite{Cvetic:2003ch}, similar formulas to the above
would hold, but without the angle dependent factors, $(1 - e^{2 i\pi
\t } ) $.  To relate the integer quantized winding numbers $p_A, \
p_B$ with the coefficients $b,\ c$ in the instanton and anti-instanton
solutions,
it is convenient to introduce the 2-d bases of holomorphic and
antiholomorphic solutions, $ X^{(i)} ,\ [i=1, 2] $ defined by \bea &&
\sqrt 2 \dh X^{(1)} (z)= b_{1 } \o _{\t ,\t '} (z),\ \sqrt 2 \dh
X^{(2)} (z)= b_{2} \o _{\t ,\t '} (z),\cr && \sqrt 2 \bar \dh X^{(1)}
( \bar z)= c_{1} \bar \o _{1-\t ,1- \t '} (\bar z) ,\ \sqrt 2 \bar \dh
X^{(2)} ( \bar z)= c_{2} \bar \o _{1-\t ,1- \t '} (\bar z)
. \label{eqmoncycl}\eea The condition that the $ X ^{(i)} $ constitute
the dual basis to the basis of cycles, $\calc _i $, is expressed by
the orthonormalization relations, \bea && 2\pi \d _{ij} = \sqrt 2 \D
_{\calc_ i} X ^{(j)} = \sqrt 2 \int _{\calc _i} dz \dh X ^{(j)} (z) +
\sqrt 2 \int _{\calc _i}d \bar z \bar \dh X ^{(j)} ( \bar z) , \ [i, j
=1,2]. \eea Expressing the general classical solutions as linear
superpositions of the two solutions in Eqs.~(\ref{eqmoncycl}), \bea &&
\sqrt 2 \dh X \equiv b \o _{\t ,\t '} (z) = \sqrt 2 (v_A \dh X ^{(1)}
+ v_B \dh X^{(2)}) \equiv (b_1 v_A +b_2 v_B) \o _{\t ,\t '} (z) \ \cr
&& \sqrt 2 \bar \dh X \equiv c \bar \o _{1-\t ,1- \t '} (\bar z) =
\sqrt 2 (v_A \bar \dh X^{(1)} + v_B \bar \dh X^{(2)} ) \equiv (c_1 v_A
+c_2 v_B) \bar \o _{1-\t ,1- \t '} (\bar z) ,\eea allows identifying
the coefficients $ b ,\ c$ as, $ b= b_1 v_A +b_2 v_B $ and $ c= c_1
v_A +c_2 v_B$, where $ v_A, \ v_B$ are the lattice vectors defined in
Eqs.(\ref{eqglomo}).  The four conditions, $ 2\pi = \sqrt 2 \D _{\calc
_{1,2} } X ^{(1), (2)} ,\ 0= \sqrt 2 \D _{\calc _{1,2} } X ^{(2), (1)}
,$ may now be used to obtain four linear equations for the four
constant coefficients $b_i, \ c_i$, whose solution is given by \bea &&
b_1 = {2\pi \over J} \int _{\calc_2} d \bar z \bar \o _{1-\t , 1- \t
'} (\bar z) , \ b_2 = -{2\pi \over J} \int _{\calc_1} d \bar z \bar \o
_{1-\t , 1- \t '} (\bar z) , \cr && c_1 = - {2\pi \over J}\int
_{\calc_2} dz \o _{\t , \t '} (z) , \ c_2 = {2\pi \over J} \int
_{\calc_1} dz \o _{\t ,\t '} ( z) , \cr &&
[J = (\int _{\calc_1} \o _{\t ,\t '}( z) ) (\int _{\calc_2} \bar \o
_{1-\t , 1- \t '} (\bar z) )- (\int _{\calc_1} \bar \o _{1-\t , 1- \t
'} (\bar z)) ( \int _{\calc_2} \o _{\t , \t '} (z) ) ]
. \label{eqcoffs} \eea

The final step involves substituting the above solutions into the
string classical action, \bea && S_{cl} = {1 \over 2 \pi \a ' } \int
_C d^2 z (\dh X \bar \dh \bar X +\bar \dh X \dh \bar X ) \cr && \equiv
{1\over 2\pi \a '} \int _C d^2 z (\vert \dh X \vert ^2 + \vert \bar
\dh X \vert ^2 = V_{11} \vert v_A \vert ^2 +V _{22} \vert v_B \vert ^2
+ 2 \Re (V_{12} v_A v_B ^\star ) , \cr && \bigg [ 4 \pi V_{ii} = \vert
b_a \vert ^2 \int _C d ^2 z \vert \o _{\t , \t '} (z) \vert ^2 + \vert
c_a \vert ^2 \int _C d ^2 z \vert \o _{ 1-\t ,1- \t '} (z) \vert ^2 ,
\ [i =1,2] \cr && 4 \pi V_{12} = b_1 \bar b_2 \int _C d ^2 z \vert \o
_{\t , \t '} (z) \vert ^2 + c_1 \bar c_2 \int _C d ^2 z \vert \o
_{1-\t , 1- \t '} (z) \vert ^2 \bigg ] \label{eqclassact}\eea and
evaluating the complex plane integrals for the holomorphic and
antiholomorphic differentials by making use of the analytic
continuation method of Kawai et al.,~\cite{kawai86}.
Using the analytic formulas for these integrals in
Eq.~(\ref{app.ampqu}) along with the previously derived formulas for
the coefficients, $ b_i, \ c_i$, in Eq.~(\ref{eqcoffs}) gives the
following results for the functions $ V_{ij}$ determining the
classical action \bea && V_{11} = {1 \over 4 I ^2 (x) } ({\sin \pi p
\t \over \pi }) ^2 [ \vert B_2 H_2 \vert ^2 ( 2 \Re (B_1 G_1 H_1 ) -
c_{\t , \t '} \vert G_1 \vert ^2 ) \cr && + \vert B_1 H_1 \vert ^2 ( 2
\Re (B_2 G_2 H_2 ) + c_{\t \t '} \vert G_2 \vert ^2 )] , \cr && V_{22}
= {1 \over 4 I ^2 (x) } 2 \Re [ G_1 G_2 ( B_2 \bar G_1 \bar H_2 + B_1
\bar G_2 \bar H_1) ] , \cr && V_{12} = {1 \over 4 I ^2 (x) } {\sin \pi
p \t \over \pi } e ^{ i \pi p \t } [B_2 G_2 \bar H_2 ( 2 \Re (B_1 G_1
\bar H_1 ) - c_{\t \t '} \vert G_1 \vert ^2 )\cr && - B_1\bar G_1 H_1
( 2 \Re (B_2 G_2 \bar H_2 ) + c_{\t \t '} \vert G_2 \vert ^2 ) ] , \cr
&&
[c_{\t , \t '}= \pi ( \cot (\pi \t ) - \cot (\pi \t ') ) ]
\label{app.V11} \eea where $I(x)$  is  given by   the 
analytic formula quoted in Eq.~(\ref{app.ampqu}).  To obtain the
classical partition factor, $ Z_{cl} $, it now remains to sum over the
2-d integer lattice generated by the pairs of branes.  The comparison
with the classical partition function for closed strings in orbifolds
in Ref.~\cite{burwick}, whose notations we have closely followed,
indeed shows that the open string correlator is the square root
truncation of the closed string correlator. With our choice of closed
contours, the parameter $p$ in the definition of Eq.(\ref{app.ampqu})
for the function $I(x)$ must be set to unity, $p=1$.

\subsubsection{Useful  limiting  formulas}

To ensure that the series for the instanton sums converge at the end
points $x=0, \ 1 $, thus avoiding the end point singularities in the
$x$-integral of string amplitudes, it is necessary to make use of the
Poisson resummation formula.  This is also needed for a proper
identification of the particle exchange pole contributions in the
field theory limit.  We record below the limiting forms of some
intermediate results at the end points of the interval, $ x\in (0,1)$.
The limiting formulas at $ x\to 0$ are given by \bea && F(\t, 1-\t; 1;
x) \to 1 ,\ F(\t, 1-\t; 1;1-x) \to {\sin (\pi \t ) \over \pi } \log
{\d (\t )\over x } , \cr && B_1 H_1 (x) \to \ln {\d _1 \over x} , \
B_2 H_2(x) \to \ln {\d _2\over x} , \cr && V_{11} \to {1\over 4} \ln
{\hat \d \over x },\ V_{12} \to - e^{i\pi p \t } {\pi ^2 (\cot (\pi \t
) - \cot (\pi \t ') ) \over 8 \sin (\pi p \t ) \ln (\hat \d /x) } ,\
V_{22}\to {\pi ^2 \over 4 \sin ^2 (\pi p \t ) \ln (\hat \d /x) } ,\cr
&& \bigg [ \ln \d (\t ) = 2 \psi (1) - \psi (\t ) - \psi (1-\t ) , \
\ln \d _1 (\t , \t ') = 2\psi (1) -\psi (1-\t ') -\psi (\t ) ,\cr &&
\ln \d _2 (\t , \t ') = 2\psi (1) -\psi (1-\t ) -\psi (\t '),\ \hat \d
(\t , \t ') = [\d _1 (\t , \t ') \d _2 (\t , \t ') ] ^\ud , \cr &&
\psi (1) =-\g , \ \psi (1-z) -\psi (z) = \pi \cot (\pi z) ,\ \G (z) \G
(1-z) = {\pi \over \sin (\pi z)}\bigg ] , \label{app.0G2} \eea where
$\psi (z) = \psi ^{(1)} (z) = {d \ln \G (z) \over dz } $ denotes the
Digamma function and $\g $ is Euler constant.
 We also quote the series representation for the Hypergeometric
function, \bea && F(\t , 1- \t : 1 ; x) \simeq {\sin \pi \t \over \pi
} \sum _{n=0} ^\infty {(\t ) _n (1-\t )_n (1-x)^n \over (n ! ) ^2} \ln
{\d _n (\t ) \over 1-x} , \cr && [\ln (\d _n (\t ))= 2 \psi (n+1) -
\psi (\t +n) - \psi (1-\t + n) ,\ (\t ) _n = {\G (\t +n) \over \G (\t
) } = \t (\t +1) \cdots (\t +n -1 )] ,\eea in order to exhibit the
singular logarithmic behavior of the $x$-integrand near the end point,
$x=1$.  The limiting formulas at $ x\to 1$ are given by \bea && G_1
(x)\to {\G (1-\t +\t ') \over \G (-\t ) \G (\t ') } + (1-x)^{\t '-\t }
{\G (\t - \t ') \over \G (\t ) \G (1-\t ') } , \ H_{1,2 } (1-x) \to 1
\cr && I(x) \to {\sin \pi (\t -\t ') \over \sin \pi \t '} (1-x)^{\t
'-\t }\bigg ({\G (\t - \t ') \over \G (\t ) \G ( 1- \t ') } \bigg ) ^2
,\cr && V_{11}\to -{ \pi \over 4 } (\cot \pi \t - \cot \pi \t ') , \
V_{22}\to \ud ( { \pi \over \sin \pi p \t } )^2 { \G (1-\t + \t ') \G
(1-\t ' + \t ) \over \G (1-\t ) \G (1-\t ') \G (\t ) \G (\t ') } , \cr
&& V_{12} \to -\pi e ^{i \pi p \t } ( { \pi \over 2 \sin \pi p \t }
)^2 (\cot \pi \t ' - \cot \pi \t ) { \G (1-\t + \t ') \G (1-\t ' + \t
) \over \G (1-\t ) \G (1-\t ') \G (\t ) \G (\t ') } .
\label{app.1G2}\eea Note that $ G_1 (x)$ and $ G_2 (x)$ differ by the
substitution, $\t \leftrightarrow \t '$.  While the above formulas are
specialized to the case, $ \t > \t '$, the results in the opposite
case, $ \t < \t ' $, are deduced by using the symmetry under the
substitution $\t \leftrightarrow \t '$.

\subsubsection{Trilinear Yukawa  coupling} 

The constraints on the four fermion correlator implied by the
conformal symmetry
can be used to infer an explicit formula for the trilinear Yukawa
couplings.  We consider the four fermion string amplitude discussed in
Subsection~\ref{sub21}.  In the limit $ x\to 1 $, the four point
correlator factorizes into the product of three point correlators as,
$ Z (x_i) < V_{-\t } (0) V_{\t } (x)V_{-\t ' } (1) V_{\t '} (X) >
\simeq < V_{\t } (x)V_{-\t ' } (1) V _{-\t +\t '} (z) > < V _{\t - \t
'} (z) V_{-\t } (0) V_{\t '} (X) >$, dominated by the $t$-channel
exchange of the mode with vertex operator, $V _{-\t +\t '} (z) $.
Retaining the leading contribution in the limit $ x\to 1 $ yields the
limiting formula for the four fermion string amplitude \bea && \cala '
\simeq - 2\pi g_s {\cals _1 \calT _1\over (-t ) } \prod _I \bigg [ \pi
\bigg ( 2 \vert \cot (\pi \t ^I ) - \cot (\pi \t ^ {'I } ) \vert \bigg
) ^\ud {\G (1-\t ^I +\t ^ {'I } ) \over \G (1-\t ^I ) \G (\t ^ {'I } )
} \bigg ] e ^{-S_{cl} (x }\vert _{x=1} .\eea To forbid in the
classical action factor the singular contribution from $ \o _{1-\t,
1-\t '} (x)$ at $ x\to 1 $, one must impose the restriction, $ 0= \bar
\dh X = c \o _{1-\t, 1-\t '} (x)$, which implies, $ 0=c = c_1 v_A +
c_2 v_B$.  Using the formulas for $c_{1,2}$ in Eq.~(\ref{eqcoffs})
reduces the classical action to the form \bea && \lim _{x\to 1} S
_{cl} (x) = \vert v_A \vert ^2 [ V_{11} + \vert {c_1\over c_2} \vert
^2 V_{22} -2 \Re (V_{12} {c_1\over c_2} ) ] \cr && = - {\pi \over 2}
\vert v_A \vert ^2 \vert \cot (\pi \t ) - \cot (\pi \t ') \vert = {1
\over 2\pi \a '} \vert 2 \pi (p_A L_A + d_A ) \vert ^2 {\sin (\pi (\t
-\t ') )\sin (\pi \t ) \over \sin (\pi \t ') } , \eea where the
coefficient of $ (2\pi \a' ) ^ {-1} $ in the last step above equals
twice the area of the triangle with base, $ 2\pi (p_A L_A + d_A ) $,
opposite angle $\pi ( \t - \t ') $ and adjacent angles $\pi \t , \ \pi
\t '$.  To illustrate in a concrete way the $t$-channel factorization
with respect to the exchange of a scalar boson mode, we consider the
$Z_3$ orbifold model discussed in Subsection~\ref{sub41}.  The
correlator and its pole factorized part, $ < q_1 q_2^\dagger
d_3^{c\dagger } d_4^ c > \simeq <q_2^\dagger d_3^{c\dagger } \bar H >
< \bar H ^\dagger q_1 d_4^ c > $, couple the conjugate localized modes
from the open string sectors, $ q= (a',b)=(b', a),\ d^c = (a, c)=(c',
a') , \ \bar H= (b', c)=(c', b)$, where the brane embedding of the
world sheet is given by the four-polygon with sides $ a', b, a',c'$.
The Dirac spinor matrix element may be simplified by using the Fierz
identity:
$ \cals _1= (\bar q_{2L} \g ^\mu q_{1L} ) (\bar d^c_{3L} \g _\mu
d^c_{4L} ) = 2 (\bar q_{2L} d_{3R} ) (\bar q_{1L} d_{4R} ) ^\star . $
The trace factor dependence on the CP factors can be put in the
factorized form , $ \calT_1 = Trace( \l_ {q _2^\dagger } \l_{d_3^ {c
\dagger } } \l _ {q _1 } \l_{d _4^{c} } ) = \calp \vert
Tr(\l_{q^\dagger } \l_{d^{c \dagger } } \l _{\bar H} ) \vert ^2 $,
where $\calp $ is a calculable coefficient that we do not attempt to
determine here.  The comparison with the $\bar H$ mode $t$-channel
exchange contribution to the amplitude, using the definition of the
trilinear Yukawa coupling, $ L_{EFF}= Trace ( \l _q \l _{d^ {c \dagger
} } \l _{\bar H } ) Y _{-\t, \t ' , \t -\t ^{'} } (\bar q_L d _R \bar
H ) + \ H. \ c. $, yields the final formula for the Yukawa coupling
constant \bea && Y _{-\t , \t ' , \t - \t ' } = \sqrt {2\pi g_s }
\sqrt { 2 \calp } \prod _I \bigg [ \pi [ 2 \vert \cot ( \pi \t ^I ) -
\cot (\pi \t ^{'I} ) \vert ] ^\ud {\G (1-\t ^I +\t ^{'I} ) \over \G
(1-\t ^I ) \G (\t ^{'I} ) } \bigg ] ^\ud \sum _{cl} e^{-\ud S _{cl} },
\cr && = (2\pi )^ {5/4} \sqrt g_s \sqrt {2\calp } \prod _I \bigg ( {
\G (\t ^{I} )\G (1- \t ^{'I} ) \G (1 -\t ^I + \t ^{'I} ) \over \G
(1-\t ^{I} )\G (\t ^{'I} ) \G (\t ^I - \t ^{'I} ) } \bigg ) ^{1\over
4} \cr && \times \sum _{p_A\in Z } \prod _I e^{ -{\vert 2 \pi (p_A L_A
+ d_A ) \vert ^2 \over 2 \pi \a ' } {\sin (\pi \vert \t ^{I}-\t ^{'I}
\vert ) \sin (\pi \t ^I) \over 2 \sin (\pi \t ^{'I} ) } } . \eea The
argument in the exponential identifies with the area of the target
space embedding given by the triangle with adjacent sides lying along
the branes $ a' $ and $ c'$.  The above result coincides with similar
ones quoted previously in Refs.~\cite{Cvetic:2003ch,cimmag04} except
for the constant normalization factor.  Besides the factor $ \sqrt 2
\calp $, the ratio of our Yukawa coupling constant to that obtained by
Cremades et al.,~\cite{cimmag04} amounts to the factor of $ (2\pi ) ^
4 $.  Identifying the prefactor of the exponential with the dependence of
the Yukawa coupling constant coming from the canonical normalization
of the kinetic energy terms in the action, $ Y _{-\t , \t ' , \t - \t
' } \propto \prod _I [K (-\t ^I) K (\t ^{'I} ) K (\t ^I - \t ^{'I} ) ]
^{-\ud } $, allows us to identify the K\"ahler potential for the
twisted modes as, $ K \propto ({\G (\t ^{I} ) \over \G (1-\t ^{I} ) }
) ^ {\ud } \ C ^\dagger _ {\t ^{I} } \ C_ {\t ^{I} } $.  This result
agrees with that obtained in Ref.~\cite{bertol05}.

The string amplitude factorization in the limit $x \to \infty $ can be
compared with the $u$-channel exchange pole contribution in order to
identify the multiplicative normalization constant $ C_\s $ in
Eq.~(\ref{eqsolz}). The limiting behavior of the string amplitude,
restricted for simplicity to the quantum partition function factor,
\bea && \cala '\simeq (2 \sin \pi \t ^I ) ^ \ud \bigg ({\sin (\pi \t ^
{'I }) \over \sin \pi (\t ^ {'I } -\t ^I ) } \bigg ) ^\ud {\G (\t ^
{'I }) \over \G (\t ^I) \G (\t ^ {'I } - \t ^I ) } {1\over (-u + \ud
\vert \t ^I -\t ^ {'I } \vert ) } .\eea The comparison with the
leading $u$-channel pole exchange contribution determines the
multiplicative normalization constant introduced in Eq.~(\ref{eqsolz})
as, $C _\s = \prod _I (2 \sin \pi \t ^I ) ^ \ud $.  One also deduces
the expected result for the string mass spectrum of modes with the
quantum numbers of vector bosons, $ \a ' M^2 = \ud \vert \t -\t '
\vert $.

\section{$ Z_3 $ orbifold models  with intersecting $D6$-branes}
\label{appz3}


We here present an encapsulated review of the construction due to
Blumenhagen et al.,~\cite{Blumenhagen:2001te} of the $Z_3$
orbifold-orientifold models with $D6$-branes.  The orbifold $ T^6 /(
Z_3 + Z_3 \O \calr ) $ uses factorisable 6-d tori symmetric under the
generator, $ \T = diag (\T^1, \T^2, \T^3),$ and the orientifold
symmetry, $\calr $, acting on the complexified orthogonal basis of
coordinate and spinor fields $ X^I, \ \psi ^I $, by the complex phase
rotations, $\T ^I = e^{ 2\pi v ^I} , \ [v ^I=({1\over 3}, {1\over 3}
,- {2\over 3} ) ]$ and the reflection about the imaginary axis, $\calr
= \calr _x :\ X ^I \to - \bar X^I$.  The orbifold Hodge numbers, $
h^{(2,1)} =0 ,\ h^{(1,1)} = 36 $, entail that all the complex
structure moduli are frozen, while the $36$ complex K\"ahler moduli
decompose into $9$ untwisted moduli, $ T_{I\bar J}$, and $27$ twisted
(blowing-up) moduli labeled by the fixed points.  The two inequivalent
solutions for the 2-d symmetric lattices are described by the pairs of
cycles, \bea && \bfA : \ e_1 = 1 , \ e_2 = e^{i\pi /3} , \ \bfB :\ e_1
= e^{-i\pi /6} , \ e_2 =e^{+ i\pi /6} , \eea having the same complex
structure moduli, $ U _{\bf A } = U _{\bf B } \equiv {e_2\over e_1} =
\ud + i{\sqrt 3 \over 2} $, but differing by the action of $\calr _x$
which acts as \bea && \bfA : \ e_1\to e_1 ,\ e_2\to e_1 -e_2 ,\ \bfB :
\ e_1\to e_2 ,\ e_2\to e_1. \eea In the sequel, we consider the
alternative solution for the lattice $ \bfB $, using the cycles, $e_1=
1, e_2= {1\over \sqrt 3} e^{i\pi /6} $, with the complex structure
moduli, $U_{\bf B}= {e_2 \over e_1} = \ud + i{ 1 \over 2 \sqrt 3} $,
and reflection symmetry realized as, $ e_1\to e_1 ,\ e_2\to e_1 -e_2
$.  The resulting $T^2$ tori lattice solutions, $ \bfA $ and $ \bfB $,
are described by the following data \bea && \bfA : \ U^A = {e_2 ^A
\over e_1 ^A} = \ud + i {\sqrt 3\over 2}, \ T^A = b + i{\sqrt 3 r^2
\over 2 }, \ [e_1 ^A = 1 ,\ e_2 ^A = \ud + i {\sqrt 3 \over 2} , \
g^A= \pmatrix{1& \ud \cr \ud & 1}] \cr && \bfB :\ U^B = {e_2 ^B \over
e_1 ^B} = \ud + i{1\over 2\sqrt 3 } ,\ T^B = b + i{r^2 \over 2\sqrt 3
}, [ e_1 ^B = 1,\ e_2 ^B = \ud + i {1 \over 2\sqrt 3 } , \ g^B=
\pmatrix{1& \ud \cr \ud & {1\over 3} }] \eea where we have displayed
the diagonal complex structure and K\"ahler  moduli as a function of
the free parameters, $r$ and $ b \simeq b +1 $, associated with the
overall radius and the NSNS field VEV; the complexified basis of
cycles, $ e_1^I = \Re (e_1 ^I ) + i \Im (e_1^I ) ,\ e_2^I =\Re ( e_2
^I) + i \Im (e_2 ^I) $, whose real and imaginary parts describe the
orthogonal components of the 2-d vielbein vectors; and the lattice
basis metric tensor in the matrix representation $ g_{ab} = \vec e _a
\cdot \vec e _b = \Re (e ^\star _a e_b )$.
The lattice basis decomposition of one-cycles, $[\Pi _a] = (n_a e _1 +
m _a e_2 ) , \ [n_a, m_a \in Z]$ defines $(n_a, m_a) $ as the 2-d
column vectors of wrapping numbers.


The orbifold and orientifold symmetries act from the left on the
column vector of wrapping numbers $(n_a, m_a) ^T $ and hence on the
right on the column vector of cycles $(e_1 ,\ e_2) ^T $, with the
matrix representatives for the $\bfA $ and $\bfB $ lattice solutions
given by \bea && \bfA : \ \T = \pmatrix{-1& -1 \cr 1 & 0} ,\ \T ^2 =
\pmatrix{0& 1 \cr -1 & -1} ,\cr && \O \calr = \pmatrix{1& 1 \cr 0 &
-1} ,\ \O \calr \T = \pmatrix{0& -1 \cr -1 & 0} ,\ \O \calr \T ^2 =
\pmatrix{-1& 0 \cr 1 & 1} \cr && \cr && \bfB : \ \T = \pmatrix{-2& -1
\cr 3 & 1} ,\ \T ^2 = \pmatrix{1& 1 \cr -3 & -2} ,\cr && \O \calr =
\pmatrix{1& 1 \cr 0 & -1} ,\ \O \calr \T = \pmatrix{1& 0 \cr -3 & -1}
,\ \O \calr \T ^2 = \pmatrix{-2& -1 \cr 3 & 2} .\eea These matrix
representations obey the operator identity, $ \O \calr \T ^g = \T
^{N-g} \O \calr .$ The wrapping numbers of the one-cycles composing
the orbits, $ [a], \ [a'] $ are obtained from those of the
representative element, $ a\sim (n_a, m_a) $, by left action with the
matrices $ \T ^g , \ \O \calr \T ^g $ appropriate to the $\bfA $ and
$\bfB $ lattices, \bea && \T ^g:\ {n _a \choose m_a } \to {n _{a_g}
\choose m _{a_g} } = \T ^g {n _a \choose m_a },\ \T ^g \O \calr :\ {n
_a \choose m_a } \to {n _{a '_g} \choose m _{a'_g} } = \O \calr \T ^g
{n _a \choose m_a } , \ [g=0,1,2] \label{eqtrfs} \eea where we use the
notational convention, $ a'_g = \O \calr \T ^g \ a \equiv (a_g)' =
(a') _{N-g} $.

 Our present conventions for the coordinate system conform to those of
Blumenhagen et al.,~\cite{Blumenhagen:2001te}. We use sideways tilted
$T^2_I$ tori with $O6$-plane lying along the real $ X_1$ coordinate
axis, and the relative $D6_{a}$-brane $O6$-plane angle evaluated by
means of the formula, $ \tan \phi _a ^I = {m _a ^I U^I_2 \over (n _a
^I + m _a ^IU^I_1 )} $.  No confusion should hopefully arise from the
fact that these conventions differ from those in the main text, where
we used upwards tilted tori, symmetric under the reflection, $\calr
_y: \ X ^I \to \bar X ^I $ with $O6$-plane also lying along the real
axis.  The two choices, $\calr _{x,y}$, of the reflection are related
by the modular transformation, $ U ^I\to -{1\over U ^I},\ n ^I\to
-m^I,\ m^I\to n^I $.

The 6-d orbifolds $ T^6/ ( Z_3 + Z_3 \O \calr ) $ are given by direct
products of the above 2-d tori solutions.  Taking the symmetry under
permutations of the three complex planes into account, one finds the
four inequivalent 6-d lattices: ${\bf AAA,\ AAB,\ ABB,\ BBB }$.  With
the reference $O6_0$-plane wrapped around the three-cycle, $ \pi
_{135}$, with wrapping numbers, $(1,0)^3$, the two other planes,
$O6_1,\ O6_2$, obtained by applying the half-angle rotations, $ O6_g =
\T ^ { - {g \over 2 } }\cdot O6_0$, form an equilateral triangle
whose sides are interchanged under $\T $.  The three-cycles are
represented as direct products of three one-cycles, $ e_{1,2}^I$ in $
T^2_{I } ,\ [I=1,2,3]$ as illustrated by the examples, $ \pi _{135}=
\pi _1\times \pi _3\times \pi _5 = e_1 ^1 \times e_1 ^2 \times e_1
^3,\ \pi _{235}= \pi _2\times \pi _3\times \pi _5 = e_2 ^1 \times e_1
^2 \times e_1 ^3$, with the intersection numbers, $ [\pi _ {2i-1}]
\cdot [\pi _{2i} ] =1,\ [i=1,2,3]$.  The total $O6$-plane RR charge is
compensated by including stacks of $D6$-branes wrapped around the
factorisable 3-cycles. We restrict consideration to setups of
$D6$-branes intersecting the $O6$-planes at the origin of the
coordinate system.  A compact representation for the wrapped cycles is
obtained by considering the orbifold invariant three-cycles belonging
to the sub-space of dimension $ b_3 = h^{(2,1)}+ 2= 2 $, with $b_3$
denoting the Betti number.  For the symplectic basis of invariant
three-cycles defined by \bea && \rho _1= \calp \pi _{135} ,\ \rho _2=
\calp \pi _{235} , \ [\calp = (1 + \T + \T ^2) ] \eea having the
intersection numbers, $\rho _i \cdot \rho _j = - \e _{ij} , \ [i, j
=1,2]$ where $\e _{12} = -\e _{21} =1$, one can define the
three-cycles effective wrapping numbers, $(\tilde Y_a, \tilde Z_a)$,
for the equivalence class, $[\Pi _a ]$, in terms of the decomposition
on invariant cycles, \bea && \Pi _a = \tilde Z_a \rho _1 + \tilde Y_a
\rho _2 = \sum _{g =0,1,2} \prod _I (n_{a_g} ^I e_1 ^I + m_{a_g} ^I
e_2 ^I ) . \eea The above equations may be used to express the
effective wrapping numbers as weighted averages of products of the
one-cycle wrapping numbers.  To adapt ourselves with the conventions
of Ref.~\cite{Blumenhagen:2001te}, we rather consider the
decomposition on effective wrapping numbers, $\Pi _a = Z_a \rho _1 +
Y_a \rho _2$, including the orientifold images, \bea && Z_a = {2\over
3} \sum _{g } \bigg ( \prod _I \tilde n ^I _{a_g} + \prod _I \tilde n
^I _{a '_g} \bigg ) ,\ Y_a = -{1\over 2} \sum _{g } \bigg ( \prod _I m
^I_{a_g} - \prod _I m ^I_{a'_g} \bigg ) ,\cr && [ \tilde n ^I _{a_g} =
(n ^I _{a_g}+m^I_{a_g} U_1^I) ,\ \tilde n ^I _{a '_g} = (n ^I _{a
'_g}+m^I_{a'_g} U_1^I)] , \eea where the wrapping numbers of the
mirror branes are calculated from Eqs.~(\ref{eqtrfs}).  The mirror
branes satisfy the relations, $Y_{a'} = -Y_a,\ Z_{a'} = Z_a $.  The
explicit formulas for $ (Y_a, Z_a)$ in terms of the one-cycles
wrapping numbers $ (n_a ^I, m_a^I)$ are listed below for the four
inequivalent 6-d lattices invariant under the $Z_3$ orbifold and
orientifold identifications.  \bea \bullet \quad {\bf AAA}:\ && Z_a=
\ud \bigg [ n_a ^1 (m_a^2 (n_a ^3 - m_a^3) + n_a ^2 (2 n_a ^3 +
m_a^3)) + m_a^1 (n_a ^2 (n_a ^3 - m_a^3) - m_a^2 (n_a ^3 + 2 m_a^3))
\bigg ] , \cr && Y_a=n_a ^2 (n_a ^3 m_a^1 + (n_a ^1 + m_a^1) m_a^3) +
m_a^2 (n_a ^3 m_a^1 + n_a ^1 (n_a ^3 + m_a^3)) \cr && \cr \bullet
\quad {\bf AAB}:\ && Z_a= \ud \bigg [ m_a^1 (n_a ^2 n_a ^3 - m_a^2
(n_a ^3 + m_a^3)) + n_a ^1 (n_a ^3 m_a^2 + n_a ^2 (2 n_a ^3 + m_a^3))
\bigg ] , \cr && Y_a= n_a ^2 (3 n_a ^3 m_a^1 + (n_a ^1 + 2 m_a^1)
m_a^3) + m_a^2 (m_a^1 (3 n_a ^3 + m_a^3) + n_a ^1 (3 n_a ^3 + 2
m_a^3)) \cr && \cr \bullet \quad {\bf ABB}:\ && Z_a={1\over 6} \bigg [
m_a^1 (3 n_a ^2 n_a ^3 - m_a^2 m_a^3) + n_a ^1 (3 n_a ^2 (2 n_a ^3 +
m_a^3) + m_a^2 (3 n_a ^3 + m_a^3)) \bigg ] , \cr && Y_a= 3 (n_a ^2 (3
n_a ^3 m_a^1 + (n_a ^1 + 2 m_a^1) m_a^3) + m_a^2 (n_a ^1 (n_a ^3 +
m_a^3) + m_a^1 (2 n_a ^3 + m_a^3))) \cr && \cr \bullet \quad {\bf BBB
}:\ && Z_a= {1\over 6} \bigg [ m_a^1 (n_a ^3 m_a^2 + n_a ^2 (3 n_a ^3
+ m_a^3)) + n_a ^1 (3 n_a ^2 (2 n_a ^3 + m_a^3) + m_a^2 (3 n_a ^3 +
m_a^3)) \bigg ] , \cr && Y_a= 9 n_a ^2 (n_a ^3 m_a^1 + (n_a ^1 +
m_a^1) m_a^3) + 3 m_a^2 (3 n_a ^1 (n_a ^3 + m_a^3) + m_a^1 (3 n_a ^3 +
2 m_a^3 )).
\label{appzy} \eea These formulas  reproduce the
results previously obtained by Blumenhagen et
al.,~\cite{Blumenhagen:2001te}, except for the single mismatch in the
formula of $Z_a$ for the lattice ${\bf AAB} $, where our above term, $
-\ud m _a^1m _a^2m _a^3$, is quoted in Eq.(A.2) of
Ref.~\cite{Blumenhagen:2001te} as, $- m _a^1m _a^2m _a^3$.

The $ N_\mu \ D6_ {\mu _g} $-brane stacks at generic angles relative
to the $O6$-planes, $\phi ^I_ {\mu _g} \ne k \pi / 3 ,\ [g=0, 1, 2 ;\
k =0, 1,2] $ realize on the 4-d world brane the gauge symmetry, $
\prod _\mu U(N_\mu ) $.  The states in the non-diagonal sectors, $ (a,
b _g) ,\ (a ', b _g),\ [a\ne b _g ]$ carry the bifundamental
representations, $I_{ab_g } (N_a, \bar N_b) + I_{a'b_g } (\bar N_a,
\bar N_b) ,\ [I_{ab} = \prod _I (n ^I_a m^I_b - n^I_b m^I_a)] $, which
we denote for convenience as, $\hat I_{ab _g} (\bar N_a, N_b) + \hat
I_{a 'b _g} (N_a, N_b) , \ [\hat I_{ab} = I_{ba} = -I_{ab} = - [\pi _a
] \cdot [\pi _b ] ] $.
From the multiplicities of the bifundamental representations, one
determines the chiral spectrum multiplicities (distinguished by the
suffix label $\chi $) by summing the intersection numbers
algebraically over the orbifold orbits.  The chiral spectrum, obtained
by summing over the distinct pairs of $D6_a/D6_b$-brane images in the
equivalence classes, $ ([a], [b]) $, without distinguishing between
the sectors related by the orbifold group, $ (a, a_{g}) \sim (a_h,
a_{gh}), \ [ g, h \in Z_3]$ only depends on the effective wrapping
numbers \bea && (\bar N_a , N_b):\ \hat I ^\chi _{ab} \equiv - \sum
_{g} I_{a b_g} = -(Y_a Z_b - Y_b Z_a) , \ ( N_a , N_b):\ \hat I
^\chi_{a'b} = - \sum _{g} I_{a' b_g} = (Y_a Z_b +Y_b Z_a ). \eea
The diagonal sectors, $(a,a _g)$, generate the adjoint representation
 of the gauge group $ U(N_a)$ with multiplicity $ I_{a a_g} {\bf Adj
 _a} $. The diagonal sector, $(a',a_g)$, includes the subset of $
 {I_{a 'a_g} \over \prod _I n ^I_{a_g} } $ two-index antisymmetric
 representations, $ {\bf A _a} $, with the remaining subset $ \ud
 I_{a' a_g} (1-{1\over \prod _I n ^I _{a_g} }) $ realizing symmetric
 and antisymmetric representations, $ {\bf A _a + S_a } $.
The resulting chiral spectrum of antisymmetric and symmetric
representations and the complete spectrum of adjoint representations
can be expressed by the following compact
formulas~\cite{Blumenhagen:2001te} \bea && {\bf A _a} :\ \hat I ^{\chi
, {\bf A} } _{a'a}= - \sum _{g} I_{a' a_g} ( \ud (1 + {1\over \prod _I
n ^I _{a_g} }) = Y_a + Y_a (Z_a-\ud ) = Y_a ( Z_a + \ud ) ,\cr && {\bf
S _a} :\ \hat I ^ {\chi, {\bf S } } _{a'a} = - \sum _{g} I_{a' a_g} (
\ud (1 - {1\over \prod _I n ^I _{a_g} })= Y_a (Z_a - \ud ) , \cr &&
{\bf Adj _a} :\ \hat I ^{Adj} _{a} =- \ud \sum _{g} I_{a a_g} = 3
^{p_\bfB } \prod _{I=1}^3 \vert L_a ^{I} \vert ^2 , \cr && [ \vert L_a
^{I} \vert _{\bfA } = (n_a ^{I2} + n_a^{I} m_a ^{I} + m_a ^{I2}) ^\ud
,\ \vert L_a ^{I} \vert _{\bfB } = (n_a ^{I2} + n_a^{I} m_a ^{I}
+{1\over 3} m_a ^{I2}) ^\ud ] \eea where $ p_\bfB $ denotes the number
of $\bfB $ tori and $\vert L_\mu ^{I} \vert ^2 = (n_\mu ^I \ m_\mu
^I)\cdot g ^I \cdot (n_a\mu ^I \ m_\mu ^I) ^T $ denotes the squared
length squared length of the $T^2_I$ one-cycles wrapped by the $D6_\mu
$-brane.  For $N_a =1$, one must set $I ^{\chi , {\bf A} } _{a'a} =0$,
reflecting the absence of the antisymmetric representation in this
case.  Note that the multiplicities are related to the cycle
intersections as, $ \hat I^\chi _{ab} = - [\pi _a ] \cdot [\pi _b ], \
\hat I^\chi _{a'b} = - [\pi _{a'} ]\cdot [\pi _b ], \ \hat I^{\chi (A,
S) } _{a'a} = -\ud ( [\pi _{a'} ] \cdot [\pi _a ]\pm 2 [\pi _{a'}
]\cdot [\pi _{O6} ] ) .$

The RR tadpole cancellation conditions, expressing the net vanishing
RR charges carried by the orbifold orbits of $D6_\mu $-branes and the
$O6_g$-planes, reduce to the unique relation: $ \sum _\mu N_\mu Z _\mu
=2 $.  The chiral and mixed gauge and gravitational anomalies
contributed by the massless fermions present in the above
representations are given by \bea && \cala _a (G_a^3) \equiv \sum
_{b\ne a} N_b (-\hat I ^{\chi } _{ab} + \hat I^{\chi } _{a'b}) +
(N_a-4) \hat I ^{\chi ,\bfA } _a + (N_a+4) \hat I^{\chi ,\bfS }_a \cr
&& = \sum _{b\ne a} 2N_b Y_a Z_b + (N_a-4) Y_a + 2 N_a Y_a (Z_a-\ud ),
\cr && \cala ^G _{a b} (U(1) _a \times G_b^2 ) \equiv N_a (-\hat
I^{\chi }_{ab} + \hat I ^{\chi }_{a'b}) =2 N_a Y_a Z_b \d _{ab},\
\cala _{a b} (U(1) _a \times U(1)_b^2 ) \equiv 2 N_a N_b Y_a Z_b , \cr
&& \cala _{grav} (U(1) _a \times R ^2 ) \equiv \sum _{b\ne a} [ N_a
N_b (-\hat I ^{\chi }_{ab} + \hat I ^{\chi }_{a'b})] + N_a (N_a-1)
\hat I^ {\chi ,\bfA } _a + N_a (N_a+1) \hat I^{\chi ,\bfS } _a = 3 N_a
Y_a . \cr && \eea

The chiral gauge anomalies are found to vanish automatically, $ \cala
_a (G_a^3)=0 $, once the RR tadpole cancellation condition is
satisfied.  The vector space of $ U(1)_\mu $ factors decomposes into
two orthogonal sub-spaces of anomaly free and anomalous abelian
symmetries, corresponding to linear combinations of $U(1)_\mu $ with
vanishing and non vanishing mixed gauge and gravitational anomalies
from the massless fermions.  The latter anomalies are cancelled, as in
the Green-Schwarz type mechanism, by including the 4-d anomalous
counterterm action descending from the 10-d Chern-Simons topological
action.  The dimensional reduction on $T^6/Z_3$ with $D6_\mu $-branes
wrapped around the three-cycles $\Pi _\mu $, produces 4-d couplings
between the gauge and gravitational fields and the dual pair of RR
scalar fields, $C_2 = \int _{ [\rho _1 ]} C_5 ,\ C_0 =\int _{ [\rho _2
]} C_3 ,\ [ d C_0 = \star d C_2 ] $ with coupling constants determined
by the decomposition of $\Pi _\mu $ on the basis of dual cycles, $
\rho _1,\ \rho _2$.  The resulting anomalous counterterm action \bea
&& \d I =\int _x \bigg [ \sum _\mu N_\mu Y_\mu C_2 \wedge F_\mu + \sum
_\nu N_\nu Z_\nu C_0 Tr (F_\nu \wedge F_\nu ) + {3\over 2} C_0 \wedge
Tr(R^2) \bigg ] , \eea with $F_\mu $ and $ R$ denoting the gauge field
strength and curvature tensor two-forms, is seen to induce the tree
level $ C_{0}\ - \ C_{2}$ exchange contributions necessary to cancel
the abelian anomalies, $\cala ^G _{a b},\ \cala _{a b} $ and $\cala
_{grav} $.  The unique anomalous $U(1)_X$ factor here is described by
the charge and gauge field strength, $ Q_X= \sum _\mu N_\mu Y_\mu
Q_\mu$ and $ F_{\rho \s } ^{X} = \sum _ \mu N_\mu Y_\mu F_{\rho \s }
^\mu$.  Finally, the uncancelled tadpole for the dilaton field, $\Phi
$, arises through the $D6$-branes scalar potential, $ V(\Phi ) = e ^{-
\Phi } (\sum _\mu N_\mu \prod _I \vert L ^I_\mu \vert -2 )$.

\section{Baryon number violation in gauge unified theories}
\label{bnvnd}

The baryon number violating processes arise in gauge unified field
theories~\cite{langarev81} from the tree level exchange of gauge and
scalar bosons gaining mass through the unified gauge symmetry
breaking.  The massive gauge bosons carrying leptoquark quantum
numbers with respect to the Standard Model group include the 
modes, $ [{X \choose Y} + {\bar Y
\choose \bar X}] \sim [(\bar 3 , 2) _{5\over 6} + (3 , \bar 2)
_{-{5\over 6}} ]$ in $SU(5)$ and $SO(10)$, and $[{X '\choose Y'} +
{\bar Y '\choose \bar X'}] \sim [(3,2) _{{1\over 6}} + (\bar 3, \bar
2) _{-{1\over 6}} ] , \ [X_s + \bar X_s] \sim [(3,1) _{{2\over 3}} +
(\bar 3, 1) _{-{2\over 3}} ] $ in   $SO(10)$.  The tree level exchange of these modes
contribute the effective Lagrangian
\bea && L_{EFF} = {g_{X,Y} ^2 \over 2
M_{X,Y} ^2} (2 O_{e_L^c} +O_{e_R^c} +O_{\nu _R^c} ) + {g_{X',Y'} ^2
\over 2 M_{X',Y'} ^2} (2 O_{\nu _L^c} +O_{e_R^c} +O_{\nu _R^c} ) + \
H.\ c.  \cr && [O_{e_L^c } = \e ^{\a \b \g } (\bar u_{j\g L}^c \g ^\mu
u_{j\b L } ) (\bar e_{iL}^c \g _\mu d_{i\a L} ) ,\ O_{e_R^c}= \e ^{\a
\b \g } (\bar u_{j \g L}^c \g _\mu u_{j\b L} ) (\bar e_{iR}^c \g ^\mu
d_{i\a R} ) ,\cr && O_{\nu _L^c}= - \e ^{\a \b \g } (\bar u_{j\g R}^c
\g _\mu d_{j\b R} ) (\bar \nu ^c _{iL} \g ^\mu d_{i\a L} ) ,\ O_{\nu
_R^c}= - \e ^{\a \b \g } (\bar u_{j\g L}^c \g _\mu d_{j\b L} ) (\bar
\nu ^c _{iR} \g ^\mu d_{i\a R} ) ]
\label{eqndop}\eea which initiates the quark and lepton subprocesses,
$u+ u\to d^c +e^+ ,\ \ u+ u\to u^c +e^+ $.  The change from gauge to
mass bases uses the transformations of quark fields, $ (q _L) _{
gauge} = V_L ^{q \dagger } (q _L) _{mass} , \ (q ^c _L) _{ gauge} =
V_R ^{q T } (q ^c _L) _{mass} ,\ (q _R) _{ gauge} = V_R ^{q \dagger }
(q _R) _{mass} $, such that the diagonalization of the quark mass
matrices, $ \l ^q _{ij} q_i q ^c _j , \ (\l^q )_{diag} = V \l V ^
\dagger ,\ [q=u,d] $ introduces the CKM flavor mixing matrix, $V = V_L
^u V_L ^{d \dagger } $.  In the minimal $SU(5)$ unification case,
assuming the mass generation from only $5,\ \bar 5$ Higgs boson
multiplets, the flavor mixing transformations,
$ d '_i =V _{ij} d_j,\ u_i ^c \to e ^{-i\phi _i} u_i ^c , \ e_i ^c \to
e_i ^ {'c } = V _{ij} e_j ^c , $ yield the simplified effective
Lagrangian for the first generation of quarks and leptons, \bea &&
L_{EFF} = {g_{X} ^2 e ^{i\phi _u} \over 2 M_X ^2} [ (1 + \vert V_{ud}
\vert ^2 ) O_{e_L^c} + O_{e_R^c} + V_{ud} O_{\nu _R^c} ] , \eea where
$ O_{e_L^c} ,\ O_{e_R^c} $ arise from the couplings, $ 10 _i\cdot {10}
^\dagger _i\cdot 10 _j\cdot {10} ^\dagger _j $ and $ 10 _i \cdot {10}
^ \dagger _i\cdot \bar 5 _j \cdot \bar 5 ^ \dagger _j $, which
initiate the proton decay processes, $ p \to \pi ^0 + e ^+ _L $ and $
p \to \pi ^0 + e ^+ _R $.  The current experimental bound, $ \tau
(p\to \pi ^0+e^+) > 5. \ 10 ^{33} $ yrs is to be compared with the
predicted proton partial decay width \bea && \G (p\to \pi ^0+e^+) =
8\pi C _\pi \kappa _ 3 ^2 A_R ^2 \vert \b _p \vert ^2 {\a _X ^2\over
M_X^4} [1 + (1 + \cos ^2\t _C ) ^2 ] \cr && \simeq {1\over 5.\ 10^{36}
\ yrs } ({M_X \over 3.\ 10^{16} \ GeV } )^{-4} ({\b _p \over -0.015 \
GeV ^3 } ) ^{2} ,\cr && [C_\pi \simeq {m_p \over 32 \pi f_\pi ^2 } ,\
\b _p u(P) = \e_{\a \b \g } <0\vert (\bar d^c _{\a L} u_{\b R} ) \bar
u ^c _{\g R} \vert P > ,\cr && \kappa _ 3 = 1+ {m_p \over m_A} (D+F)
\simeq 2 ,\ A_R \simeq 3.6 ,\ f_\pi \simeq 139 \ MeV] \eea where we
use self-explanatory standard notations.  
The comparison with the experimental   lower limit on  $M_X$   from  
the nucleon decay rates  has long been known to favor the conventional
supersymmetric version of grand unification over the non-supersymmetric one.  However,
new possibilities  have been opened  by  the current
studies~\cite{nath06},    using   the
alternative schemes for the Yukawa coupling constant unification and
quark and lepton flavor mixing effects, especially  those involving
orbifold   compactification of  the extra  space dimensions   to realize
grand unification models~\cite{hallnomura02,buch04,wiesenfeldt04}, or  TeV
scale  models~\cite{appelquist01} with variant selection rules
on the $B, \ L$ number non-conservation. 

In supersymmetric grand unification, additional baryon number
violating contributions arise from tree level exchange of color
triplet matter fermions between pairs of matter fermions and
sfermions.  The largest contributions arise from the $ B-L$ conserving
F term operators, $QQQL ,\ U^cD^c U^c E^c$.
Accounting for the quark flavor mixing, leads to several quartic
couplings between fermions and sfermions of which a few illustrative
examples are  \bea && L_{EFF}= {1\over M_X} \bigg [ {\kappa ^{(1)}
_{ijkl}} \e ^{\a \b \g } [(Q_{i\a } Q_{j\b })( Q_{k \g } L_l ) ]_F +
{\kappa ^{(2)} _{ijkl} } \e ^{\a \b \g } [U^c _{i\a } E^c_{j } U^c_{k
\b } D^c_ {l \g } ]_F \bigg ] +\ H.\ c. \cr && = {1\over M_X} \bigg [
{ \kappa ^{(1)} _{ijkl} } \e ^{\a \b \g } [(u_{i\a } d _{j\b} - d _
{i\a } u_{j\b}) (\tilde u_{k \g } \tilde e_l - \tilde d_{k \g } \tilde
\nu _l ) ] + {\kappa ^{(2)} _{ijkl} } \e ^{\a \b \g } [ (u^c _{i\a }
e^c_{j } ) (\tilde u ^c _{k \b } \tilde d^c_ { l \g } ) +\cdots ]\bigg
] +\ H.\ c.  \eea 

For minimal $SU(5)$, the trilinear superpotential coupling the matter
supermultiplets to the color triplet Higgs supermultiplets, $ H_c,\
\bar H_c$, is described in terms of the Yukawa coupling constants of 
quarks and leptons, $\l ^{u,d,e}$, as, $ W= \l _{ij} ^u ( \ud Q_iQ_j + u_i ^c
e_j^c ) H_c + (\l _{ij} ^{d,e} Q_iL_j + \l _{ij} ^{d,e} u^c_id^c_j )
\bar H_c ,$ where the choice of one or the other of the down quarks
and charged leptons Yukawa coupling constants, $ \l _{ij} ^{d,e} $, reflects on the
ambiguity in  describing the unification of Yukawa interactions at the scale
$M_X$,  and  hence of  the  quark and lepton flavor mixing.  
For instance, the choice of the
couplings, $ ( \l _{ij} ^{e} Q_iL_j + \l _{ij} ^{d} u^c_id^c_j ) \bar
H_c ,$ yields the effective coupling constants for the dimension
$\cddd =5$   operators, 
\bea && {\kappa
^{(1)} _{ijkl} \over M_X} = \d _{ij} { \l_ i ^u \l_ k ^e \over 2 M
_{H_c} } e ^{i\phi _i} V ^\star _ {kl},\ {\kappa ^{(2)} _{ijkl} \over
M_X} = {\l_ i ^u \l_ l ^d \over M_{H_c} } e ^{-i\phi _k } V _{ij}
. \eea Analogous results obtain in the flipped $SU(5)\times U(1)
_{fl}$ case.  The four
fermion amplitudes are obtained through the one-loop electroweak
gaugino dressing of the fermion-sfermion couplings using the Yukawa
gauge couplings, $ L_{EFF}= g_2 [ \tilde u _L ^\star (\tilde w ^+ d) +
\tilde d _L ^\star (\tilde w ^- u ) + \cdots ] + \ H.\ c. $, and the
analogous electroweak higgsino dressing.  Since these calculations are
standard ones~\cite{hisano92}, they will not be reviewed here.

The current experimental bound, $ \tau
(p\to K^+ + \bar \nu _l) > 1.6 \ \times 10 ^{33} \ yrs $, should be
compared with the partial decay width for the dominant decay mode \bea
&& \G (p\to K^+ + \bar \nu _l) = C _K A_R ^{'2} \vert \b _p \vert ^2
({\a _X \cos \t _C \over \pi M_X^2})^2 \vert \kappa ^{(1)} _{221l}
\sin \t _C \kappa _3 [f(c,e_l) +f(c,d ') ] \cr && + \kappa ^{(1)}
_{112l} \cos \t _C \kappa _2 [f(c,e_l) +f(u,d ') ] \vert ^2 \cr &&
\simeq {1 \over ( {1/ 3 } \ - \ 3 )\ \times 10^{34} \ yrs } ({M_{H_c}
\over 2.\ 10^{16} \ GeV } )^{-2 } ({\b _p \over -0.015 \ GeV ^3 }
)^{2} , \cr && [f (u,d) = {\tilde w ^2 \tilde u ^2 \over (\tilde u^2 -
\tilde d^2)(\tilde u^2 -\tilde w ^2) } \log {\tilde u^2 \over \tilde w
^2 } + (\tilde u\leftrightarrow \tilde d ) ,\ A'_R \simeq 10.5,\ \cr
&& \kappa _ 2 = 1+ {m_p \over 3 m_A} (D+ 3F) \simeq 1.6 ,\ C_K = {m_p
(1-{m_K^2\over m_A^2} )^2 \over 32 \pi f_\pi ^2 } ]
\label{eqndke} \eea 
 where we use self-explanatory standard notations.    The large
hierarchy in the mass ratio, $ M_ {H_c} /m_Z$, needed to suppress the
$\cddd =5 $ operators can be explained by several
mechanisms~\cite{raby04,murapierce01,babubarr02}.  The currently
favoured proposals use the realization of Higgs bosons as Goldstone
bosons~\cite{csaki96}, discrete global
symmetries  not  commuting with  the  unified  gauge 
group~\cite{goodman,witten02,dineshad02}  or the flavor
physics~\cite{murakap94}.   
The contributions to nucleon decay from the $\cddd =5$ dangerous operators
are  strongly to the  threshold corrections to the gauge coupling
constants~\cite{raby04}.   Thus, in the minimal $SU(5)$ unification  with 
the  choice of couplings, $ ( \l _{ij} ^{d}
Q_iL_j + \l _{ij} ^{d} u^c_id^c_j ) \bar H_c ,$  the experimental limit from the nucleon
lifetime yields the upper bound, $ M_{H_c} > 2.\ 10 ^{17} $ GeV, which
clashes with the lower bound deduced from the high energy
extrapolation of the Standard Model gauge coupling
constants~\cite{murapierce01}.    By  contrast, the alternative
choice   for the couplings, $ ( \l _{ij} ^{d} Q_iL_j + \l _{ij} ^{e}
u^c_id^c_j ) \bar H_c  ,$  suitably adjusted to the fermion masses, suppresses
the predictions for the nucleon decay rate to values below the
experimental upper bound~\cite{wiesenfeldt04}.

The interest in the higher order baryon number violation, initiated by
dangerous operators of dimension $\cddd \geq 7$, is motivated by the
sensitivity of the nucleon decay processes to mass scales
significantly lower than those of grand unification and by the variant
selection rules.  The $ \cddd =7$ operators involve four matter
fermions and gauge or Higgs doublet bosons, obeying the selection
rules, $ \D B = - \D L = -1$.  Two illustrative examples are given by
the operators, $ (\bar q _R^c q_L)(\bar l_L d_R) \phi ^ \dagger $ and
$ (\bar d _L^c d_R)(\bar e^c _R q_L) \phi ^ \dagger $, with $\phi $
denoting the Higgs boson electroweak doublet.

The baryon number violating six quark operators of dimension $ \cddd
=9$ are especially interesting in view of their impact on the $ N-\bar
N$ oscillation and two nucleon desintegration processes.
The local operators in the quark fields of fixed chirality and color
quantum numbers, $\chi $ and $\a $, enter in three types involving two
pairs of spin-flavor and color structures \bea && (O_1) _{\chi _1,
\chi _2, \chi _3 } = (u^T_{ \a _1 \chi _1} C ^\dagger u _{\a _2} \chi
_1, ) (d^T_{\a _3 \chi _2} C ^\dagger d_{ \a _4 \chi _2} ) (d^T_{\a _5
\chi _3 } C ^\dagger d _{\a _6 \chi _3 } ) T^S _{ \a _1 \a _2 \a _3 \a
_4 \a _5 \a _6 } , \cr && (O_{[2, 3]} ) _{\chi _1, \chi _2, \chi _3 }
= (u^T_{\a _1 \chi _1 } C ^\dagger d _{\a _2 \chi _1 } ) (u^T_{\a _3
\chi _2 } C ^\dagger d _{ \a _4 \chi _2 } ) (d ^T_{ \a _5 \chi _3 } C
^\dagger d _{ \a _6 \chi _3} ) T^{[S,A]} _{ \a _1 \a _2 \a _3 \a _4 \a
_5 \a _6 },\cr && \bigg [T^S _{ \a _1 \a _2 \a _3 \a _4 \a _5 \a _6 }
= [ \e _{\a _1 \a _3 \a _5} \e _{\a _2 \a _4 \a _6} + (\a _5
\leftrightarrow \a _6 ) ] + [ \a _1 \leftrightarrow \a _2 ] ,\cr &&
T^A_{ \a _1 \a _2 \a _3 \a _4 \a _5 \a _6 } = \e _{\a _1 \a _2 \a _5}
\e _{\a _3 \a _4 \a _6} + \a _5 \leftrightarrow \a _6 \bigg ] \eea
where $\chi _1, \ \chi _2,\ \chi _3 \in [ L,R ] $, $C$ denotes the 4-d
charge conjugation matrix and the parentheses enclose the pair
contractions of the anticommuting Weyl spinor fields with respect to
the $SL(2,C)$ spin group indices.
The corresponding representation using the Dirac spinor fields is
given by, $ (O_1) _{\chi _1, \chi _2, \chi _3 } = - (\bar u^c _{ -
\chi _1} u _{\chi _1} )(\bar d^c _{ - \chi _2} d_{\chi _2} ) (\bar d^c
_{ - \chi _3} d_{\chi _3} ) T^S,\ (O_{[2,3]} ) _{\chi _1, \chi _2,
\chi _3 } = - (\bar u^c _{ - \chi _1} d _{\chi _1} )(\bar u^c _{ -
\chi _2} d_{\chi _2} ) (\bar d^c _{ - \chi _3} d_{\chi _3} )
T^{[S,A]},$ where $ - \chi $ denotes the opposite chirality to $\chi
$, and we have omitted the color indices for notational convenience.
In the electroweak symmetry limit, the allowed
contributions~\cite{rao82} arise for the list of six operators: $ (O_1
)_{RRR} , \ (O_2)_{RRR} , \ (O_3 )_{RRR} , \ (O_3 )_{LRR} , \ (O_3
)_{LLR} , \ (O_1 )_{LLR} - (O_2)_ {LLR} . $

The current experimental observability limit on the $N-\bar N$
oscillation parameter, $ \G _{\D B} \equiv G _{N\bar N } = \tau ^{-1}
_{osc} \leq 10^{-7.5} \ \text{sec} ^{-1} = 10^{-31.5} $ GeV, is to be
compared with the approximate theoretical prediction, $ \G _{\D B} =
\d m \simeq g_{EFF} \vert \psi _N (0) \vert ^4 $, where the model
dependent coefficient $g_{EFF} $  is defined through the effective
Lagrangian as, $ L_{EFF} = g_{EFF} O_ {\cddd =9} $, and the quark wave
function of the nucleon  is given by the estimate, $\vert \psi _N (0)
\vert ^4 = < \bar N \vert O_ {\cddd =9} \vert N> \simeq 10^{-5} \
\text{GeV} ^{6} $.

The simplest case where a tree level contribution to the $ \D (B-L)
=-2 $ processes takes place in the left-right symmetric gauge
theory~\cite{mohasenja} with gauge group, $G_{3221} = SU(3)_c \times
SU(2)_L\times SU(2)_R \times U(1)_{B-L} $, by including scalar
multiplets with color triplet diquark and dilepton quantum numbers, $
\D _L ^q \sim (3,3, 1) _{2/3} , \ \D _R ^q \sim (3,1, 3) _{2/3} , \ \D
_L ^l \sim (1,3, 1) _{-2} , \ \D _R ^l \sim (1,1, 3) _{-2} $, coupled
through the renormalizable interaction superpotential, $ W= q^c q^c \D
_R ^q + l^c l^c \D _R ^l + (\D _R^q )^3 \D _R^l$.  The prediction for
the coefficient, $ g_{EFF} \sim <\D _R ^l > / (m^2 _{\D _R ^q} ) ^3, $
along with the natural assumption that the spontaneous breaking mass
scale of the $ U(1)_{B-L}$ symmetry group and the color triplet scalar
mass are of same magnitude, $ v_{BL} = <\D _R ^{l0} > \simeq m _{\D _R
^q} $, yields the estimate for the oscillation rate, $ \G _{\D B} \sim
v_{BL}^{-5} \vert \psi _N (0) \vert ^4 ,\ [v_{BL} = <\D _R ^{l0} >
\simeq m _{\D _R ^q} ] .$ However, the resulting experimental bound, $
m _{\D _R ^q} \sim v_{BL} \geq 10^5 $ GeV, clashes with the range of
values, $v_{BL} \sim 10^{11}\ - \ 10^{14}) $ GeV, generating an
acceptable see-saw mechanism for neutrino masses.

More favorable conditions are offered in the supersymmetric
version~\cite{babudb2,dutta06} of the Pati-Salam gauge
theory~\cite{patisalam74} with gauge group, $G_{422} = SU(4) \times
SU(2)_L\times SU(2)_R $, by exploiting the possibility to compensate a
large mass scale $ v_{BL}$ by a suppressed loop induced contribution
to the $\cddd =10$ operators.  This is illustrated by the model in
Ref.~\cite{dutta06} which includes the matter and Higgs boson chiral
supermultiplets, $ f = (4,2,1) , \ f^c = ( \bar 4,1,2) ,\ \phi _1 \sim
(1, 2,2) , \ \phi _{15} \sim (15, 2,2) ,\ \D ^c \sim ({10 } , 1, 3) ,
\ \bar \D ^c \sim ( \overline {10 } , 1, 3 )$, interacting through the
superpotential, $ W= f f^c \phi _1 + f f^c \phi _{15} + f^c f^c \D ^c
+ \D ^c \bar \D ^c + ( \D ^c \bar \D ^c) ^2 + \phi _1 \phi _1 + \phi
_{15} \phi _{15}+ \phi _1 \D ^c \bar \D ^c \phi _{15} $. The singlet
modes are omitted for simplicity.  The see-saw mechanism for the
neutrino mass requires the VEV, $ v_{BL} = O(10 ^{11} ) $ GeV, whereas
the loop induced contribution to the coefficient, $ g_{EFF} \sim
v_{BL} ^{-2} v^{-3} $, obtained by assigning the mass value, $ M_{\D
^c} \sim 1 $ TeV, leads to an $N-\bar N$ oscillation rate, $ \G _{\D
B} \sim v_{BL} ^{-2} m_W^{-3}\vert \psi _N (0) \vert ^4 $, which is
compatible with the experimental observability limit.





\newpage 

\squeezetable
\begin{table}  
\begin{center}
\caption{\it Matter content  for the  $Z_3$ orbifold  model with gauge
group $SU(3) \times SU(2) \times U(1)_Y \times U(1) _{B-L} .$} 
\vskip 0.5cm
\begin{tabular}{|c |cccccc|cc |}   
\hline  Mode & $ q $ & $ u^c $ & $ d^c $ & $ l $ & $ e^c $& $ \nu ^c $
& $K_a$ & $K_b$ \\ &&&&&&&& \\ \hline &&&&&&&& \\ Sector & $ (a',b) $
& $ (a',a) $ & $ ( a,c) $ & $ (b,c) $ & $ (b',b) $ &$(c,c') $
& $(a,a)$ & $(b,b)$ \\ &&&&&&&& \\ Irrep & $ 3(3,2)_{1,1,0} $ & $
3(\bar 3,1)_{2,0,0} $ & $ 3(\bar 3,1)_{-1,0,1}$ & $ 3(1,\bar
2)_{0,-1,1}$ & $ 3(1,1) _{0,2,0}$& $ 3(1,1) _{0,0,-2}$
& $ I_{a}^{Adj} (8,1) $ & $ I_{b}^{Adj} (1,3) $ \\ \hline
\end{tabular}
\label{tablesmz3} 
\end{center}
\end{table} 
\vskip 0.5 cm

\squeezetable
\begin{table} 
\begin{center}
\caption{\it Solutions for the $SU(5)$ unified $Z_3$ orbifold model in
the lattices ${\bf A A A }$ and ${\bf B B B}$.  The column entries
give in succession the solution number, the  wrapping
numbers  for the three-cycles  $ a,\ c $ 
in the order, $ [(n_\mu ^1, m_\mu ^1), (n_\mu ^2, m_\mu ^2),
(n_\mu ^3, m_\mu ^3)] $, and the  branes intersection numbers 
$ \hat I_ {a'a_g } \ \hat I_{ac_g} \ \hat I_{cc'_g} \ \hat I_{a'c_g} $
in  the  four relevant open string sectors, $ (a' ,a_g ),\  (a,c_g),\  (c,c'_g),\
(a',c_g) $,  with the last three columns     corresponding to   
the elements  of  the orbifold group orbit, $ g=0, 1, 2$.}
\vskip 0.5cm
\begin{tabular}{|c |c |c |c  |c  |c |}   
\hline Solution & $ (n_a^I, m_a ^I) $ & $ (n_c^I, m_c ^I) $ & $ \hat I_
{a'a_0} \ \hat I_{ac_0} \ \hat I_{cc'_0} \ \hat I_{a'c_0} $ & $ \hat
I_ {a'a_1}\ \hat I_{ac_1}\ \hat I_{cc'_1} \ \hat I_ {a'c_1} $& $\hat
I_{a'a_2} \ \hat I_{ac_2} \ \hat I_ {cc'_2} \ \hat I_{a'c_2} $ \\
\hline  ${\bf A A A }$ &&&&& \\  I & $-3 2-1
0-1 1 $ & $-1-2-1 0 0 1 $ & $ 0 0 0 0 $ &$ 0 3 3 7 $ & $ 3 0 0-7 $ \\
II & $-3 2 0-1 0 1 $ & $ \qquad \qquad .. \qquad \qquad $ & $ 8 0 0 0
$ &$-5 3 3 0 $ &$ 0 0 0 0 $ \\ III & $ 0-1 0-1 3-2 $ & $ \qquad \qquad
.. \qquad \qquad $ & $ 8-3 0-3 $ &$-5 6 3 0 $ &$ 0 0 0 3 $ \\ IV & $ 0
1 0 1 3-2 $ & $ \qquad \qquad .. \qquad \qquad $ & $ 8-3 0-3 $ &$-5 6
3 0 $ &$ 0 0 0 3 $ \\ V & $ 0 1 1 0 2 1 $ & $ \qquad \qquad .. \qquad
\qquad $ & $ 0 0 0 0 $ &$ 3-3 3-2 $ &$ 0 6 0 2 $ \\ VI& $-3 2-1 0-1 1
$ & $-1-2 0-1 1 0 $ & $ 0 8 0 0 $ &$ 0 0 3 0 $ &$ 3-5 0 0 $ \\ VII &
$-3 2 0-1 0 1 $ & $ \qquad \qquad .. \qquad \qquad $ & $ 8 0 0 0 $
&$-5 3 3 0 $ &$ 0 0 0 0 $ \\ VIII & $-3 2 1-1 1 0 $ & $ \qquad \qquad
.. \qquad \qquad $ & $ 0 0 0 0 $ &$ 0 3 3 7 $ &$ 3 0 0-7 $ \\ IX &
$-2-1 0-1 1 0 $ & $ \qquad \qquad .. \qquad \qquad $ & $ 0 0 0 0 $ &$
3-5 3 0 $ &$ 0 8 0 0 $ \\ X & $-1 3 1 0 1 0 $ & $ \qquad \qquad
.. \qquad \qquad $ & $ 0 0 0 0 $ &$ 8 0 3 0 $ &$-5 3 0 0 $ \\ XI & $
0-1 0-1 3-2 $ & $-1-2 0-1 1 0 $ & $ 8 0 0 6 $ &$-5-3 3-6 $ &$ 0 6 0 0
$ \\ XII & $ 0 1 0 1 3-2 $ & $ \qquad \qquad .. \qquad \qquad $ & $ 8
0 0 6 $ &$-5-3 3-6 $ &$ 0 6 0 0 $ \\ XIII & $-1 0-1 0-1 3 $ & $-1 0 1
0 2 -3 $ & $ 0 0 0 0 $ &$ 8-5-5 7 $ &$-5 8 8-7 $ \\ XIV & $-1 0 0-1 2
1 $ & $ \qquad \qquad .. \qquad \qquad $ & $ 0 0 0 0 $ &$ 3 3-5 0 $ &$
0 0 8 0 $ \\ XV & $ 0-1 0-1 3-2 $ & $ \qquad \qquad .. \qquad \qquad $
& $ 8-5 0-7 $ &$-5 8-5 0 $ &$ 0 0 8 7 $ \\ XVI & $ 0 1 0 1 3-2 $ & $
\qquad \qquad .. \qquad \qquad $ & $ 8-5 0-7 $ &$-5 8-5 0 $ &$ 0 0 8 7
$ \\  \hline 
${\bf B B B }$ &&&&& \\  
I & b$-1-1 1-2 1-2 $ & $-1 1 0 1 2-1 $ & $ 0 6 3 3 $ &$-5-3 0-3 $ &$ 8 0 0
0 $ \\ II & $-1 2 0 1 2-5 $ & $ \qquad \qquad .. \qquad \qquad $ & $ 0
0 3-7 $ &$ 0 0 0 0 $ &$ 3 3 0 7 $ \\ III & $ 0-1 1-2 2-5 $ & $ \qquad
\qquad .. \qquad \qquad $ & $ 0 8 3 0 $ &$ 0-5 0 0 $ &$ 3 0 0 0 $ \\
IV & $ 0 1 1-1 1 1 $ & $ \qquad \qquad .. \qquad \qquad $ & $ 3 3 3 0
$ &$ 0 0 0 7 $ &$ 0 0 0-7 $ \\ V & $-2 5 0 1 1-2 $ & $ 0-1 1-1 2-1 $ &
$ 0 6 3 0 $ &$ 0-3 0-2 $ &$ 3 0 0 2 $ \\ VI & $-1-1-1 1 0 1 $ & $
\qquad \qquad .. \qquad \qquad $ & $ 3 0 3-2 $ &$ 0 6 0 2 $ &$ 0-3 0 0
$ \\ VII & $-1-1-1 2-1 2 $ & $ \qquad \qquad .. \qquad \qquad $ & $
0-3 3-6 $ &$-5 0 0 0 $ &$ 8 6 0 6 $ \\ VIII & $-1-1 1-2 1-2 $ & $
\qquad \qquad .. \qquad \qquad $ & $ 0-3 3-6 $ &$-5 0 0 0 $ &$ 8 6 0 6
$ \\ IX & $-1 1 0-1 1 1 $ & $ \qquad \qquad .. \qquad \qquad $ & $ 3 3
3 0 $ &$ 0 0 0 7 $ &$ 0 0 0-7 $ \\ 
X & $ 0-1 1-2 2-5 $ & $ \qquad
\qquad .. \qquad \qquad $ & $0 0 3-7 $ &$ 0 0 0 0$ & $ 3 3 0 7 $ \\
\hline  
\end{tabular}
\label{tablez3}
\end{center} 
\end{table}

\squeezetable
\begin{table}  
\begin{center}
\caption{\it Solutions  in the  broken  $SU(5)$  unified  symmetry     
case  with   three   non-parallel  stacks, $ a, \ b, c$ 
for the $Z_3$ orbifold  model  with lattices  ${\bf AAA }$ and 
${\bf BBB}$.  The  three  column entries  give 
the   three-cycles wrapping numbers in the order, 
$ [(n_\mu  ^1,  m_\mu  ^1), (n_\mu  ^2,  m_\mu  ^2), (n_\mu  ^3,
m_\mu  ^3)] $.  The  Cases I and II  refer to the 
fully and partially  constrained   searches.}  \vskip 0.5 cm
\begin{tabular}{|c|c|c||c|c|c|}    
\hline &&&&& \\ $ (n_a^I, \ m_a^I ) $ & $ (n^I_b, \ m^I_b ) $ & $
(n^I_c, \ m^I_c ) $ & $ (n^I_a, \ m^I_a ) $ & $ (n^I_b, \ m^I_b ) $ &
$ (n^I_c, \ m^I_c ) $ \\ \hline ${\bf A A A }\ (Case\ II) $ &&&&& \\
$-3 2-1 0-1 1$ &$-2-1 0-1 1 0$ &$-1-2 0-1 1 0$ & $ 0 1 1 0 2 1$ &$-3
2-1 0-1 1$ &$-1-2-1 0 0 1$ \\ $-1 3 1 0 1 0$ &$-3 2 1-1 1 0$ &$-1-2
0-1 1 0$ & $-1 0-1 0-1 3$ &$ 0-1 0-1 3-2$ &$-1 0 1 0 2-3$ \\ $ 0 1 0 1
3-2$ &$ 0-1 0-1 3-2$ &$-1 0 1 0 2-3$ &&&\\ \hline $ {\bf BBB }\ (Case
\ II)$ &&&&& \\ $-1-1 1-2 1-2$ &$ 0-1 1-2 2-5$ &$-1 1 0 1 2-1$ & $-1 2
0 1 2-5$ &$ 0-1 1-2 2-5$ &$-1 1 0 1 2-1$ \\ $-1 2 0 1 2-5$ &$ 0 1 1-1
1 1$ &$-1 1 0 1 2-1$ & $ 0 1 1-1 1 1$ &$ 0-1 1-2 2-5$ &$-1 1 0 1 2-1$
\\ $-2 5 0 1 1-2$ &$ 0-1 1-2 2-5$ &$ 0-1 1-1 2-1$ & $-1-1-1 1 0 1$
&$-2 5 0 1 1-2$ &$ 0-1 1-1 2-1$ \\ $-1-1-1 2-1 2$ &$ 0-1 1-2 2-5$ &$
0-1 1-1 2-1$ & $ 0-1 1-2 2-5$ &$-2 5 0 1 1-2$ &$ 0-1 1-1 2-1$ \\ $ 0-1
1-2 2-5$ &$-1-1-1 1 0 1$ &$ 0-1 1-1 2-1$ & $ 0-1 1-2 2-5$ &$-1 1 0-1 1
1$ &$ 0-1 1-1 2-1$ \\ \hline $ {\bf BBB }\ (Case\ I)$ &&&&& \\ $-1-1-1
1 0 1$ &$-2 5 0 1 1-2$ &$ 0-1 1-1 2-1$ & $-1-1-1 2-1 2$ &$-2 5 0 1
1-2$ &$ 0-1 1-1 2-1$ \\ $-1-1 1-2 1-2$ &$-2 5 0 1 1-2$ &$ 0-1 1-1 2-1$
& $ 0-1 1-2 2-5$ &$-2 5 0 1 1-2$ &$ 0-1 1-1 2-1$ \\ \hline
\end{tabular}
\label{tabz3scan}  
\end{center}
\end{table}
\vskip 0.5 cm

\squeezetable
\begin{table} 
\begin{center}
\caption{\it Solutions for the  $Z_3$ orbifold  model realizing the  flipped  $SU(5)$  
unified gauge  symmetry  model with   three   stacks, $ a, \ b, c$.}  
\vskip 0.5 cm
\begin{tabular}{|c|c|c||c|c|c|}    
\hline &&&&& \\ $ (n^I_a, \ m^I_a )$ & $ (n^I_b, \ m^I_b ) $ & $
(n^I_c, \ m^I_c ) $ & $ (n^I_a, \ m^I_a ) $ & $ (n^I_b, \ m^I_b ) $ &
$ (n^I_c, \ m^I_c ) $ \\ \hline ${\bf ABB}\ ( CASE \ I) $ &&&&& \\
$-2 1 1-2 1-1$ &$ 0-1-1 2 0-1$ &$ 1 0-1 2 0 1$ & $-2 1 1-2 1-1$ &$ 0
1-1 2 0 1$ &$ 1 0-1 2 0 1$ \\ $-1 1 1-2 1 0$ &$-1 1 0 1 0 1$ &$ 0-1
1-2 1-1$ & $ 0-1-1 0 1-1$ &$ 0-1-1 2 0-1$ &$ 1 0-1 2 0 1$ \\ $ 0-1-1 0
1-1$ &$ 0 1-1 2 0 1$ &$ 1 0-1 2 0 1$ & $ 1 0-1 3 1-1$ &$-1 1 0 1 0 1$
&$ 0-1 1-2 1-1$ \\ \hline ${\bf BBB}\ ( CASE \ II) $ &&&&& \\ $-1-1-1
1 0 1$ &$ 0-1 1-1 1 0$ &$-1 1-1 2 1-2$ & $ 0-1 1-2 2-5$ &$-1 1 0 1 1
0$ &$-1 2 0 1 0 1$ \\ \hline
\end{tabular}
\label{scanz3flip} 
\end{center} 
\end{table}


\vskip 1 cm \leavevmode

\vskip 1 cm \leavevmode 

\vskip 0.5 cm \leavevmode

\vskip 0.5 cm \leavevmode 

\vskip 0.5 cm \leavevmode 

\vskip 0.5 cm \leavevmode

\vskip 0.5 cm \leavevmode 

\begin{center}  
\begin{figure}  
\epsfxsize =6in
\epsffile{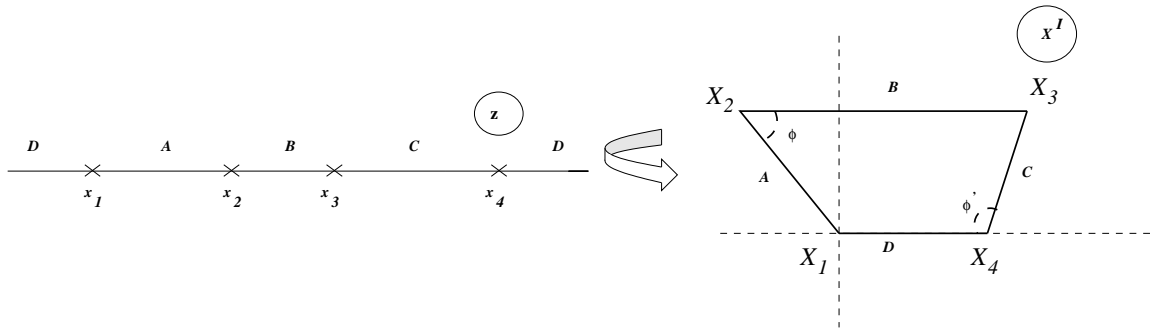}
\vskip 0.5cm
\caption{\it Embedding of the world sheet disk complex plane, $z$
(left),  in the 2-torus complex plane, $ X^I\in T^2_I$ (right). 
The    four-polygon edges, $ A,B,C, D$ are identified with the $D$-branes 
and vertices $ X_1, \cdots , X_4$ with their intersection
points $ X_i$.} 
\label{mapbr} \end{figure} \end{center}

\end{document}